%% file: main.tex
\documentclass[12pt,letterpaper]{lsuetd}
\usepackage{setspace,graphics,dsfont,verbatim,paralist,indentfirst}
\usepackage{graphicx}
\usepackage{amssymb}
\usepackage{comment}
\usepackage{amsmath}
\usepackage{csquotes}

\usepackage{epstopdf}
\usepackage{physics}
\usepackage{subfigure}
\usepackage{xcolor}
\usepackage{hyperref}
\hypersetup{
    colorlinks=true,
    citecolor=red,
    filecolor=black,
    linkcolor=black,
    urlcolor=black,
    linktoc=all
}

\makeatletter

\begin{document}
\renewcommand\@pnumwidth{1.55em}
\renewcommand\@tocrmarg{9.55em}
\renewcommand*\l@chapter{\@dottedtocline{0}{1.5em}{2.3em}}
\renewcommand*\l@figure{\@dottedtocline{1}{0em}{3.1em}}
\let\l@table\l@figure
\renewcommand{\bibname}{References}
\pagenumbering{roman}
\thispagestyle{empty}
\begin{center}

\textbf{\fontsize{15}{12} \selectfont SMART QUANTUM TECHNOLOGIES USING PHOTONS}

\vfill
\doublespacing
A Dissertation\\
\singlespacing
Submitted to the Graduate Faculty of the\\
Louisiana State University and\\
Agricultural and Mechanical College\\
in partial fulfillment of the\\
requirements for the degree of\\
Doctor of Philosophy\\
\doublespacing
in\\
                                       
The Department of Physics \& Astronomy\\
\singlespacing
\vfill

by\\
Narayan Bhusal\\
M.S., Louisiana State University, USA, 2018\\
B.Sc. \& M.Sc., Tribhuvan University, Nepal, 2013\\
May 2021
\end{center}
\pagebreak
\singlespacing
\tableofcontents
\pagebreak
\renewenvironment{abstract}{{\hspace{-2.2em} \huge \textbf{\abstractname}} \par}{\pagebreak}
\addcontentsline{toc}{chapter}{\hspace{-1.5em} ABSTRACT}

\begin{abstract}
\doublespacing
The technologies utilizing quantum states of light have been in the spotlight for the last two decades. The emerging field of quantum technology has attracted great attention of researchers from multiple disciplines. In this regard, quantum metrology, quantum imaging, quantum-optical communication are some of the important quantum technologies that exploit fascinating quantum properties like quantum superposition, quantum correlations, and nonclassical photon statistics. However, as of now, the state-of-art photonic technologies operating at the single-photon level are not efficient and fast enough to truly realize a reliable application. The major motivation of this Ph.D. dissertation is to explore faster, and smarter alternatives to the existing methods utilized in current photonic technologies.

In Chapter \ref{ch1}, I present a historical account of the light field and photon-based technologies. Furthermore, I discuss recent efforts and encouraging developments in the field of quantum-photonic technologies. In addition, I briefly discuss the major challenges for the experimental realization of reliable quantum technologies utilizing photons, setting up a stage for unveiling our smart methodologies to cope with them. Similarly, in Chapter \ref{ch2}, I review the fundamental concepts of quantum optics and photonics such as states of light, spatial modes of light, qubits, etc. In addition, I present a brief overview of artificial intelligence and machine learning techniques towards the end of the chapter.  

In Chapter \ref{ch3} of this dissertation, I present a theoretical work on a nonlinear quantum metrology scheme in which I lay out our efforts to improve the sensitivity of parameter estimation using the displaced-squeezed light. I show a sub-shot-noise limited phase estimation with both the parity and on/off detection strategy. In addition, I discuss a camera-based squeezed-light detection technique that can be a smart and time-efficient alternative to conventional balanced-homodyne detection. The proposed detection scheme is very simple and effortless compared to the conventional homodyne measurement. This will undoubtedly be beneficial to various future quantum technologies.
  
In Chapter \ref{ch4}, I discuss our efforts to incorporate artificial intelligence in a quest to improve the efficiency of discriminating thermal light sources from coherent light sources. The conventional identification technique requires a large number of measurements and yields a very limited efficiency. We utilize self-learning and self-evolving features of artificial neural networks to dramatically reduce the number of measurements required to distinguish the thermal light and coherent light. As a result, the amount of time and complexity of measurement substantially decreases which will be beneficial to a large variety of quantum technologies such as light detection and ranging (LIDAR), quantum microscopy, super-resolution imaging, etc. operating at the low-photon regime.

In Chapter \ref{ch5}, I present a communication protocol that utilizes the spatial modes of light. Despite being valuable resources for a wide variety of quantum technologies such as quantum communication, quantum imaging, quantum cryptography, quantum sensing, spatial modes of light are extremely vulnerable to the random phase fluctuation which presents a significant hurdle to the realistic implementation of these technologies. The effect of phase distortion is even more detrimental at the single-photon level. The conventional techniques to cope with these challenges include adaptive optics, optical nonlinearities, and quantum correlations. However, these techniques are very slow and relatively inefficient. I discuss our recent efforts to utilize self-learning and self-evolving features of convolutional neural networks to perform the spatial mode correction of single photons which results in a near-unity fidelity and a substantial boost in channel capacity of the communication channel.

I wrap up my dissertation in Chapter \ref{ch6} by summarizing the historical context of quantum technologies and the challenging problems facing state-of-art quantum technologies. In addition, I briefly discuss the importance of our efforts in introducing artificial intelligence to deal with these challenging problems.

\end{abstract}

\pagenumbering{arabic}
\addtocontents{toc}{\vspace{12pt} \hspace{-1.8em} CHAPTER \vspace{-1em}}
\singlespacing
\setlength{\textfloatsep}{12pt plus 2pt minus 2pt}
\setlength{\intextsep}{6pt plus 2pt minus 2pt}

\chapter{Introduction}
\doublespacing
\input{chapter1}

\pagebreak
\singlespacing
\chapter{Review of Fundamentals}
\doublespacing
\input{chapter2}

\pagebreak
\singlespacing
\chapter{Novel Theoretical Methods in Quantum Metrology}
\doublespacing
\input{chapter3}

\pagebreak
\singlespacing
\chapter{Smart Light Source Identification}
\doublespacing
\input{chapter4}

\pagebreak
\singlespacing
\chapter{Smart Spatial Mode Correction}
\doublespacing
\input{chapter5}

\pagebreak
\singlespacing
\chapter{Conclusion}
\doublespacing
\input{chapter6}

\pagebreak
\singlespacing
\addtocontents{toc}{\vspace{12pt}}
\addcontentsline{toc}{chapter}{\hspace{-1.6em} REFERENCES}
\bibliographystyle{vancouver}
\bibliography{main.bib}


\end{document}

%% file: chapter1.tex
\label{ch1}
Our current understandings of photons are the results of historically contentious and vigorous debates about the nature of light. In ancient times, scientists and philosophers characterized light as rays. Sir Issac Newton, in 1672, put forward his work on the famous corpuscular theory of light, which, however, failed to describe the prominent physical phenomena like refraction, diffraction, and interference \cite{newton1952opticks,sabra1981theories,feynman2006qed}. Soon after, in 1678, Christian Huygens proposed a theory that characterized light as waves, not particles \cite{houstoun1915treatise,achinstein1987light}. But in the 19th century, James Clerk Maxwell pacified these contentions with four famous equations (in 1862 but listed explicitly in 1865 in \enquote{A dynamical theory of the electromagnetic field}) that opened the door for treating light as an electromagnetic field \cite{maxwell1865viii,maxwell1996dynamical}. 

The notion of discretization of the electromagnetic field was started by Max Planck in 1900 with his theory of blackbody radiation \cite{planck2013theory,klein1961max,planck1901law}. However, Albert Einstein, in 1905, formalized the modern concept of photons as quantized packets of energy in his Nobel-prize-winning work on the photoelectric effect \cite{einstein1965concerning}. Neil Bohr in 1913 expanded on the idea of quantization which was remarkably successful and accurate in predicting the atomic spectra \cite{bohr1913constitution,aaserud2013love,whitaker2006einstein}. Then Werner Heisenberg, Erwin Schroedinger, Enrico Fermi, and Paul Dirac laid a solid foundation of modern quantum mechanics and the most important fundamental ideas like quantum state and unitary evolution in a short period before 1926 \cite{dirac1927quantum, gerry2005introductory, roychoudhuri2017nature}. Louis de Broglie, Max Born are lauded for their remarkable contribution in introducing the idea of quantum mechanical probability amplitude and wave-particle duality which led to a series of successful theoretical models to describe various physical systems and processes \cite{de1923waves, broglie1924xxxv,born1927physical,gerry2005introductory}. At the time, Albert Einstein was one of the biggest critics of quantum mechanics. His famous quote, “God does not play dice with the universe” is believed to have come in strong opposition to the probabilistic approach of quantum mechanics to describe the physical phenomena \cite{natarajan2008einstein}. Despite his opposition to the probabilistic description, Albert Einstein along with his two postdoctoral researchers Boris Podolsky and Nathan Rosen in 1935 introduced the idea of nonlocality and quantum entanglement through their seminal work entitled “Can Quantum Mechanical Description of Physical Reality Be Considered Complete?”, which is now famously known as EPR paper \cite{einstein1935can}. These works collectively inspired the development of new fields like quantum optics, quantum information, and quantum photonics. The development of the theory of quantum mechanics and quantum field theory furthered our understanding of photons and their intrinsic properties.

As the debates about photons and their peculiar properties began to subside, a new field of quantum optics and photonics emerged. Hanbury Brown and Twiss demonstrated intensity correlations of the two independent random light fields in 1956 which is now widely known as the HBT experiment \cite{brown1956correlation}. This invention is considered a breakthrough for quantum optics. The invention of the laser in 1960 by Theodore H. Maiman is a momentous event that revolutionized the field of quantum optics \cite{maiman1960stimulated}. Indeed, these inventions of mid 20\textsuperscript{th} century were followed by remarkable studies like quantum theories of coherence (1963 by Roy Glauber and ECG Sudarshan) \cite{glauber1963quantum, sudarshan1963equivalence}. A few years later, Emil Wolf, Leonard Mandel, and ECG Sudarshan followed up with their work on optical coherence and photodetection \cite{mandel1964theory, mandel1965coherence, mandel1995optical}, which made it possible to describe light in phase space. Hong-Ou-Mandel interferometry (1987) \cite{hong1987measurement} and Franson interferometry (1989) \cite{franson1989bell, franson1991two} are two other remarkable inventions that substantially advanced the field of quantum optics and photonics.

As a result of these remarkable works, we now know a lot more about photons beyond just the quanta of energies. With the turn of 20\textsuperscript{th} century, the focus is largely shifted towards developing the technologies utilizing the various properties of photons. The photons constitute key resources for a large variety of technologies ranging from imaging, sensing, and communication to cryptography and computing \cite{o2009photonic, pirandola2018advances, wang2020integrated, magana2013compressive}. In addition, photonic technologies now play a crucial role in a wide range of other fields like manufacturing, medical science, security, microelectronics, energy \cite{slusher1999laser, callaway2013introduction}. Historically, photons have always been the cornerstone of the majority of medical diagnostic imaging \cite{callaway2013introduction}. The photons also play an important role in laser printing, 3d scanning, radiotherapy \cite{haleem20193d}. Similarly, there are numerous other applications that utilize photons as the major resource. With the significant advancement in photon detection technologies towards the end of 20\textsuperscript{th} century, multiple quantum photonic technologies started to explode, which is growing even more rapidly now. Recognizing the importance of light in our lives, UNESCO declared the year 2015 as a \enquote{year of light} and light-based technologies \cite{daukantas20152015}. 

Quantum metrology, quantum sensing, quantum communication, and quantum computation are some of the fascinating examples of quantum technologies that utilize photons, often at an ultra-low intensity or single-photon level. More precisely, the majority of quantum technologies are designed to exploit the intriguing properties of quantum states of light such as quadrature squeezing, quantum superposition, and quantum entanglement. In this regard, quantum metrology aims to push the limits of measurement precision utilizing the reduced quantum noise and/or quantum entanglement \cite{giovannetti2011advances, birchall2020quantum, you2020multiphoton, quiroz2019exceptional}. Similarly, in quantum sensing, a quantum state is used to probe an object. The quantum sensors utilize quantum mechanical resources such as quantum coherence and quantum entanglement to detect the properties of an object beyond our classical capabilities \cite{degen2017quantum}. Schemes for quantum cryptography such as quantum key distribution (QKD) exploits the quantum mechanical properties of photonic states \cite{gerry2005introductory, liao2017satellite}. Moreover, quantum-optical communication utilizes the non-classical properties of photons to establish a very secure link between Alice and Bob \cite{acin2018quantum}. To enable such communication protocols, there have been some efforts towards developing a large-scale global quantum network like quantum internet \cite{khatri2021spooky, liao2017satellite,yin2017satellite, liao2018satellite}. Similarly, Quantum computing utilizes the quantum superpositions instead of 0 (low) and 1 (high) in classical computers. In the last two decades, significant strides have been made towards all-optical quantum computing, which utilizes photonic qubits, and linear optical elements \cite{o2007optical, o2003demonstration}.  Even though the photonic approach hasn't been as successful as superconducting qubits \cite{cho2020ibm, arute2019quantum}, and trapped ions \cite{wright2019benchmarking}, the recent demonstration of \enquote{quantum supremacy} through boson sampling experiment has once again brought the photonic quantum computing and other photonic applications to the spotlight \cite{zhong2020quantum}.


Nevertheless, these technologies depend on our ability to efficiently prepare, control, and measure quantum states or quantum bits (qubits). A photonic qubit can be prepared using one or more degrees of freedom such as spatial modes, spin, orbital angular momentum, time, energy, etc. Despite having tremendous potential, quantum photonic technologies suffer from low preparation efficiency, low detection efficiency, and high error rates. Quantum state preparation, quantum control, and quantum measurements are crucial components of any quantum photonic technology that require significant improvements. In other words, the performances of current state-of-art technologies to build these components are not robust enough to realize a truly reliable quantum photonic technology. There have been numerous efforts to cope with these challenging problems. In the last two decades, these efforts have substantially advanced quantum photonic technologies \cite{slussarenko2019photonic, o2009photonic}.

First of all, preparing photons in desired quantum states is an important task in implementing quantum technologies. Ideally, we want the photonic state preparation process to be deterministic and precisely on-demand with great accuracy. However, the current state-of-art technologies to prepare photonic qubits are far from perfect. Moreover, the challenges regarding the realization of deterministic multi-qubit gates using photons are considered a major hurdle \cite{slussarenko2019photonic}. Improving the preparation of quantum states largely depends on engineering innovations. For example, the efficiency of preparing or initializing a qubit in the Nitrogen-vacancy center heavily depends on the purity of the diamond samples. The presence of impurities and other isotopes of Carbon in the sample causes the rapid decay of the qubits \cite{liu2018quantum, acosta2013nitrogen}. Similarly, generating highly efficient single-photon photon sources with the perfect correlations is a challenging task. In addition to the nonlinear crystals such as ppKTP, there are multiple alternative techniques developed to generate single-photon sources: quantum dot, silicon carbide, Rydberg atoms, polarized microcavities \cite{senellart2017high, reimer2019quest, wang2018bright, wang2019towards}. Lately, some efforts have been devoted to looking into the optimization of quantum circuits to generate desired photonic states \cite{arrazola2019machine}. Secondly, photonic quantum states are vulnerable to effects like decoherence, loss, scattering, cross-talk or interference, etc. Preserving the coherence of a prepared quantum state for a long time is undoubtedly one of the most challenging tasks. The simplest way to deal with this problem would be to isolate qubits from unwanted interactions, but it is not simple. Cooling a system to low cryogenic temperatures is one of the methods that has been used in quantum computing platforms. Similarly, dynamical decoupling is utilized to correct the decoherence of spin qubits. Furthermore, quantum information encoded in the spatial degree of freedom has been conventionally restored using methods like adaptive optics, nonlinear processes, and quantum correlations. Last but not least, measuring quantum states of light is a crucial task of any quantum-photonic technology. In addition to preparing and preserving quantum states, our ability to make measurements on them actually enables us to harness the true potential of quantum states. In other words, the advantages that quantum states have to offer over the classical ones depend on how well we can measure them. Conventionally, making measurements on quantum states of light involves a large number of measurements \cite{slussarenko2019photonic}. The characterization of quantum state and reconstruction of the corresponding density matrix is a resource-intensive and cumbersome process \cite{daniel:2001, weinbub2018recent}.

In addition to all the developments in preparing, controlling, and measuring quantum states of light, the $21^\text{st}$ century has already seen an unprecedented level of growth in our computational capabilities. An important question is, are we fully utilizing available computing power to enhance the performance of quantum photonic technologies? Is this the maximum we can achieve with quantum technologies using state-of-art devices? The simple answer is, no. Lately, artificial intelligence, machine learning, in particular, has attracted attention as an excellent alternative to conventional techniques to improve the performance of quantum technologies that operate at the single-photon level \cite{dunjko:2018,mohseni2017commercialize}. In this regard, neural networks are one of the widely used models of machine learning algorithms \cite{torlai2018neural, carleo2019machine, dunjko:2018}. Neural networks have remarkable abilities to draw inferences from a large volume of data without knowing the theoretical models.  These self-learning features make them very suitable to handle complex data structures. Most importantly, these smart algorithms have outperformed conventional techniques for the generation, control, and measurement of quantum states of light. In addition, the introduction of artificial intelligence in the technologies designed to improve the existing technologies has been proven to be remarkably successful \cite{gao:2018, torlai2018neural, krenn2020computer}. This is the major motivation of our efforts towards the smartification of quantum technologies in part or full.  

The contents of this dissertation are organized as follows. In Chapter \ref{ch1}, I presented a historical context of photons, photonic technologies, and quantum mechanics in general. In addition to the historical context, I discuss our motivations for incorporating smart algorithms such as machine learning in various quantum technologies utilizing photons as the major resources. In Chapter \ref{ch2}, I review the fundamentals of quantum optics and discuss in detail various quantum states of light. Since I utilize the spatial modes of light to encode the quantum information in this thesis, I also review orbital angular momentum (OAM) and other spatial modes of light namely the Laguerre-Gaussian and Hermite-Gaussian modes. Finally, I discuss fundamental ideas of machine learning and artificial intelligence and their application in various quantum photonic technologies. In Chapter \ref{ch3}, I begin by discussing our recent efforts to improve quantum metrology and detection of squeezed light. In the former, we utilize displaced-squeezed states of light as inputs and novel detection strategies. In the latter, we exploit quantum correlations on high-efficiency camera images to measure the squeezing strength of squeezed light. This represents a time-efficient and smart alternative to the tedious balanced-homodyne detection technique. Similarly, in Chapter \ref{ch4}, I discuss our efforts to introduce a smart quantum technology that efficiently discriminates different sources of light based on their intrinsic photon statistics. This technology utilizes neural networks to identify light sources attenuated to the single-photon level. Our machine learning based approach dramatically reduces the number of measurements required to discriminate light sources by orders of magnitude. Likewise, in Chapter \ref{ch5}, I introduce another smart quantum technology that exploits the self-learning and self-evolving features of artificial neural networks to correct the distorted spatial profile of single photons. Our technique to correct spatial modes of single photons substantially improves the amount of information that can be exchanged in an optical communication protocol. Last but not least, in Chapter \ref{ch6}, I conclude my Ph.D. dissertation by summarizing our efforts to introduce and/or improve various smart quantum technologies.

%% file: chapter2.tex
\label{ch2}
Photons, elementary excitation modes of a quantized electromagnetic field, have been used as an information carrier, sensing probe, imaging resource. There are various states of light with unique intrinsic characteristics that can be defined either classically or quantum-mechanically. Recently, there have been significant efforts to utilize non-classical states of light to improve the performance of existing photonic technologies. Before we get into the specifics of various quantum photonic technologies in Chapter \ref{ch3}, \ref{ch4}, and \ref{ch5}, this chapter introduces various states of light, and spatial modes. Toward the end of the chapter, I discuss the basics of machine learning and artificial intelligence.

\section{States of Light}
\begin{figure}[ht!]
    \centering
    \includegraphics[width=0.75\textwidth]{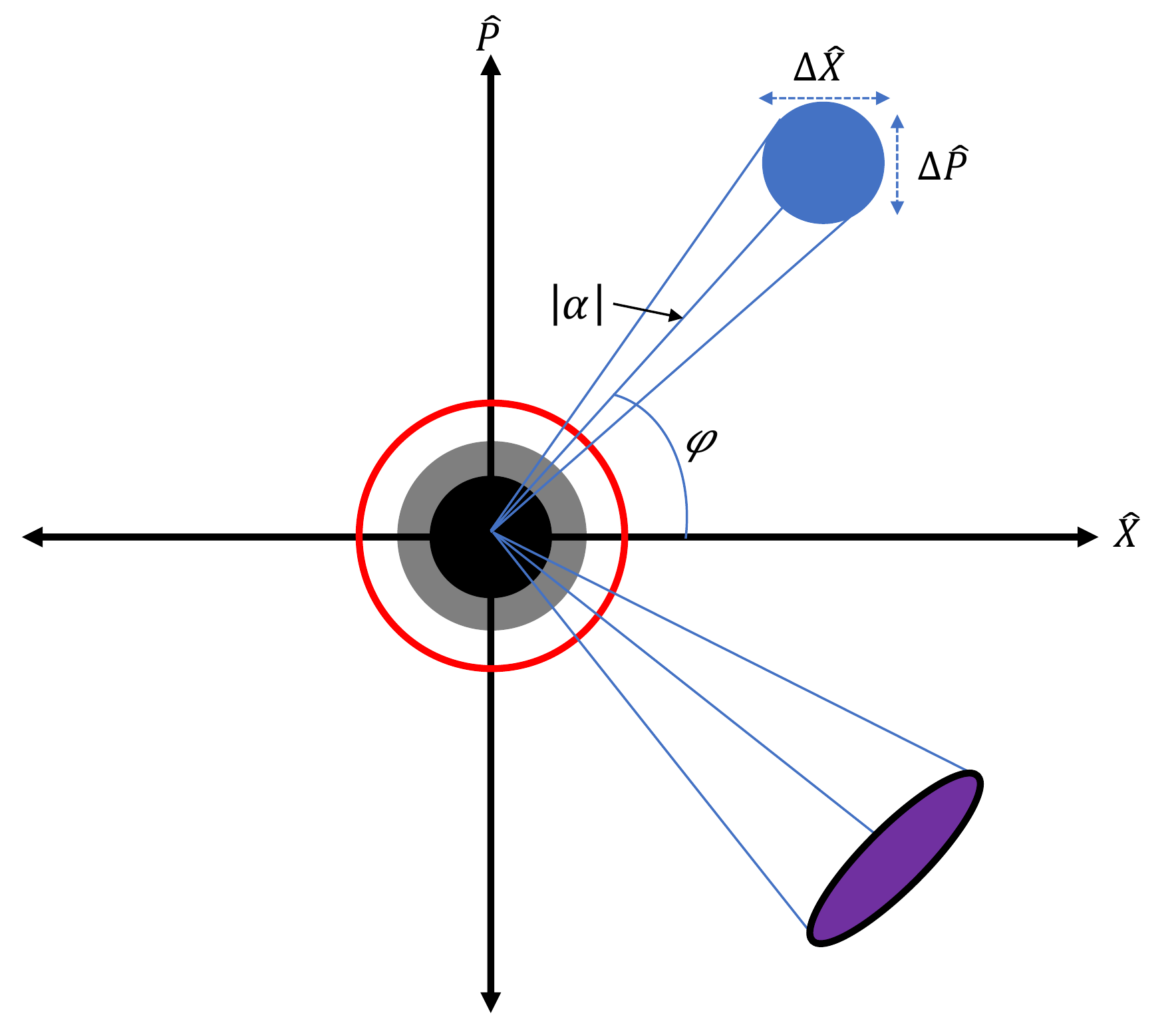}
    \caption{Phase-space representation of various states of the electromagnetic field. The shaded area denotes the possible combinations of quadrature values for a particular state of light. The extent of these shaded circles/ellipses along $\hat{X}$ and $\hat{P}$ axes denote the quadrature uncertainties of the position and momentum quadrature respectively.  The black circle at the origin represents the vacuum state. Similarly, the gray circle centered on origin represents a thermal state. Likewise, the red circle (not the area inside) depicts the possible quadrature values for a particular Fock state $\ket{n}$. Coherent states are represented by the blue circle. The purple ellipse represents a displaced-squeezed vacuum (DSV) state.}
    \label{ch2_phaseD}
\end{figure}

The concept of the electromagnetic field began with the four famous Maxwell's equations. The quantization of the electromagnetic field was a paradigm shift. As a result, the light field can now be treated as a quantum mechanical state as well as an operator in a Hilbert space of finite or infinite dimension \cite{maxwell1865viii, gerry2005introductory}. There are several different ways to represent the states of light in quantum optics. The most common representations include density matrix, Wigner function, state vector $\ket{\psi}$, photon statistics, quadrature mean and variance, etc. In this dissertation, I will be extensively utilizing the continuous variable phase-space representation with position ($\hat{X}$) and momentum ($\hat{P}$) quadrature, as shown in Figure \ref{ch2_phaseD}. These quadratures represent the electric field amplitude oscillations in orthogonal phase, i.e. $\hat{E}(t)=E_0[\hat{X}\cos(\omega t) + \hat{P}\sin(\omega t)]\sin(kz)$, where $\omega$ and $z$ represent the angular frequency and propagation axis. The quadrature operators satisfy the commutation relation $[\hat{X},\hat{P}]=\frac{i}{2}$. In addition, the Wigner function representation of states of light and their photon fluctuation properties are used in this thesis. Therefore, I will briefly discuss various states of light along with their Wigner functions, photon statistics, and quadrature fluctuations. Let's begin by defining the quadrature operators in terms of the creation ($\hat{a}^\dag$) and annihilation ($\hat{a}$) operators of the electromagnetic field. The creation and annihilation operators are also called raising and lowering operators as they mathematically increase or decrease a photon from a single-mode field. These mode operators satisfy the standard commutation relation, $[\hat{a},\hat{a}^{\dagger}]=\hat{\mathds{1}}$.
\begin{equation}
    \hat{X}=\frac{\hat{a}+\hat{a}^\dag}{2} \hspace{5mm} \textrm{and} \hspace{5mm} \hat{P}=\frac{\hat{a}-\hat{a}^\dag}{2i}
\end{equation}
The Heisenberg uncertainty principle dictates that the variance of quadrature fluctuations in a single-mode field has to be at least $\frac{1}{16}$.
\begin{equation}
    \langle(\Delta \hat{X})^2 \rangle  \langle(\Delta \hat{P})^2 \rangle \geq \frac{1}{16}
\end{equation}
Therefore, it is possible to squeeze the uncertainty on one of the quadratures, but only at the expense of other quadrature. Squeezing of a quadrature below the threshold of classical states of light and negativity of the quasi-probability function, Wigner function \cite{wigner1997quantum}, are some of the key signatures of \enquote{quantumness}.

\subsection{Fock State}
\begin{figure}[ht!]
    \centering
    \includegraphics[width=1.0\textwidth]{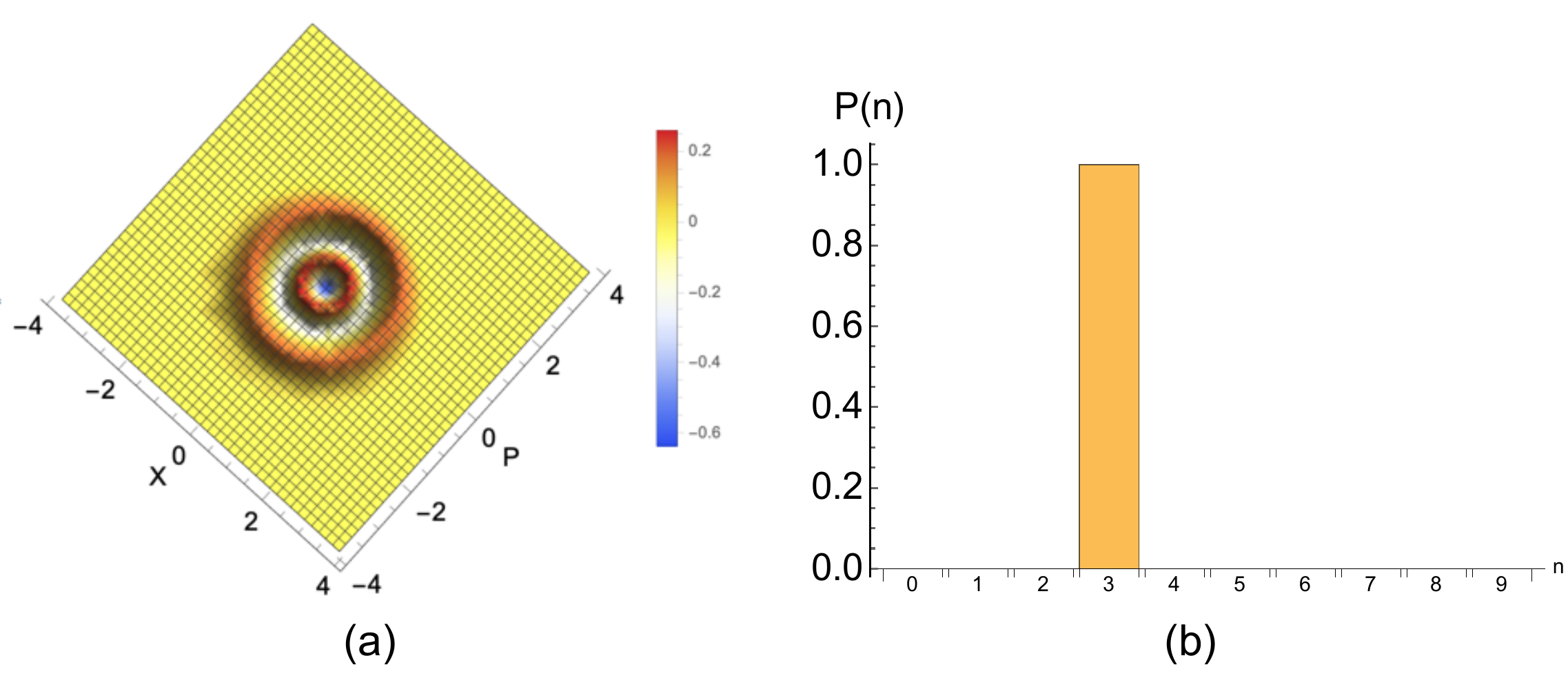}
    \caption{Fock state. (a) The Wigner function of Fock state for $n=3$. The Fock states of non-Gaussian Wigner function excent for $n=0$. The negative values of Wigner function indicates the non-classicality of the state. (b) The photon statistics for $\ket{n=3}$ Fock state. This shows that we measure three photons 100\% of the time, indicating no photon fluctuation.}
    \label{ch2_fock}
\end{figure}
Fock state, denoted by $\ket{n}$, is considered one of the interesting quantum states of light for various reasons. First of all, these are the eigenstates of the optical field Hamiltonian $\hat{H}$. Secondly, Fock states are the definite photon number states with no fluctuation in photon number. Moreover, these states yield negative Wigner functions except for the vacuum state, $n=0$, as shown in Figure \ref{ch2_fock}(a). For a single mode field, the Hamiltonian is expressed as $\hat{H}=\hbar\omega(\hat{a}^\dag\hat{a}+1/2)$, where $\hat{a}^\dag\hat{a}$ is the photon number operator $\hat{n}$. Fock states are eigenstates of the number operator i.e. $\hat{n}\ket{n} = n\ket{n}$ for all integer values $n \in [0,\infty]$. It is important to note that even though it is a definite photon number state, the product of quadrature variances is large: $\langle(\Delta \hat{X})^2 \rangle  \langle(\Delta \hat{P})^2 \rangle = \frac{1}{16}(2n+1)^2$. The minimum uncertainty Fock state is actually the vaccuum state $\ket{n=0}$. Individually, both the quadratures have equal uncertainties, $\langle \Delta \hat{X} \rangle= \langle \Delta \hat{P} \rangle = \frac{\sqrt{2n+1}}{2}$, as shown in the phase diagram, Figure \ref{ch2_phaseD}. The Wigner function for Fock state $\ket{n}$ is given by,
\begin{equation}
    W_\textrm{Fock}(x,p,n)=\frac{2}{\pi}(-1)^n  L_n\left[4 \left(x^2+p^2\right]\right) e^{-2 \left(x^2+p^2\right)},
\end{equation}
which is plotted in Figure \ref{ch2_fock}(a) for $n=3$. The photon statistics ($P_n$), which describes the probability of detecting $n$ photons, is depicted in Figure \ref{ch2_fock}(b). 

\subsection{Vacuum State}
The vacuum state is a very interesting quantum state from a physical perspective. This is one of the important discoveries in the field of quantum optics. This state is very counter-intuitive since it has no photons but the quadrature fluctuation and fluctuation of electric field amplitude is non-zero, which is known as vacuum fluctuations. In fact, vacuum fluctuation is one of the hurdles that limit measurement sensitivities in quantum metrology. Therefore, scientists are looking for ways to replace the vacuum input of the optical interferometer with the states having reduced quadrature noise. I will discuss one of such metrology schemes in Chapter \ref{ch3}. The most interesting fact about the vacuum state is every other state of light reduces to the vacuum state in certain limiting conditions. Indeed, a Fock state reduces to the vacuum state when $n=0$, a coherent state reduces to the vacuum state when $\alpha=0$, a thermal state reduces to the vacuum state when $n_\textrm{th}=0$, and a squeezed vacuum state reduces to the vacuum state when $r=0$. Therefore, we can safely say that the vacuum state is the most fundamental state of light.
\begin{figure}[ht!]
    \centering
    \includegraphics[width=1.0\textwidth]{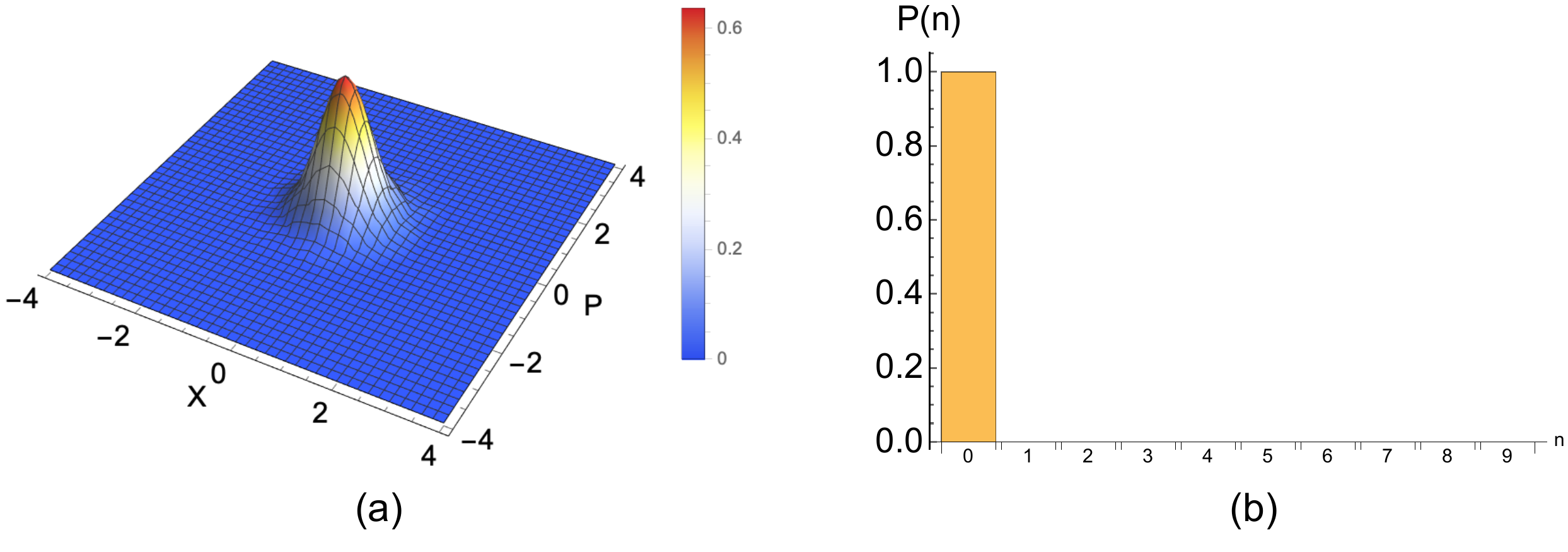}
    \caption{Vacuum state. (a) The Wigner function of the vacuum state. The vacuum state has a Gaussian Wigner function with only positive values. (b) The histogram shows the deterministic photon statistics with $\ket{0}$ measured 100\% of the time.}
    \label{ch2_vac}
\end{figure}
The Wigner function of the vacuum state is given by,
\begin{equation}
    W_\textrm{vac}(x,p)=\frac{2}{\pi} e^{-2 \left(p^2+x^2\right)},
\end{equation}
which is plotted in Figure \ref{ch2_vac}(a). The Figure \ref{ch2_vac}(b) shows the photon statistics of the vacuum state. The figure shows that no photons are detected in the field.

\subsection{Coherent State}
A coherent state is a quantum state which exhibits classical statistical properties, which is why it is regarded as a borderline state between the classical and quantum. The coherent state is an eigenstate of annihilation operator $\hat{a}$. The existence of such a basis for $\hat{a}$ can be inferred from the fact that the expectation value of $\hat{a}^\dag\hat{a}$ vanishes for any photon number state regardless of its magnitude. Assuming the coherent state amplitude $\alpha= \abs{\alpha}\exp(i\varphi)$, the eigenfunction equation for the coherent state can be written as, 
\begin{equation}
    a\ket{\alpha}=\alpha\ket{\alpha},
\end{equation}
where $\varphi$ represents the phase angle of the coherent state. This equation is also valid for the creation operator with eigenvalues of $\alpha^*$, i.e. $\bra{\alpha}a^\dag=\alpha^*\bra{\alpha}$.
\begin{figure}[ht!]
    \centering
    \includegraphics[width=1.0\textwidth]{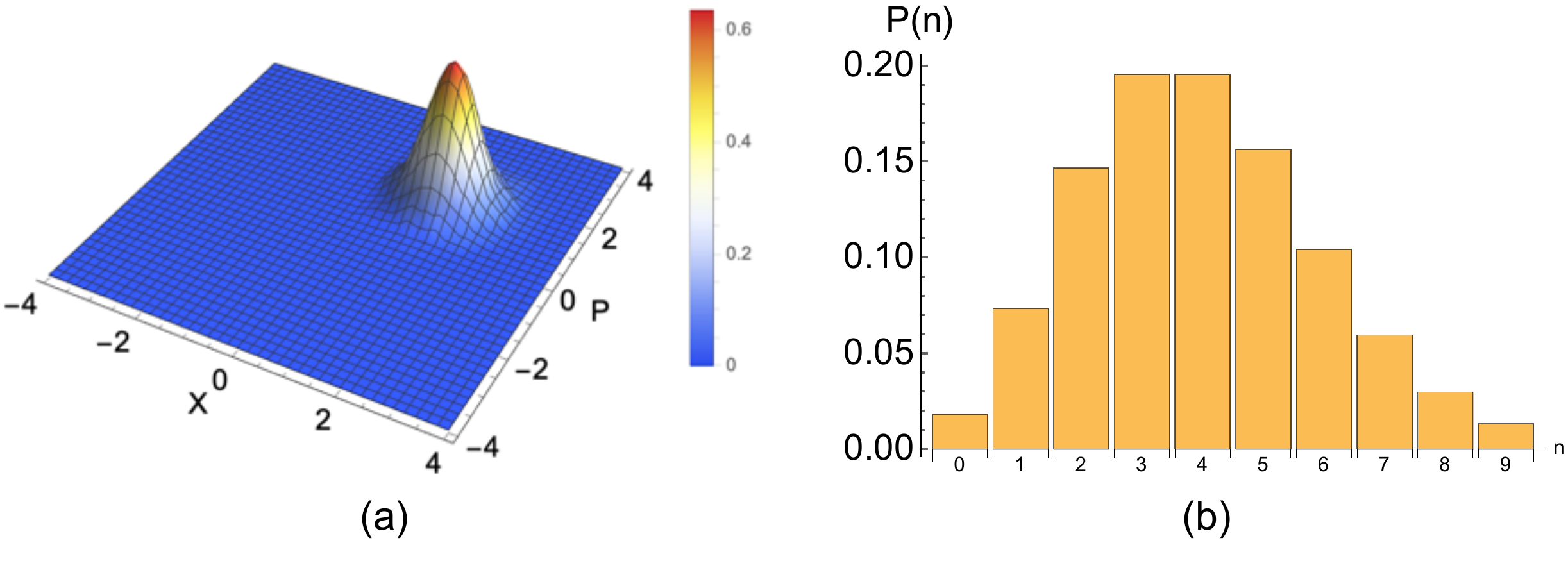}
    \caption{Coherent state. (a) The Wigner function of the coherent sate having amplitude $\alpha=2$  and phase $\varphi=\pi/4$. The Wigner function is Gaussian in shape and is strictly positive. The Wigner function of this coherent state can also be  obtained by displacing the vacuum state Wigner function by $\abs{\alpha}^2$ along the phase angle $\varphi=\pi/4$ in the phase space. (b) The photon statistics of coherent light follows Poissonian distribution with mean $\abs{\alpha}^2$ and variance $\abs{\alpha}^2$.}
    \label{ch2_coh}
\end{figure}
Coherent states can also be obtained by displacing the vacuum in the phase space. Therefore, coherent states are sometimes referred to as a displaced vacuum states. Coherent states, therefore, are sometimes written as following
\begin{equation}
	\ket{\alpha}= \hat{D}(\alpha)\ket{0},
\end{equation}
where $\hat{D}(\alpha)= \exp(\alpha \hat{a}^{\dagger} - \alpha^{*}\hat{a})$ represents the displacement operator. As the eigenstates of hamiltonian operator ($\hat{H}$ of the electromagnetic field, it is desirable to express coherent states in terms of photon number states or Fock states becuase it provides us a clearer physical picture and, sometimes, simpler mathematical forms. The Fock state representation of the coherent state is given by,
\begin{equation}
	\ket{\alpha} = \sum_{n=0}^{\infty} \frac{e^{-|\alpha|^2/2 }\alpha^n}{\sqrt{n!}} \ket{n}.
\end{equation}
Similarly, the Wigner function of the coherent state is Gaussian in shape. Therefore, it is categorized as a Gaussian state. The coherent sate Wigner function given by
\begin{equation}
    W_\textrm{coh}(x,p,\alpha,\varphi)=\frac{2}{\pi} e^{-2 \left[(x-\alpha  \cos \varphi)^2 + (p-\alpha  \sin \varphi )^2\right]}
\end{equation}
is plotted in Figure \ref{ch2_coh}(a). Experimentally, coherent states are produced by a laser which are characterized by Poissonian photon statistics,
\begin{equation}
	P_\textrm{coh}(n) = \frac{(|\alpha|^{2})^n}{n!}e^{-|\alpha|^2 }.
\end{equation}
The photon statistics of the coherent state is shown in Figure \ref{ch2_coh}(b), which is centered around $\abs{\alpha}^2$ with the standard deviation of $\abs{\alpha}$.

\subsection{Thermal State}
\begin{figure}[ht!]
    \centering
    \includegraphics[width=1.0\textwidth]{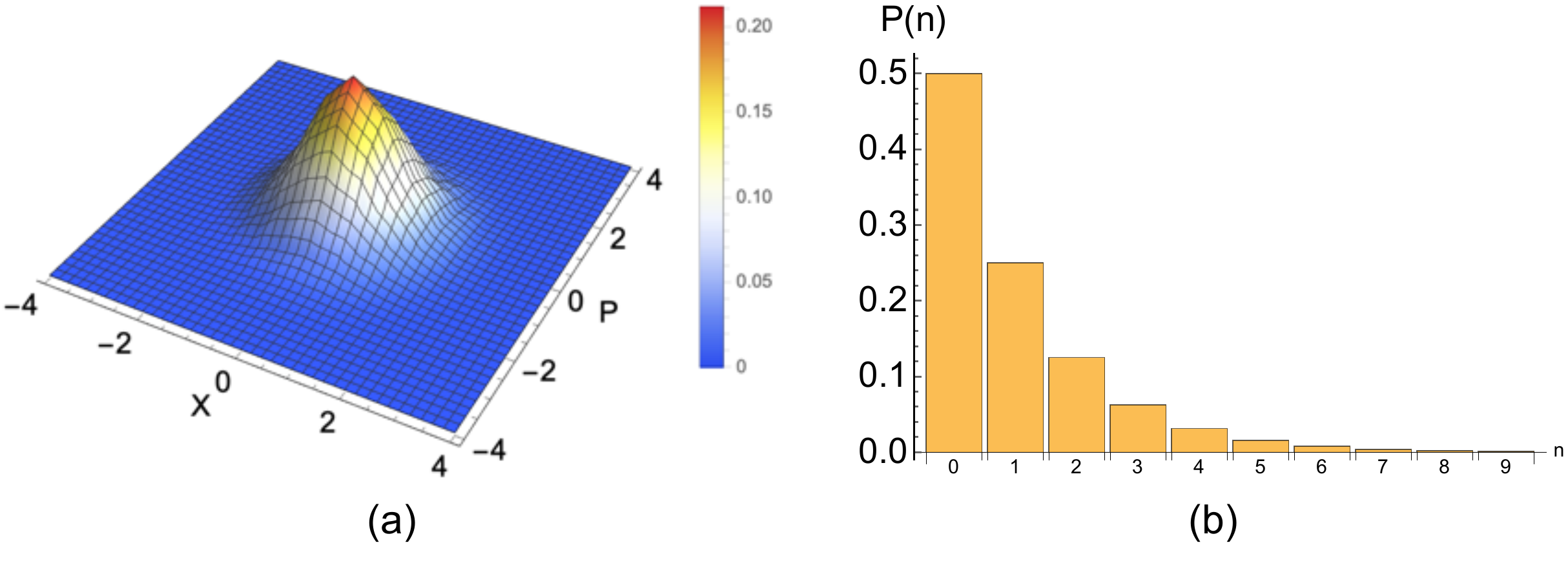}
    \caption{Thermal state. (a) The Wigner function for thermal state with mean photon number $\Bar{n}_{th}=1$. The Wigner function of thermal state is Gaussian in shape with strictly non-negative values. This function is wider than minimum uncertainty states such as vacuum state and coherent state. However, the Wigner function of thermal state reduces to the vacuum state Wigner function for $\Bar{n}_{th}=0$. (b) The photon statistics of thermal light is super-Poissonian (variance greater the mean photon number $(\Delta n)^2 > \bar{n}_\textrm{th}$).}
    \label{ch2_therm}
\end{figure}
A thermal state is a statistical mixture of thermal radiations. The most common example of thermal light is sunlight. In most quantum optical and quantum photonic applications, thermal radiation act as a noise, limiting the performance of various photonic technologies. A thermal state is a mixed state that can not be represented as a superposition of the pure states. Therefore, we express the thermal state as a density operator in the Fock basis
\begin{equation}
	\hat{\rho}_\textrm{th} = \frac{1}{1+\bar{n}_\textrm{th}} \sum_{n=0}^{\infty} \Big( \frac{\bar{n}_\textrm{th}}{1+\bar{n}_\textrm{th}} \Big)^n |n \rangle \langle n|,
\end{equation}
where $\Bar{n}_\textrm{th}$ denotes the mean photon number in the thermal light field. The mean photon number can be expressed in terms of temperature ($T$) as $\frac{1}{e^{\hbar \omega/k_B T}-1}$ with the photon fluctuation of $ \Delta n= \sqrt{\Bar{n}_\textrm{th}+\Bar{n}_\textrm{th}^2}$, where $\hbar$, $k_B$, and $\omega$ denote the reduced Planck constant, Boltzmann constant, and frequency of light, respectively. The photon statistics of thermal light is super-Poissonian, meaning the photon variance is higher than the mean photon number. The photon statistics of the thermal light is given by
\begin{equation}
	P_\textrm{th}(n) = \frac{\bar{n}_\textrm{th}^n}{(1+\bar{n}_\textrm{th})^{n+1}} = e^{-n\hbar\omega/K_B T}\left[1-e^{-\hbar\omega/K_B T}\right],
\end{equation}
which is depicted in Figure \ref{ch2_therm}(b) for $\Bar{n}_\textrm{th}=1$. This distribution is also called Bose-Einstein distribution. Similarly, the expression of the Wigner function is given by
\begin{equation}
    W_\textrm{th}(x,p,\Bar{n}_\textrm{th})=\frac{2}{\pi (2 \Bar{n}_\text{th}+1)} e^{-\frac{2 \left(x^2+p^2\right)}{2 \Bar{n}_\text{th}+1}},
\end{equation}
which is shown in Figure \ref{ch2_therm}(a) for $\Bar{n}_\textrm{th}=1$. The Wigner function is non-negative and Gaussian in shape.

\subsection{Squeezed Vacuum State}
\begin{figure}[ht!]
    \centering
    \includegraphics[width=1.0\textwidth]{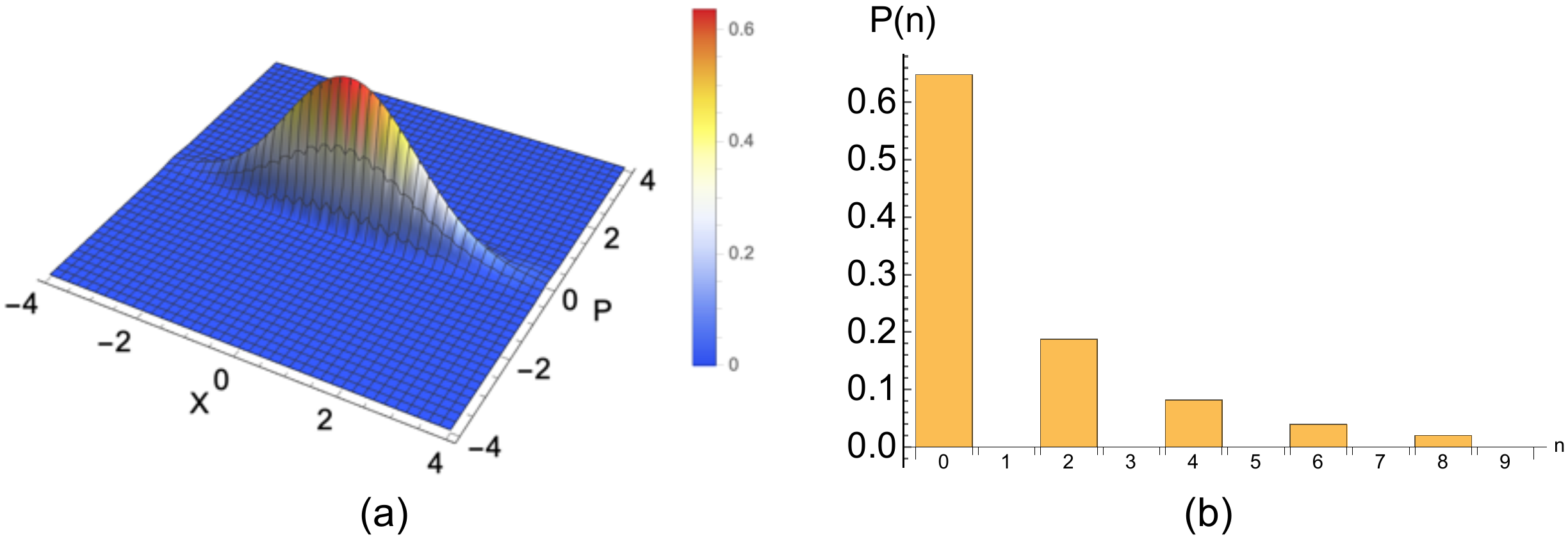}
    \caption{Squeezed vacuum (SV) state. (a) The Wigner function of a squeezed vacuum state having squeezing parameter $r=1$ and squeezing angle $\theta=\pi$. The Wigner function is of Gaussian shape with one of the sides squeezed, indicating the reduced quadrature noise along $\hat{P}$ quadrature (this particular case). (b) The photon statistics of SV state. The photon statistics shows that only even-photon events are permissible.}
    \label{ch2_sqz}
\end{figure}
Squeezed vacuum (SV) state is one of the most important non-classical states of light. The key feature of this kind of light is the unequal quadrature fluctuation $\Delta \hat{X}\neq \Delta \hat{P}$. As the name suggest, it is indeed a squeezed version of the vacuum state of light. There are different kind of squeezed states such as quadrature squeezed light, number squeezed light (has sub-Poissonian statistics), phase squeezed light \cite{gerry2005introductory}. Mathematically, squeezed states $\ket{\xi}$ are generated by applying the squeezing operator $\hat{S}(\xi)$ on vacuum state.
\begin{equation}
\ket{\xi} = \hat{S}(\xi) \ket{0}=
\textrm{exp}\Big[\frac{1}{2}(\xi^*\hat{\textrm{a}}^2-\xi\hat{\textrm{a}}^{\dagger 2})\Big]\ket{0},
\end{equation}
where $\xi = r e^{i \theta}$ represents the squeezing amplitude with phase angle $\theta$.
Experimentally, squeezed light is produced using nonlinear interactions, optomechanical interactions, four-wave-mixing in Rb vapour, etc. Interestingly, it was shown that two phase-matched single-mode squeezed vacuum states mixed in a 50:50 beamsplitter generates highly entangled two-mode states called two-mode squeezed-vacuum (TMSV). However, the TMSV reduces to a thermal state having mean photon number $\sinh^2r$ if one of the modes of TMSV is traced out. The effective temperature of such a thermal source can be shown to be $\frac{\hbar \omega}{2k_B\ln(\coth{r})}$. The squeezed state can be represented as a infinite superposition of the Fock states as
\begin{equation}
	\ket{\xi=re^{i\theta}} = \frac{1}{\sqrt{\cosh{r}}} \sum_{n=0}^{\infty} (-1)^n \frac{\sqrt{{(2n)!}}}{ 2^{n}n!} e^{i n \theta} (\tanh{r})^n|2 n \rangle,
\end{equation}
which can be used to derive the expression for photon statistics of the state. The probability of finding the odd number of photons in SV field is zero as shown in Figure \ref{ch2_sqz}(b). Therefore, the following expression for photon statistics is valid only for even values of $n$.
\begin{equation}
    \textrm{Only for even values of $n$: } P_\textrm{sv}(n)=\frac{n! \tanh^nr}{2^n \left(\frac{n}{2}!\right)^2 \cosh r}.
\end{equation}
Similarly, the Wigner function for squeezed state is given by expression
\begin{equation}
    W_\textrm{sv}(x,p,r,\theta)=\frac{2}{\pi} e^ {-2 \left[\left(x^2+p^2\right) \cosh (2 r)+\left\{ \left(x^2-p^2\right)\cos\theta+2 x p \sin\theta \right\} \sinh (2 r) \right]},
\end{equation}
which is plotted in Figure \ref{ch2_sqz}(a). The Wigner function of SV state is Gaussian in shape with one of the quadrature noise reduced below $\frac{1}{2}$, which is the classical limit.

\subsection{Displaced Squeezed State}
\begin{figure}[ht!]
    \centering
    \includegraphics[width=1.0\textwidth]{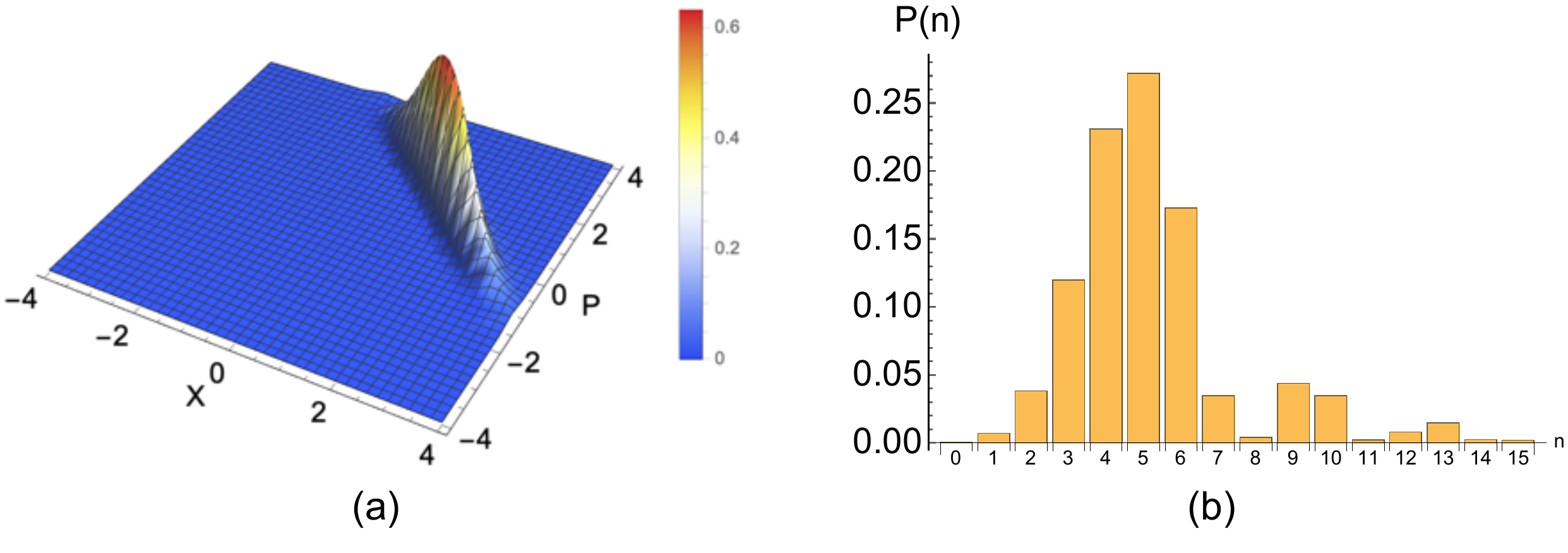}
    \caption{Displaced squeezed vacuum state (DSV). (a) The wigner function of DSV having parameters $\alpha=2$, $\varphi=\pi/4$, $r=1$, and $\theta=\pi/2$. This state also has a Gaussian Wigner function which is displaced from origin by distance $\abs{\alpha}^2$ along phase angle $\varphi$ in phase space. The amount of squeezing and the direction of squeezing is defined by squeezing parameter $r$ and squeezing angle $\theta$. (b) The photon statistics of DSV state. The photon statistics is centered around the Poissonian distribution for given value of $\alpha$, but the photon statistics contains rapid oscillations towards the tail due to the asymmetrical or squeezed quadrature fluctuations. The photon statistics of DSV stae can be sub-Poissonian ($(\Delta n)^2 < \bar{n}$) and super-Poissonian ($(\Delta n)^2 > \bar{n}$) depending on the values of phase angles $\varphi$, and $\theta$.}
    \label{ch2_dsqz}
\end{figure}
As the name suggests, displaced-squeezed vacuum (DSV) states are obtained mathematically by applying the displacement and squeezing operator sequentially on the vacuum state. Experimentally, this operation can be realized by displacing the squeezed-vacuum state by a local oscillator (coherent state) in a beamsplitter. Unlike squeezed vacuum states, the DSV states are bright. But they possess the squeezed quadrature noise like squeezed vacuum states. It is represented by the the purple ellipse in the phase space diagram in Figure \ref{ch2_phaseD}. It is both displaced from the origin by distance $\abs{\alpha}^2$ in phase space and squeezed along the direction dictated by the squeezing angle $\theta$. The DSV state can be expressed as a linear superposition of Fock state by 
\begin{equation}
\begin{aligned}
|\alpha e^{i\varphi}, \xi=re^{i\theta}\rangle=& \frac{1}{\sqrt{\cosh r}} \exp \left[-\frac{\alpha ^2}{2}  \left(1+e^{i (\theta -2 \varphi )} \tanh (r)\right)\right] \\
& \times \sum_{n=0}^{\infty} \frac{\left[\frac{1}{2} e^{i \theta} \tanh r\right]^{n / 2}}{\sqrt{n !}} H_n\left[\frac{e^{i \varphi } \alpha  \cosh (r)+e^{i (\theta -\varphi )} \alpha  \sinh (r)}{\sqrt{e^{i \theta } \sinh (2 r)}}\right]|n\rangle
\end{aligned}
\end{equation}
where $\alpha$, $\varphi$, $r$, and $\theta$ represent the displacement amplitude, displacement angle, squeezing parameter (strength), and squeezing angle, respectively. The photon statistics of the DSV state is given by 
\begin{equation}
\label{dsvPS}
    P_\textrm{dsv}(n)=\frac{\left(\frac{\tanh (r)}{2}\right)^n \exp \left\{-\alpha^2(1+ \tanh (r) \cos (\theta - 2 \varphi))   \right\} \left| H_n\left(\frac{e^{i \varphi } \alpha  \cosh (r)+e^{i (\theta -\varphi )} \alpha  \sinh (r)}{\sqrt{e^{i \theta } \sinh (2 r)}}\right)\right|^2}{n! \cosh (r)},
\end{equation}
which has some properties of coherent light and some properties of squeezed light, leading to the oscillatory tails on the photon distribution as shown in Figure \ref{ch2_dsqz}(b). The photon statistics looks more Poissonian if the term $\alpha$ is dominant compared to $r$, and it looks more oscillatory when the squeezing parameter $r$ is large. Interestingly, DSV state can be made both sub-Poissonian and super-Poissonian by setting the four parameters appropriately. Similarly, the Wigner function of the DSV is given by
\begin{equation}
\begin{aligned}
    W_\textrm{dsv}(x,p,\alpha,\varphi,r,\theta)&=\frac{2}{\pi}e^{-2 \cosh (2 r) \left[x^2+p^2+\alpha^2-2 \alpha (x\cos\varphi+p \sin\varphi)\right]}\\
    & \times e^{-2 \sinh (2 r) \left[\alpha ^2 \cos (\theta -2 \varphi )+\cos\theta \left(x^2-p^2\right)-2 \alpha  p \sin (\theta -\varphi )+2 x p \sin \theta-2 \alpha  x \cos (\theta -\varphi )\right]},
\end{aligned}
\end{equation}
which is plotted in Figure \ref{ch2_dsqz}(a). The Wigner function is non-negative and Gaussian in shape.

\section{Spatially Structured Light}
So far, I have discussed various properties of light photons but the spatial modes. Structuring the spatial profile of light beams is very important for a variety of photonic technologies such as imaging, communication, sensing, cryptography, etc. The structuring of the light field is typically done through various physical processes such as diffraction, interference. In this thesis, however, we produce structured light by controlling the dynamic phase profile of the beam through devices such as spatial light modulator (SLM), digital micromirror device (DMD). Our ability to control and manipulate phases enables interesting applications like Optical tweezers and 3D scanning. There are various other utilities of structured light. For the scope of this thesis, I will limit the discussion to orbital angular momentum (OAM) modes, Laguerre-Gaussian modes, and Hermite-Gaussian modes.

\subsection{Orbital Angular Momentum Mode}
\begin{figure}[ht!]
    \centering
    \includegraphics[width=1.0\textwidth]{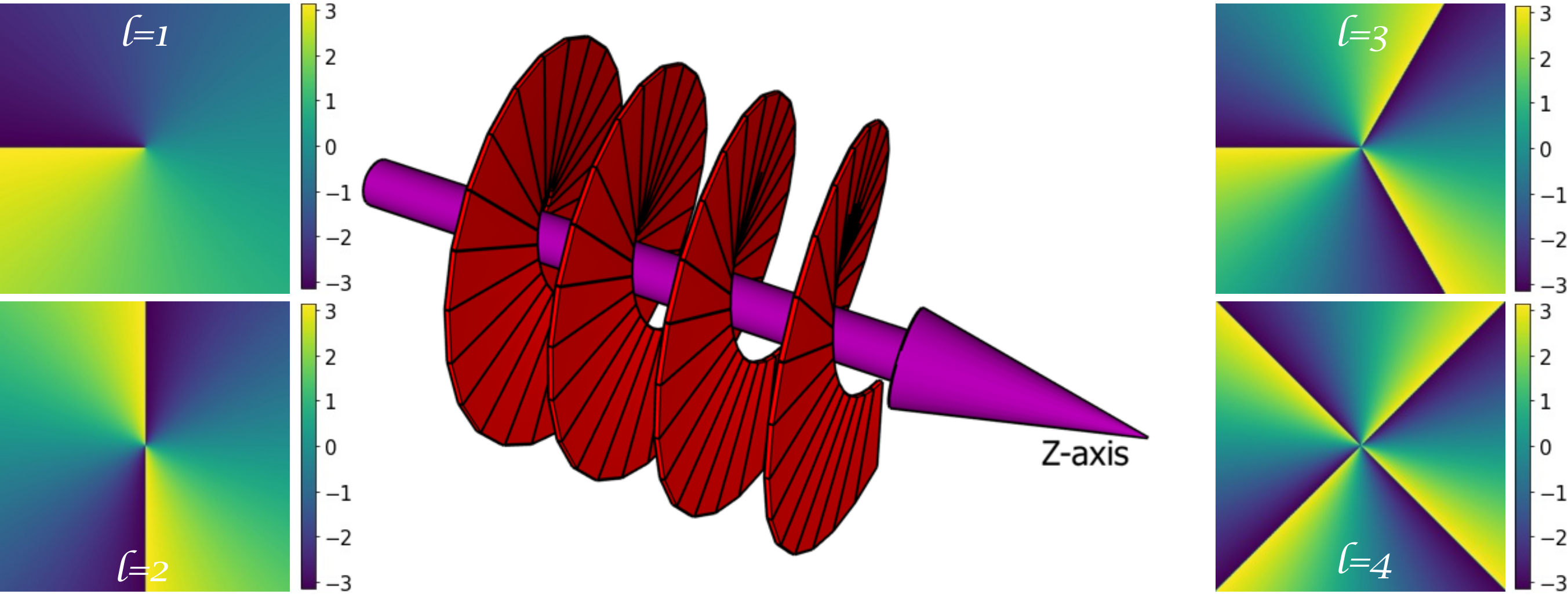}
    \caption{Helical phase front of a light beam carrying orbital angular momentum (OAM). The phase masks for the first four orbital angular momentum modes ($\ell=1, 2, 3, 4$) are plotted in (a), (b), (c), and (d), respectively. The phase values make $\abs{\ell}$ complete rotations for an OAM mode $\ell$.}
    \label{ch2_oam}
\end{figure}
A photon not only carries energy as an intrinsic property but also carries momentum. The radiation pressure is one of the prominent manifestations of the linear momentum which has been exploited in several optomechanical applications to generate the squeezed light and entangled source of photons. Likewise, the spin angular momentum (SAM), which takes values (-1, 0, +1)$\hbar$, is associated with the polarization property of light. The linearly polarized light, right circularly polarized (RCP), and left circularly polarized (LCP) light carry SAM $0\hbar$, $+1\hbar$, and $-1\hbar$, respectively. The orbital angular momentum (OAM) is associated with the spatial profile of a light beam. A spatial mode of light having a helical phase profile of the form $e^{i\ell \phi}$, is known as an OAM mode. Each photon in an OAM mode carries an orbital angular momentum of value $\ell\hbar$. Helical phase fronts and optical singularities or optical vortices at the center of the beam profile are the characteristic features of the OAM modes. The diameter of the vortex is proportional to the $\abs{\ell}$ values of OAM modes, which are also called topological charges. In fact, optical vortices are ubiquitous in natural optical fields. In the field of quantum optics and photonic quantum technologies, it is viewed as a degree of freedom to encode quantum information having tremendous potential. An important distinction of this new degree of freedom is that this gives access to infinite-dimensional Hilbert space, which enables us to encode multiple bits of information per photon. The electric field amplitude of an OAM mode carrying topological charge $\ell$ is given by, 
\begin{equation}
    U(r, \phi, z)= u_0(r,z)\times e^{ikz} \times e^{i\ell\phi}
\end{equation}
where, $u_0(r, z)$ represents the slowly varying amplitude profile of the light propagating along $z$-axis. Similarly, $\phi$ and $k$ describe the azimuthal angle and wavenumber, respectively.

The helical phase profile of the OAM modes upon propagation is schematically depicted in Figure \ref{ch2_oam}. The figure also shows some example phase profiles of OAM modes in a plane perpendicular to the propagation $z$-axis. There are several different ways the OAM modes can be generated. Some of the tools to generate the OAM modes include spatial light modulator, digital micro-mirror device, spiral phase plate, q-plate, specially designed fibers, etc. In this dissertation, we will rely on phase modulation and fork holograms \cite{mirhosseini2013rapid} to generate these modes.

\subsection{Laguerre-Gaussian Mode}
\begin{figure}[hb!]
    \centering
    \includegraphics[width=0.90\textwidth]{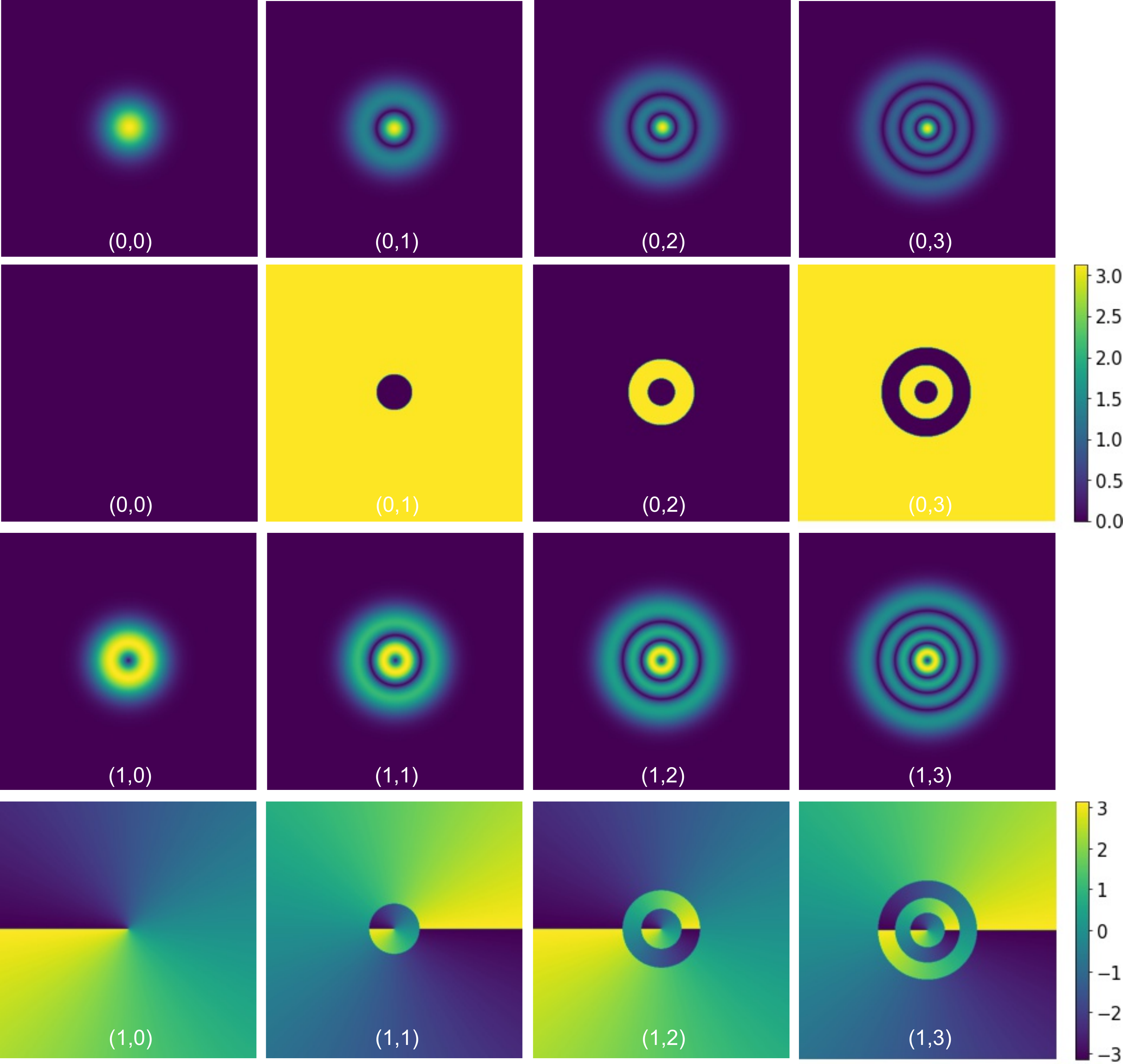}
    \caption{Some examples of Laguerre-Gaussian (LG) modes. The first row depicts the LG modes with $\ell=0$ and $p=0,1,2,3$. The corresponding phase masks are provided in the second row. Similarly, the beam profiles of LG modes for $\ell=1$, and $p=0,1,2,3$ are shown in the third row. The corresponding phase masks are displayed in the bottom row.}
    \label{ch2_lgmode}
\end{figure}
Laguerre-Gaussian (LG) modes represent an important family of spatial modes which carry OAM. Therefore, an object interacting with photons prepared in LG modes experience the torque in addition to the radiation pressure. These modes are solutions to the paraxial Helmholtz equation in cylindrical coordinate system. The paraxial Helmholtz equation (modified wave equation) in cartesian coordinate system is expressed as,
\begin{equation}
    \left(\frac{\partial^{2}}{\partial x^{2}}+\frac{\partial^{2}}{\partial y^{2}}+2 i k \frac{\partial}{\partial z}\right) U(x, y, z)=0.
\end{equation}
The LG modes satisfy this differential equation in cylindrical coordinate system. Naturally, the LG modes have the azimuthal symmetry on the transverse plane. In addition to the azimuthal degree of freedom ($\ell$), the LG modes provide access to the radial degree of freedom ($p$). Furthermore, these modes form a complete orthonormal basis set with respect to the azimuthal ($\ell$) and the radial ($p$) degrees of freedom \cite{siegman:1986}, meaning any arbitrary spatial mode of light field can be expressed as some linear combination of LG modes. These elegant properties of LG modes make them good candidates to construct an unbiased basis set in cryptographic protocols. In addition, they allow us to prepare a higher dimensional qubits called qudits. This property is very important for several different applications. The optical field amplitudes of the LG modes can be written as \cite{allen1992orbital, siegman:1986}
\begin{equation}
\begin{aligned}
\text{LG}_{\ell, p}(r, \phi, z) = & C_{\ell p}^{L G} \frac{w_{0}}{w(z)}\left(\frac{r \sqrt{2}}{w(z)}\right)^{|\ell|} \exp \left(\frac{-r^{2}}{w^{2}(z)}\right) L_{p}^{|\ell|}\left(\frac{2 r^{2}}{w^{2}(z)}\right) \exp \left(\frac{-i k r^{2}}{2 R(z)}\right)\\ 
& e^{i \ell \phi} \times e^{i(2p + \left| \ell \right| + 1)\xi (z)},
\end{aligned}
\end{equation}
where $L_p^{\left| \ell \right|}$ is the generalized Laguerre polynomial, $C_{\ell p}^{L G}=\sqrt{\frac{2p!}{\pi(\abs{\ell +p})!}}$ is the constant term, $\xi(z)=\arctan(z/z_R)$ is the Gouy phase. Similarly, $R(z)=z\left[1+ (z_R/z)^2\right]$ represents the radius of curvature at distance $z$, where $z_R=\pi w_0^2/\lambda$ is the Rayleigh range. Also, $w(z)=w_0 \sqrt{1+(z/z_R)^2}$ represents the beam waist at distance $z$ from the source. Similarly, $k=2\pi/\lambda$ is the wave number. Some examples of Laguerre-Gaussian (LG) modes with $\ell=0, 1$ and $p=0,1,2,3$ are shown in Figure \ref{ch2_lgmode}. The corresponding phase masks are provided right below the intensity profiles. As expected, the diameter of the optical vortices increase with the OAM values or topological charge.

\subsection{Hermite-Gaussian Mode}
HG modes are the solutions of the paraxial Helmholtz equation in the cartesian coordinate system. This family of spatial modes does not carry OAM but is extremely important to represent an arbitrary optical field in a simple cartesian coordinate system. As expected, the LG modes have rectangular symmetry on the transverse plane orthogonal to the propagation $z$-axis. Like LG modes, the HG modes also form a complete orthonormal set. Therefore, any arbitrary intensity profile of an optical field can be decomposed into the HG modes. The electric field amplitude of HG mode in the cartesian coordinate system is given by \cite{beijersbergen1993astigmatic, walborn2005conservation}
\begin{figure}[hb!]
    \centering
    \includegraphics[width=0.90\textwidth]{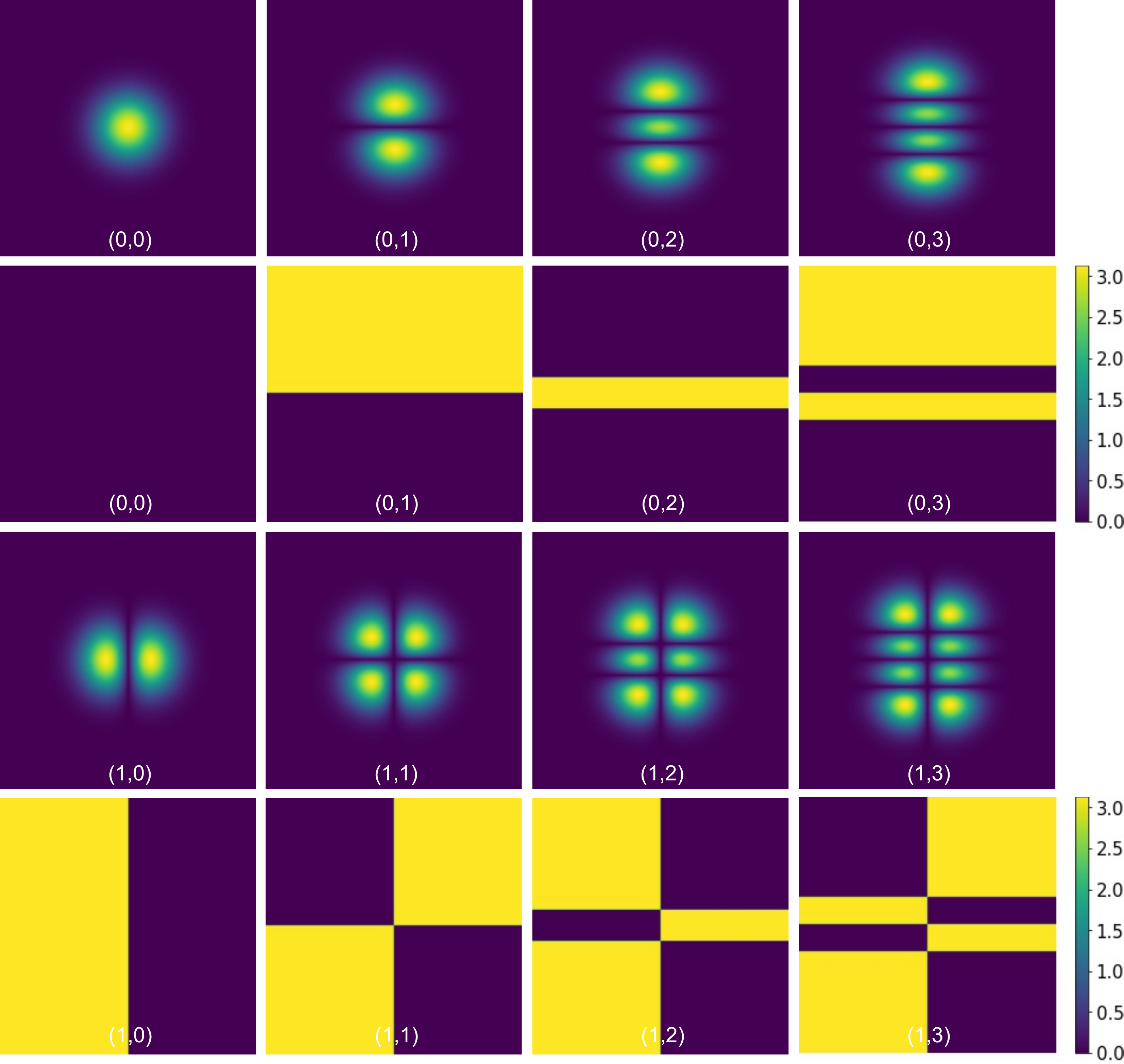}
    \caption{Example beam profiles of the Hermite-Gaussian (HG) modes. The first row depicts the HG modes with $n=0$ and $m=0,1,2,3$. The corresponding phase masks are provided in the second row. Similarly, the beam profiles of HG modes for $n=1$, and $m=0,1,2,3$ are shown in the third row. The corresponding phase masks are displayed in the bottom row.}
    \label{ch2_hgmode}
\end{figure}
\begin{equation}
\begin{aligned}
\text{HG}_{n, m}(x, y, z)=& C_{n m}^{H G} \frac{w_{0}}{w(z)} H_{n}\left(\frac{\sqrt{2} x}{w(z)}\right) H_{m}\left(\frac{\sqrt{2} y}{w(z)}\right) \exp \left(-\frac{x^{2}+y^{2}}{w^{2}(z)}\right) \exp \left(\frac{-i k\left(x^{2}+y^{2}\right)}{2 R(z)}\right)\\
& \times e^{-i(n+m+1)\xi(z)},
\end{aligned}
\end{equation}
where $H_n$ and $H_m$ are the Hermite polynomial of order $n$ and $m$, respectively. Similarly, $k=2\pi/\lambda$ is the wave number. Also, $R(z)=z\left[1+ (z_R/z)^2\right]$ represents the radius of curvature at distance $z$, where $z_R=\pi w_0^2/\lambda$ is the Rayleigh range. Likewise, $w(z)=w_0 \sqrt{1+(z/z_R)^2}$ represents the beam waist at distance $z$ from the source. $\xi(z)=\arctan(z/z_R)$ is the Gouy phase. The constant quantity is defined as $C_{n m}^{H G}=\sqrt{\frac{2}{\pi 2^{(n+m)}n!m!}}$. In Figure \ref{ch2_hgmode}, some examples of Hermite-Gaussian (HG) modes with $n=0, 1$ and $m=0,1,2,3$ are shown along with their corresponding phase masks right below the intensity profiles.

\section{Machine Learning}
We humans learn every day through our experiences. We recognize people we meet, voices we hear, and learn to perform our tasks better as we practice more. Can we program our machines to mimic this human trait? The answer is yes, and it is called machine learning. The term \enquote{machine learning} was formally introduced by Arthur Samuel in 1959 \cite{samuel1959some}. Machine learning is a branch of artificial intelligence (AI) that deals with computer algorithms that improve their performance automatically by learning from their past experiences. Machine learning is a powerful computational tool for dealing with a large volume of data where there is no explicit function or set of instructions to guide the algorithm. Moreover, the data itself contains many complex features or patterns, and several degrees of freedom, making it almost impossible to keep track of through a closed-form analytical function \cite{buchanan2019the}. Self-learning is the most important distinguishing feature that separates other computer algorithms from machine learning. Machine learning algorithms build their own \enquote{functions}, called models, by drawing inferences from a large data set. \textit{Carleo et al.} \cite{carleo2019machine} write, \enquote{models are agnostic and machine provides the intelligence by extracting it from data}. This dynamic feature of machine learning makes them extraordinarily successful in applications like object recognition, speech recognition, medical diagnosis \cite{lecun:2015}. The input data set utilized by a machine-learning algorithm to build a model and implicit set of rules are called training data. The quality and quantity of the training data are at the heart of every machine learning algorithm that determines their efficacies. With the rapid technological advances, lately, machine learning algorithms have taken an important spot in multiple areas of science to deal with complex problems of classification and optimization. Dramatic improvement in performances and outcomes has been reported in a wide range of fields including medical science, computer vision, automation, and physical science researches \cite{carleo2019machine, butler2018machine}. I will now briefly discuss various learning methods, artificial neural networks, and the activation function and optimization algorithms that are commonly used in neural networks.

\subsection{Supervised Learning}
As the name suggests, supervised learning requires supervision, meaning the labeling of training data is necessary. A good example of supervised learning would be a machine-learning algorithm to classify different OAM modes in which a large set of intensity profiles of OAM modes along with their labels are supplied as the training data set. In this particular case, the labels are discrete integer values, therefore this would be a classification problem. However, if the data labels are continuous such as the noise strength in the OAM modes, it is called a regression problem. Nevertheless, machine learning algorithms, utilizing supervised learning, attempt to find a correlation between the input data and data labels by extracting the features or trends that exist in the provided data set. Therefore, the training data is always considered the heart of the machine learning algorithm. The trained machine learning algorithm relies on the mysterious \enquote{model} it extracted from the data to perform the prediction during the test.

In a typical supervised classification scheme having $n$ categories, the user provides a large volume of labeled data. The machine learning algorithm outputs a vector having the highest score on the element representing the true class of the test object. During the training, the machine learning algorithm optimizes the parameters such as the weight of each node. The optimization is performed by minimizing the error between the expected and predicted output vector. In a multi-layer neural network, there are numerous weights that are adjusted during the training process to optimize artificial neurons so as to minimize the cost function or error.

\subsection{Unsupervised Learning}
In the unsupervised learning method, the training data supplied as the input doesn't contain any labels. This is also a powerful learning method that is widely used to cluster the large volume of data into different groups according to the prominent features extracted from the training data. Even though the algorithm can not tell what the data represents, it can discretize and break the large volume of data into smaller clusters according to common features. For example, if we train a model on biographical information, buying habits, online activity of millions of users, the unsupervised learning algorithm can cluster the users according to common features. However, due to the unavailability of labels, the actual meanings of the clusters have to be interpreted later. 

\subsection{Reinforcement Learning}
Unlike supervised and unsupervised learning methods, reinforcement learning relies on trial and error. This doesn't require the labeled training data set. The basic structure of reinforcement learning contains an environment that works like a Markovian Decision Process. Reinforcement learning creates an artificial agent that takes some actions to the environment. The goal of this artificial agent is to maximize the reward. As the name suggests, there is an element of reinforcement in this type of machine learning algorithm. The actions taken by the agent perturb the state of the environment. In addition to the state of the environment, the artificial agent receives the rewards generated through the last action. As mentioned earlier, this method is based on trial and error. This is the training process that helps the algorithm to build a model based on the reinforcement received. If the process is repeated a sufficient number of times, the algorithm is able to find a good strategy to interact with the environment, maximizing the reward. The reinforcement learning method is very successful in control theory, game theory, and simulation-based optimization schemes.

\subsection{Artificial Neural Networks}
\begin{figure}[ht!]
    \centering
    \includegraphics[width=1.0\textwidth]{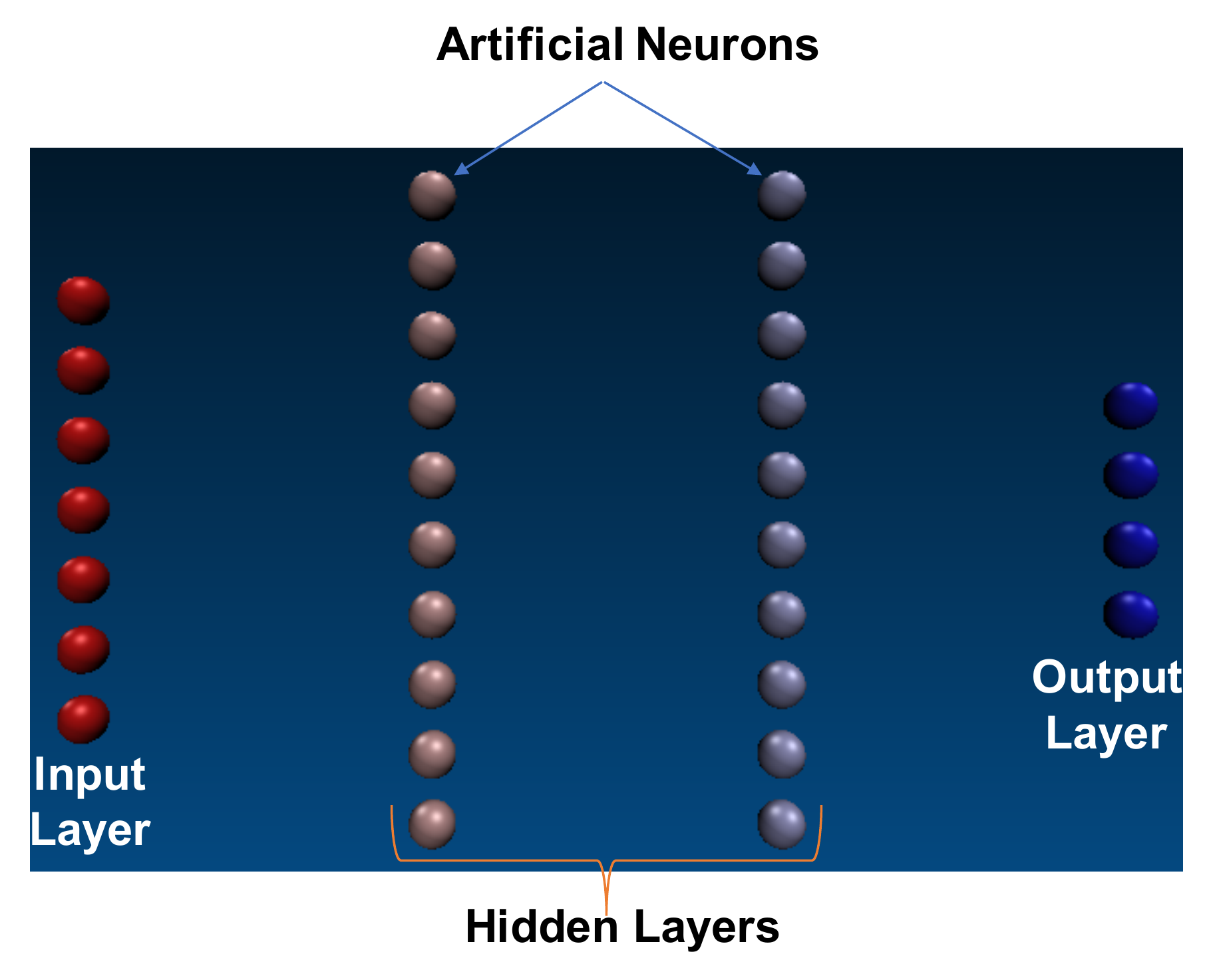}
    \caption{Schematic diagram of ANN. The first layer of artificial neurons called the input layer accepts the input from the supplied data set. The layers sandwiched between input and output layers, called hidden layers, are the key to \enquote{machine intelligence}. The hidden layers are connected in one of several ways according to the computational need.}
    \label{ch2_nnSch}
\end{figure}
Artificial neural networks are one of the most successful models of machine learning that are used in a variety of quantum photonic technologies \cite{carleo2019machine}. This machine learning approach utilizes the features of more than one of three learning methods. The working mechanism of ANNs is similar to the human nervous system. As in our nervous system, the building block of ANN is referred to as neurons (also called perceptrons). The neurons are arranged in several layers which are usually assigned weight before being wired to form complex networks of neurons. The neurons are also supplied with bias values to avoid the situation of trivial convergence. Following the biological nomenclature, the weights assigned are called synaptic weights in artificial neural networks. The simplest artificial neural network contains just two layers of artificial neurons -- input and output. However, realistic neural networks designed to perform complex tasks consist of a number of hidden layers of neurons as shown in Figure \ref{ch2_nnSch}. The series of hidden layers process the information supplied by the input layer with the help of activation functions to draw statistical inferences from the training data set \cite{nielsen:2015}. 

\subsection{Activation Functions and Optimization Methods}
Some of the commonly used activation functions in neural networks are hyperbolic sine, hyperbolic cosine, hyperbolic tangent, Sigmoid function, ReLU (rectified linear unit). Any neural network consisting of three or more hidden layers is known as a deep neural network or deep network. In this case, the machine learning algorithm is referred to as deep learning. Convolutional neural networks (CNN) is an example of a deep network that is efficient in recognizing patterns in images and videos. The image filters designed to extract a series of features in an image are the key elements of CNNs. A pooling layer is used after the convolution with the filters to reduce the dimension of the matrices. The photonic applications discussed in this dissertation utilize CNNs. Finally, the output layer, equipped with a function like Softmax, reduces the available output classes to a probability distribution.

\section{Summary}
In this chapter, I reviewed fundamental concepts of quantum optics. I began the chapter by discussing basic concepts behind the quantization of electromagnetic fields. Then I discussed various states of light that will be used in the next chapters. More specifically, I discussed Fock states, vacuum states, coherent states, thermal states, squeezed-vacuum states, and displaced-squeezed-vacuum states. For each case, the Fock state decomposition, Wigner function representation, photon statistics, and other relevant information are discussed. Furthermore, I discussed spatial modes of light and other structured fields of light. In particular, I discussed orbital angular momentum modes of light, Laguerre-Gaussian, and Hermite-Gaussian modes. Last but not least, I discussed basic ideas behind artificial intelligence and machine learning. Moreover, different machine learning approaches, common activation functions, optimization methods are discussed. Finally, I discussed the artificial neural networks as a computational model that utilizes the machinery described previously. We will be utilizing artificial neural networks, convolutional neural networks in particular, in a number of smart quantum photonic technologies.

%% file: chapter3.tex
\label{ch3}
The advantages that quantum technologies have to offer rely on the efficient generation and measurement of various quantum states. The precision of parameter estimation in optical interferometry, in particular, strongly depends on the input states and measurement strategies. As discussed in Chapter \ref{ch2}, different states of light have their own photon statistics, quadrature fluctuations, photon number fluctuations, and other noise characteristics. In this chapter, first, I discuss the phase sensitivity of an SU(1,1) interferometer with coherent and dispaced-squeezed vacuum states as inputs. Parity and on-off detection schemes are utilized to achieve the sub-shot-noise limited phase sensitivity. Towards the end of this chapter, I discuss a novel squeezed light detection technique that uses the spatial correlation on camera images. This new detection technique can be viewed as an alternative to the experimentally tedious balanced homodyne detection. 

\section{Background and Motivation}
\begin{figure}[ht!]
    \centering
    \includegraphics[width=0.8\textwidth]{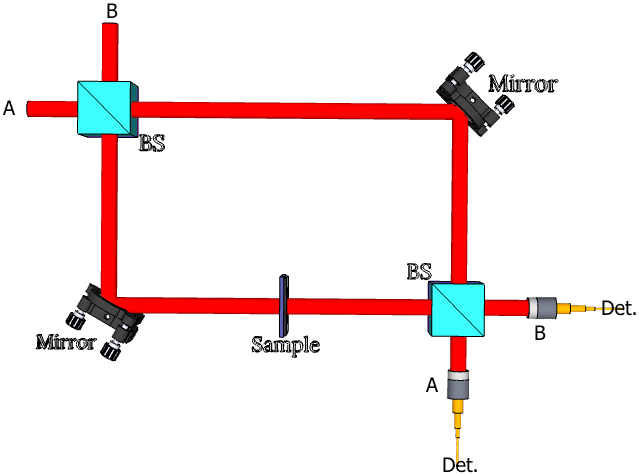}
    \caption{Schematic diagram of the Mach-Zehnder Interferometer. It consists of two optical arms A and B which are designed to enable the detection and estimation of the phase shift or small change in arm length with the help of interference of two input electromagnetic fields. Two optical fields are mixed in the first beamsplitter. One of the two beams emerging from the first beamsplitter interacts with the sample or parameter of interest before mixing up in the second beamsplitter with the other field.}
    \label{ch3_mzi}
\end{figure}

The most common scheme used in quantum metrology is the Mach-Zehnder interferometer (MZI) which uses the interference of two light fields as a tool to gain access to the parameter phase shift. It consists of two 50:50 beamsplitters as shown in Figure \ref{ch3_mzi}. Two input states at port A and port B get mixed in the first beamsplitter before they encode the phase shift ($\phi$) represented by the sample. The parameter $\phi$, in this case, is the relative phase shift between the two arms. The second beamsplitter, combined with a suitable detection strategy, is also viewed as a decoder of information.

The Jordan-Schwinger representation \cite{schwinger1965quantum} is an easy way to understand how an MZI works, which describes the phase estimation process in terms of angular momentum operators $\hat{J}_{x} = \frac{1}{2}(\hat{a}^\dagger b + \hat{b}^\dagger a ),  \hat{J}_{y} = \frac{1}{2}(\hat{b}^\dagger a - \hat{a}^\dagger b ),  \hat{J}_{z} = \frac{1}{2}(\hat{a}^\dagger a - \hat{b}^\dagger b )$. These operators satisfy the usual angular momentum commutation relations \cite{edmonds1996angular}. The interaction of input states with the linear optical elements is interestingly defined as a rotation in abstract spin space. The MZI is also called a SU(2) interferometer because the operations are equivalent to SU(2) rotation group \cite{yurke19862}. For example, action of beamsplitter is characterized by $\frac{\pi}{2}$ rotation around $x-$axis i.e. $\exp(-i \frac{\pi}{2}\hat{J}_x)$, and phase shifter is described as the roation around $z-$axis  by angle $\phi$ i.e. $\exp(-i \phi \hat{J}_z)$. With this definition of intensity difference measurement $\hat{n}_a-\hat{n}_b=2\hat{J}_z$, we can use simple error propagation technique to obtain the phase sensitivity: $(\Delta{\phi})^2 = \frac{(\Delta{{\hat{J}}_{z}})^2}{\Big|{\frac{d\langle{{\hat{J_z}}\rangle}}{d\phi}}\Big|^2}$.

If we inject coherent state ($\ket{\alpha}$) in mode A and vacuum ($\ket{0}$) state in mode B, the minimum phase sensitivity of MZI is $\frac{1}{\sqrt{\bar{n}}}$ where $\bar{n}$ represents the average number of input photons ($\bar{n}=\abs{\alpha}^2$). This limit is called the shot-noise limit (SNL) which is the ultimate limit for classical measurements. This results from the Poissonian nature of photon number distribution in the coherent state. On the surface, it seems that we can achieve any phase sensitivity just by cranking up the power of the laser. However, experimentally, higher power results in larger photon fluctuations than the improvement in phase sensitivity. Therefore, it is very important to squeeze the inherent quantum noise of the state of light \cite{walls2007quantum, dowling1998correlated}. Nevertheless, we can beat this classical limit by using quantum resources like squeezed states. Caves \cite{caves1981quantum} in 1981 showed that it is possible to bring down the sensitivity to $\frac{1}{\bar{n}^{2/3}}$ if we inject displaced squeezed vacuum (DSV) with average photon number $\frac{\bar{n}}{2}$ in the port B of MZI. This led to a surge of attention towards the sub-shot-noise limited measurement schemes. It was later in 1984 \cite{bondurant1984squeezed}, it was shown that the two-mode-squeezed vacuum in MZI gives the sensitivity of $\frac{1}{\bar{n}}$, which is called Heisenberg limit (HL). This is named so because of the fact that this limit is derived from uncertainty relation $\Delta n \Delta \phi\geq 1$. Furthermore, a dual Fock state input was shown to achieve Heisenberg limited phase sensitivity in 1993 \cite{holland1993interferometric}.
In subsequent papers \cite{dowling1998correlated, boto2000quantum, lee2002quantum}, a two-mode special Fock state which is highly entangled called N00N state was shown to achieve Heisenberg sensitivity. However, none of these limits are the ultimate quantum limits. The shot-noise limit and Heisenberg limit are dependent on how we decode the encoded information through measurement strategies. Therefore, it is important to discuss quantities like Fisher Information and Cram\`er-Rao bound (CRB) which allow us to discover the ultimate limits for a given set of inputs and the interferometric scheme.

Let us assume that a measurable quantity $\hat{M}$ is a function of the parameter $\phi$ that we want to estimate as precisely as possible. That means, in some way the information about $\phi$ is stored in the measurable quantity  $\hat{M}$. Assuming $x$ the possible measurement outcomes, $p(x|\phi)$ denotes the probability of measuring value $x$, given the unknown parameter is $\phi$. The major goal in estimation theory is to construct an unbiased estimator $\tilde{\phi}(x)$ which converges to any actual value of $\phi$ when the number of trials is large, i.e. $\int p(x|\phi) \tilde{\phi}(x) = \phi$. The lower bound on the uncertainty of the parameter $\phi$ is given by Cram\`er-Rao Bound CRB \cite{cramer1999mathematical}. It is defined at some optimal value $\phi_0$ as
\begin{equation}
\label{CRB}
CRB=(\Delta\tilde\phi)^2 |_{\phi_{0}} \geq \frac{1}{ F_{cl}|_{\phi_0}},
\end{equation}
where $F_{cl}$ is the Classical Fisher Information (CFI) \cite{braunstein1994statistical, kay1993fundamentals} which is defined as,
\begin{equation}
F_{cl} = \int \frac{1}{p(x|\phi)} \Big( \frac{dp(x|\phi)}{d\phi} \Big)^2 dx = \int \Big( \frac{d}{d\phi} \ln p(x|\phi) \Big)^2 dx.
\end{equation}
The problem with CRB is  that it is dependent on the kind of input states and detection scheme being used. CRB gives a lower bound on the uncertainty, however, it is unclear when the bound is attainable. That is why sometimes CRB is called \enquote{too ambitious} bound. The most important bound in the quantum parameter estimation is the Quantum Cram\`er-Rao Bound (QCRB). QCRB is the ultimate bound, which is attainable with at least some measurement scheme.
\begin{equation}
QCRB=(\Delta \tilde{\phi})^2 \geq \frac{1}{F_{Q}},
\end{equation}
where, $F_Q$ is called the Quantum Fisher Information (QFI) \cite{helstrom1969quantum, paris2009quantum} which is defined as, 
\begin{equation}
F_{Q} = \textrm{Tr}(\rho_\phi L_{\phi}^2),
\end{equation}
where, $L_{\phi}$ denotes the symmetric logarithmic derivative (SLD). For pure states, QFI takes a simpler form: 
$F_{Q} = 4(\langle\dot{\psi}_{\phi}|\dot{\psi}_{\phi}\rangle - |\langle\dot{\psi}_{\phi}|\psi_{\phi}\rangle|^2)$.

In this chapter, I discuss a nonlinear interferometric scheme utilizing coherent and displaced squeezed vacuum as the input states. Furthermore, parity measurement and on-off detection strategies are used in the scheme. The sensitivities of the measurements are compared with the limits such as SNL, HL, and QCRB, discussed in this section.

\section{SU(1,1) Metrology with Displaced Squeezed Light}
\label{ch31}
The central task of quantum metrology is to estimate a physical parameter as precisely as possible using quantum mechanical resources like squeezing and entanglement. Improving the sensitivity of phase measurement has huge significance in applications ranging from low photon phase-contrast microscopy to large-scale applications like LIGO (Laser Interferometer Gravitational-Wave Observatory) \cite{barish1999ligo, chua2014quantum}. In this section, I present an SU(1,1) interferometric scheme with coherent state and displaced-squeezed-vacuum (DSV) state as inputs, and parity detection and on-off detection or click detection as the measurement strategies. The SU(1,1) metrology scheme with parity detection approaches the Heisenberg limit with the increase in squeezing strength. Furthermore, a sub-shot-noise limited sensitivity of phase estimation is reported for on-ff detection as well for some values of squeezing strength.
\begin{figure}[ht!]
    \centering
    \includegraphics[width=1.0\textwidth]{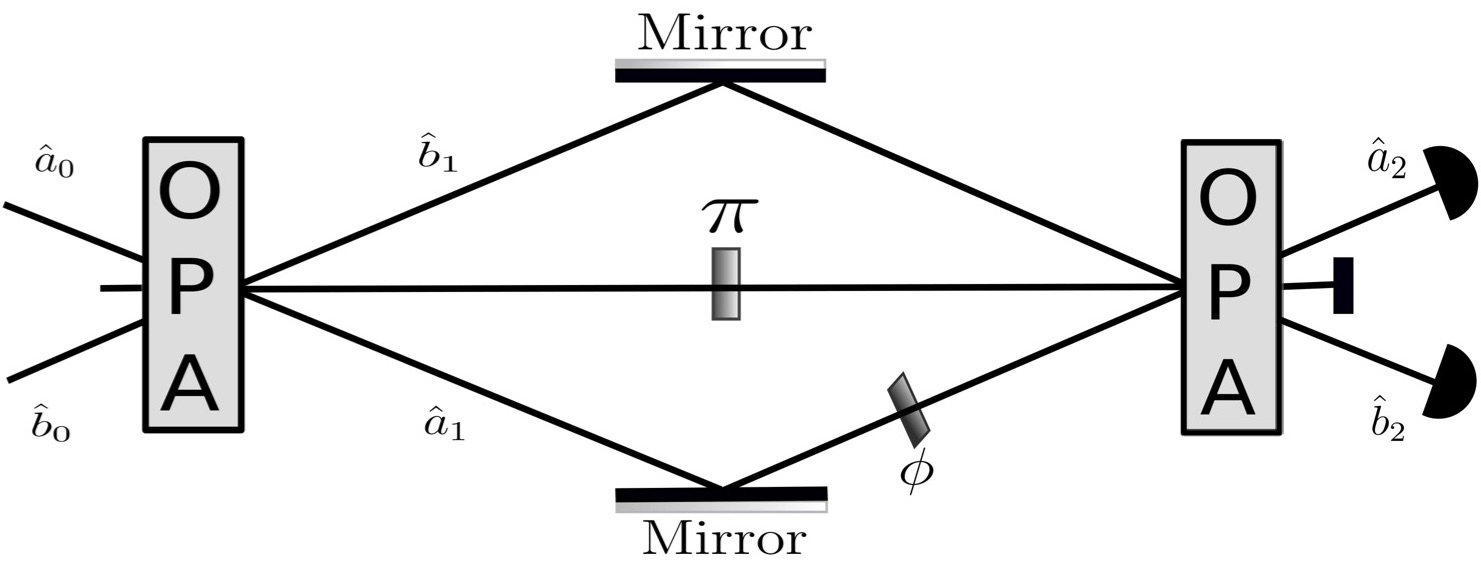}
    \caption{Schematic diagram of SU(1,1) interferometer. It consists of two optical parametric amplifiers (OPA) which differ in phase by $\pi$. The goal of this interferometric scheme is to estimate the parameter $\phi$ as precisely as possible. This figure is reprinted from \textit{OSA Continuum. 2018 Oct 15;1(2):438-50} \cite{adhikari2018phase}, with the permission of Optical Society of America Publishing.} 
    \label{ch3_su11}
\end{figure}

A SU(1,1) interferometer is constructed by replacing the beamsplitters with nonlinear crystals. Physically,  the nonlinear crystals represent the four-wave-mixing (FWM) or optical parametric amplification (OPA) process. The naming of this interferometer comes from the fact that the transformations involved in this scheme characterize the SU(1,1) group. SU(1,1) interferometer is very similar to MZI in the sense that the first OPA encodes the phase information $\phi$ and the second OPA helps to decode the information through measurements. This type of interferometry was first proposed by  Yurke et al. in 1986 \cite{yurke19862} building upon the works published in the preceding year \cite{wodkiewicz1985coherent}. They showed that the SU(1,1) interferometer, as shown in Figure \ref{ch3_su11}, can achieve phase sensitivity as high as $1/\sqrt{\bar{n}(\bar{n}+2)}$ with just the vacuum states on each input arm. However, the low signal-to-noise ratio presented a significant experimental challenge at the time, which was later remedied by modifying the input states and measurement strategies. Furthermore, SU(1,1) interferometry has gained considerable theoretical and experimental attention in the last few years. Plick et al. \cite{plick2010coherent} used a modified scheme using coherent states in both input arms of the interferometer and showed a sub-shot-noise scaling of the phase sensitivity, which was later experimentally demonstrated in reference \cite{ou2012enhancement}. Lately, there have been a series of efforts to improve the measurement sensitivity of the SU(1,1) interferometer through various detection strategies and input states of light. Li et al. used a combination of a coherent state and a squeezed vacuum state, with homodyne measurement, to reach HL sensitivity  \cite{li2014phase}. Later, Li et al. modified the detection strategy to parity measurement and showed Heisenberg-like scaling in the optimal case  \cite{li2016phase}. There are few other variants of SU(1,1) interferometers such as truncated SU(1,1) are proposed that utilize a single nonlinear crystal and apply homodyne detection in both the output modes \cite{gupta2018optimized, prajapati2019polarization}. In this chapter, I discuss a symplectic theoretical model to describe the SU(1,1) interferometer and dive deeper into our SU(1,1) scheme utilizing coherent and DSV state as input states, and parity and on-off as detection strategies. The on-off detection is relatively more resilient compared to the parity detection against experimental photon loss. In addition, on-off detection is easier to implement into an experiment as it doesn't require a photon number resolving detector.

\section{Symplectic Formalism for SU(1,1)}
Symplectic formalism is also called characteristic-function formalism. It utilizes the field quadrature transformations instead of the transformation of annihilation and creation operators or so-called mode operators. This technique is very useful for the states of light that are Gaussian in phase-space representation. The Gaussian states include vacuum state, coherent state, squeezed vacuum state, displaced-squeezed vacuum state, thermal state, etc. For Gaussian states, transforming the first and second moment of the quadrature operators is sufficient to describe the evolution of the states of light in an optical system. Nevertheless, it is possible to handle non-Gaussian states through an inverse transformation but is not as intuitive as it is for Gaussian states.

Let us denote the quadrature operators representing the input states by $\hat{X}_{a0}$, $\hat{P}_{a0}$, $\hat{X}_{b0}$, $\hat{P}_{b0}$, which are mathematically expressed as,
\begin{equation}
\label{eqch3quad}
{{\hat X}_{{a_k}}} = \frac{1}{2}({{\hat a}_k} + \hat a_k^\dag), \ {{\hat P}_{{a_k}}} =  \frac{1}{2i}({{\hat a}_k} - \hat a_k^\dag ),
\end{equation}
with $\hat{a}$, $\hat{a^\dag}$ representing the annihilation and creation operators for the corresponding modes. Similarly, $k=0,1,2$ in the subscript represent the initial, intermediate, and final stages in the interferometer as shown in Figure \ref{ch3_su11}. Likewise, $a$ and $b$ in the subscript represent the arms A and B of the interferometer respectively. For the purpose of matrix transformation of mean and covariance of the initial quadrature operators, they are represented in matrix form as below:
\begin{equation}
\label{eqnch3Quadvec}
    \bar{X}_0=\left(
    \begin{matrix}
        \bar{X}_{a0} \\ \bar{X}_{b0}
    \end{matrix}\right)
    \hspace{5mm} \text{and} \hspace{5mm}
    \Gamma_0=\left(
    \begin{matrix}
        \Gamma_{a0} & 0 \\ 0 & \Gamma_{b0}
    \end{matrix}\right).
\end{equation}

Individually, the mean and covariance of a single-mode field (mode A) are simply represented as $\bar{X}_{a}=\left(\begin{smallmatrix}\langle\hat{X}_a\rangle \\ \langle\hat{P}_a\rangle\end{smallmatrix}\right)$, and  $\Gamma_a=\left(\begin{smallmatrix}2(\langle\hat{X}_a^2\rangle-\langle\hat{X}_a\rangle^2) & \langle\hat{X}_a\hat{P}_a+\hat{P}_a\hat{X}_a\rangle-2\langle\hat{X}_a\rangle\langle\hat{P}_a\rangle \\ \langle\hat{P}_a\hat{X}_a+\hat{X}_a\hat{P}_a\rangle-2\langle\hat{P}_a\rangle\langle\hat{X}_a\rangle & 2(\langle\hat{P}_a^2\rangle-\langle\hat{P}_a\rangle^2) \end{smallmatrix}\right)$, respectively. For a two mode optical system, the mean is $4\times1$ matrix and covariance is a $4\times4$ matrix. The simplicity of symplectic transformation allows us evolve these matrices through our optical system to calculate various output quantities. I will, first, define mean and covariance matrices for five main states of light that are relevant to our calculations in this chapter.

\noindent 1) Vacuum state:
\begin{equation}
    \bar{X}=\left(
    \begin{matrix}
        0 \\ 0
    \end{matrix}\right)
    \hspace{5mm} \text{and} \hspace{5mm}
    \Gamma=\frac{1}{2}\left(
    \begin{matrix}
        1 & 0 \\ 0 & 1
    \end{matrix}\right).
\end{equation}
2) Thermal state:
\begin{equation}
    \bar{X}=\left(
    \begin{matrix}
        0 \\ 0
    \end{matrix}\right)
    \hspace{5mm} \text{and} \hspace{5mm}
    \Gamma=\frac{1}{2}\left(
    \begin{matrix}
        (2n_{th}+1) & 0 \\ 0 & (2n_{th}+1)
    \end{matrix}\right).
\end{equation}
3) Coherent state:
\begin{equation}
    \bar{X}=\left(
    \begin{matrix}
        \alpha\cos(\varphi) \\ \alpha\sin(\varphi)
    \end{matrix}\right)
    \hspace{5mm} \text{and} \hspace{5mm}
    \Gamma=\frac{1}{2}\left(
    \begin{matrix}
        1 & 0 \\ 0 & 1
    \end{matrix}\right).
\end{equation}
4) Squeezed vacuum state:
\begin{equation}
    \bar{X}=\left(
    \begin{matrix}
        0 \\ 0
    \end{matrix}\right)
    \hspace{1mm} \text{and} \hspace{1mm}
    \Gamma=\frac{1}{2}\left(
    \begin{matrix}
        \cosh (2 r)-\cos (\theta ) \sinh (2 r) & -\sin (\theta ) \sinh (2 r) \\ -\sin (\theta ) \sinh (2 r) & \cosh (2 r)+\cos (\theta ) \sinh (2 r)
    \end{matrix}\right)
\end{equation}
5) Displaced squeezed vacuum state:
\begin{equation}
    \bar{X}=\left(
    \begin{matrix}
        \alpha\cos(\varphi) \\ \alpha\sin(\varphi)
    \end{matrix}\right)
    \hspace{1mm} \text{and} \hspace{1mm}
    \Gamma=\frac{1}{2}\left(
    \begin{matrix}
        \cosh (2 r)-\cos (\theta ) \sinh (2 r) & -\sin (\theta ) \sinh (2 r) \\ -\sin (\theta ) \sinh (2 r) & \cosh (2 r)+\cos (\theta ) \sinh (2 r)
    \end{matrix}\right)
\end{equation}

\noindent So far, I have discussed how to define and prepare the input mean and covariance matrices. To carry out the theoretical calculations using the phase-space variables, the operators like beamsplitter, phase-shifter, two-mode-squeezer need to be defined in the phase-space.

\noindent Beamsplitter with transmittivity $T$:
\begin{equation}
BS(T)=
\left(
\begin{array}{cccc}
 \sqrt{T} & 0 & \sqrt{1-T} & 0 \\
 0 & \sqrt{T} & 0 & \sqrt{1-T} \\
 \sqrt{1-T} & 0 & -\sqrt{T} & 0 \\
 0 & \sqrt{1-T} & 0 & -\sqrt{T} \\
\end{array}
\right) 
\end{equation}
Two-mode-squeezer:
\begin{equation}
 TMS(g,\psi)=
 \left(
\begin{array}{cccc}
 \cosh (g) & 0 & \cos (\psi ) \sinh (g) & \sin (\psi ) \sinh (g) \\
 0 & \cosh (g) & \sin (\psi ) \sinh (g) & -\cos (\psi ) \sinh (g) \\
 \cos (\psi ) \sinh (g) & \sin (\psi ) \sinh (g) & \cosh (g) & 0 \\
 \sin (\psi ) \sinh (g) & -\cos (\psi )\sinh (g) & 0 & \cosh (g) \\
\end{array}
\right) 
\end{equation}
Phase-shift in upper mode (A) or lower mode (B):
\begin{equation}
PS_a(\phi)=
\left(
\begin{array}{cccc}
 \cos (\phi ) & -\sin (\phi ) & 0 & 0 \\
 \sin (\phi ) & \cos (\phi ) & 0 & 0 \\
 0 & 0 & 1 & 0 \\
 0 & 0 & 0 & 1 \\
\end{array}
\right) 
\textrm{ and }
PS_b(\phi)=
\left(
\begin{array}{cccc}
 1 & 0 & 0 & 0 \\
 0 & 1 & 0 & 0 \\
 0 & 0 & \cos (\phi ) & -\sin (\phi ) \\
 0 & 0 & \sin (\phi ) & \cos (\phi ) \\
\end{array}
\right)
\end{equation}
Phase-shift in both modes A and B:
\begin{equation}
PS(\phi)=
\left(
\begin{array}{cccc}
 \cos (\phi ) & -\sin (\phi ) & 0 & 0 \\
 \sin (\phi ) & \cos (\phi ) & 0 & 0 \\
 0 & 0 & \cos (\phi ) & -\sin (\phi ) \\
 0 & 0 & \sin (\phi ) & \cos (\phi ) \\
\end{array}
\right)
\end{equation}

The nonlinear crystal OPA in SU(1,1) interferometer is mathematically represented by $TMS$ matrix (operator). Note: the second OPA is phase-shifted by $\pi$ relative to the first OPA. The matrix $S$ represents the evolution operator which connects the mean and covariance of input states to the mean and covariance of output states.
\begin{equation}
S=S_{\rm{OPA2}}\cdot S_{\phi} \cdot S_{\rm{OPA1}}=TMS(g, \psi \rightarrow 0)\cdot PS_a(\phi)\cdot TMS(g, \psi \rightarrow \pi)
\end{equation}
The output mean and covariance matrices are given by 
\begin{equation}\label{eq14}
{\bar X_2} = S\cdot{\bar X_0}\textrm{  and  }{\Gamma _2} = S\cdot{\Gamma _0}\cdot{S^\dag},
\end{equation} 
which in fact represent the output states of light. We utilize these pair of evolved matrices to perform measurements at the output port. In order to decode the phase-shift information $\phi$ encoded in the evolved states of light, a variety of measurement strategies are employed. The sensitivity with which the parameter can be decoded also depends on the choice of measurement strategy.  

\section{Sensitivity of Phase Estimation}
After the theoretical formalism, it is time to discuss the results of our proposed metrology scheme. First of all, I will discuss the phase sensitivity of the SU(1,1) interferometer using parity measurement. Since parity is a single-mode parameter, we obtain the Wigner function of the output state at the second output mode, denoted as mode $\hat{b}_2$ in Figure \ref{ch3_su11}. The parity, by definition, refers to the probability of detecting an even number of photons. The eigenvalue of the parity operator is $1$ for an even number of photons detected and $-1$ for the odd number of photons detected at the output port. In general, parity signal gives the statistics about the odd and even number of photons $\hat{\Pi}=(-1)^{\hat{n}}=(-1)^{\hat{b}_2^\dag\hat{b}_2}$. As such, this measurement strategy requires a photon-number-resolving detection. However, there is a simpler way to measure the parity signal using the Wigner function approach. For this particular metrology scheme, we calculate the Wigner function at the output mode B as,
\begin{equation}
\label{eqwigF}
W({X_b,P_b})  = \frac{\exp \left(-({\overrightarrow X} - {\bar X_2})^T \cdot ({\Gamma_2})^{ - 1} \cdot ({\overrightarrow X} -{\bar X_2})\right)}{ \pi^N\sqrt {\text{Det}\left(\Gamma_2 \right)}},
\end{equation}
where N denotes the number of modes (N=1 in this case). Note: the evolved mean and covariance used in this equation are the reduced one-mode matrices. The reduction is physically equivalent to tracing out one of the modes. In phase-space representation, the first mode (mode A) can be easily traced out by extracting only the fourth quadrant submatrix of $\Gamma_2$. Similarly, the first quadrant is extracted to trace out the second mode (mode B).
\begin{figure}[ht!]
    \centering
    \includegraphics[width=1.0\textwidth]{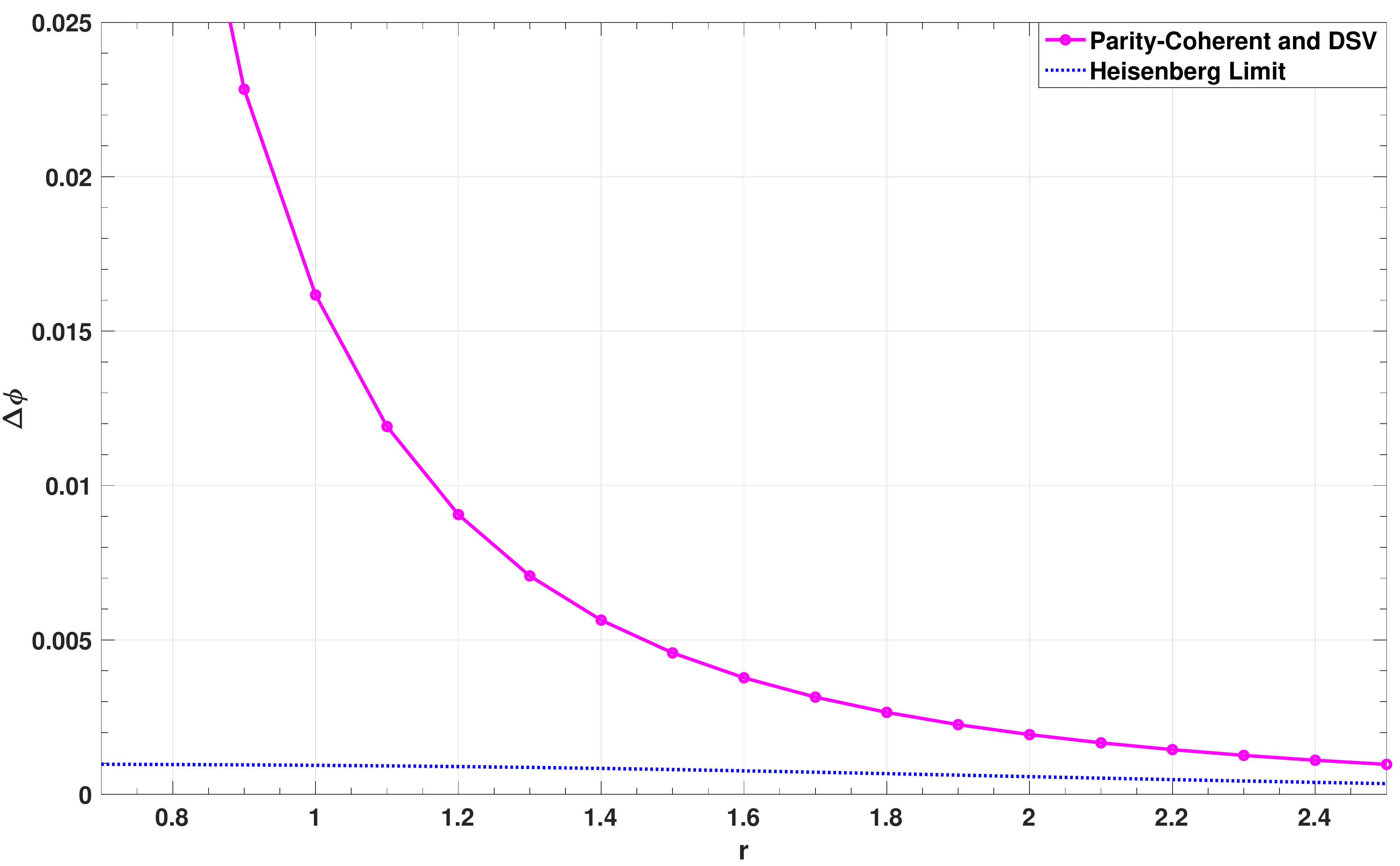}
    \caption{The effect on the phase sensitivity with the increase in the squeezing parameter $r$ (of the DSV). The Heisenberg Limit (HL) (blue) is also plotted for the comparison. Plotted with $n_1 = 16$ (mean photon number of the coherent state in mode A), $n_2 = 4$ (mean photon number of the local oscillator used to produce the DSV), $g = 2$ (squeezing strength of the either OPA). This figure is reprinted from \textit{OSA Continuum. 2018 Oct 15;1(2):438-50}, with the permission of Optical Society of America Publishing.}
    \label{ch3_parity}
\end{figure}
The Wigner function, thus obtained, can be utilized to calculate the expectation value of the parity operator as $\langle{\hat{\Pi}_{b}} \rangle = \frac{\pi}{2} W(0,0)$ \cite{plick2010parity, gerry2010parity, seshadreesan2011parity}. This elegant mathematical tool allows us to access the parity signal relatively easily in theories and experiments both. The expectation value of $\langle\hat{\Pi}_b^2\rangle$ is always 1. This property of the parity operator further simplifies our task of estimating the phase sensitivity of SU(1,1) interferometer with parity detection.
\begin{equation}
\Delta \phi =\frac{\langle \Delta \hat{\Pi}_{b} \rangle}{|\frac{\partial \langle \hat{\Pi}_{b} \rangle}{\partial \phi}|}=\frac{\sqrt{\langle \hat{\Pi}_{b}^2 \rangle  - \langle \hat{\Pi}_{b}\rangle^2}}{|\frac{\partial \langle \hat{\Pi}_{b} \rangle}{\partial \phi}|}=\frac{\sqrt{1 - \langle \hat{\Pi}_{b} \rangle^2}}{|\frac{\partial \langle \hat{\Pi}_{b} \rangle}{\partial \phi}|}
\end{equation}

Before discussing the results of our interferometric scheme, I will first define the SNL and HL applicable to this particular case. In equations \ref{SNL} and \ref{HL}, $\bar{n}_{1}$ is the average photon number in  coherent state in the first arm. Similarly, $\bar{n}_{2}$ and $\bar{n}_{\xi}=\sinh^2{(r)}$ are the average photon number in the displaced and squeezed part of the DSV, with $r$ being the squeezing parameter. And lastly, $\bar{n}_{opa}=2\sinh^2{(g)}$ is the average photon number of the two-mode squeezer, or equivalently, the OPA, and $g$ is the squeezing strength. We assume the internal phase $\varphi=0$ for both the coherent and DSV state. In addition, the squeezing angles $\theta=0$, and $\psi=0$ were chosen in this entire calculation. The shot-noise and Heisenberg limits for our metrology scheme are given by
\begin{equation}
\label{SNL}
 {\Delta {\phi_{\text{SNL}}} = \frac{1}{\sqrt{\bar{n}_{\rm{Total}}}} = \frac{1}{\sqrt{(\bar{n}_{1} + \bar{n}_{2} + \bar{n}_{\xi}) (1+\bar{n}_{\rm{opa}}) + \bar{n}_{\rm{opa}}+ 2\sqrt{\bar{n}_{1} \bar{n}_{2} \bar{n}_{\rm{opa}}(\bar{n}_{\rm{opa}}+2) }}}},
\end{equation}
\begin{equation}
\label{HL}
 {\Delta {\phi_{\text{HL}}} = \frac{1}{\bar{n}_{\rm{Total}}} = \frac{1}{(\bar{n}_{1} + \bar{n}_{2} + \bar{n}_{\xi}) (1+\bar{n}_{\rm{opa}}) + \bar{n}_{\rm{opa}}+ 2\sqrt{\bar{n}_{1} \bar{n}_{2} \bar{n}_{\rm{opa}}(\bar{n}_{\rm{opa}}+2) }}},
\end{equation}
where the symbols represent the quantities defined above.

\begin{figure}[ht!]
    \centering
    \includegraphics[width=0.95\textwidth]{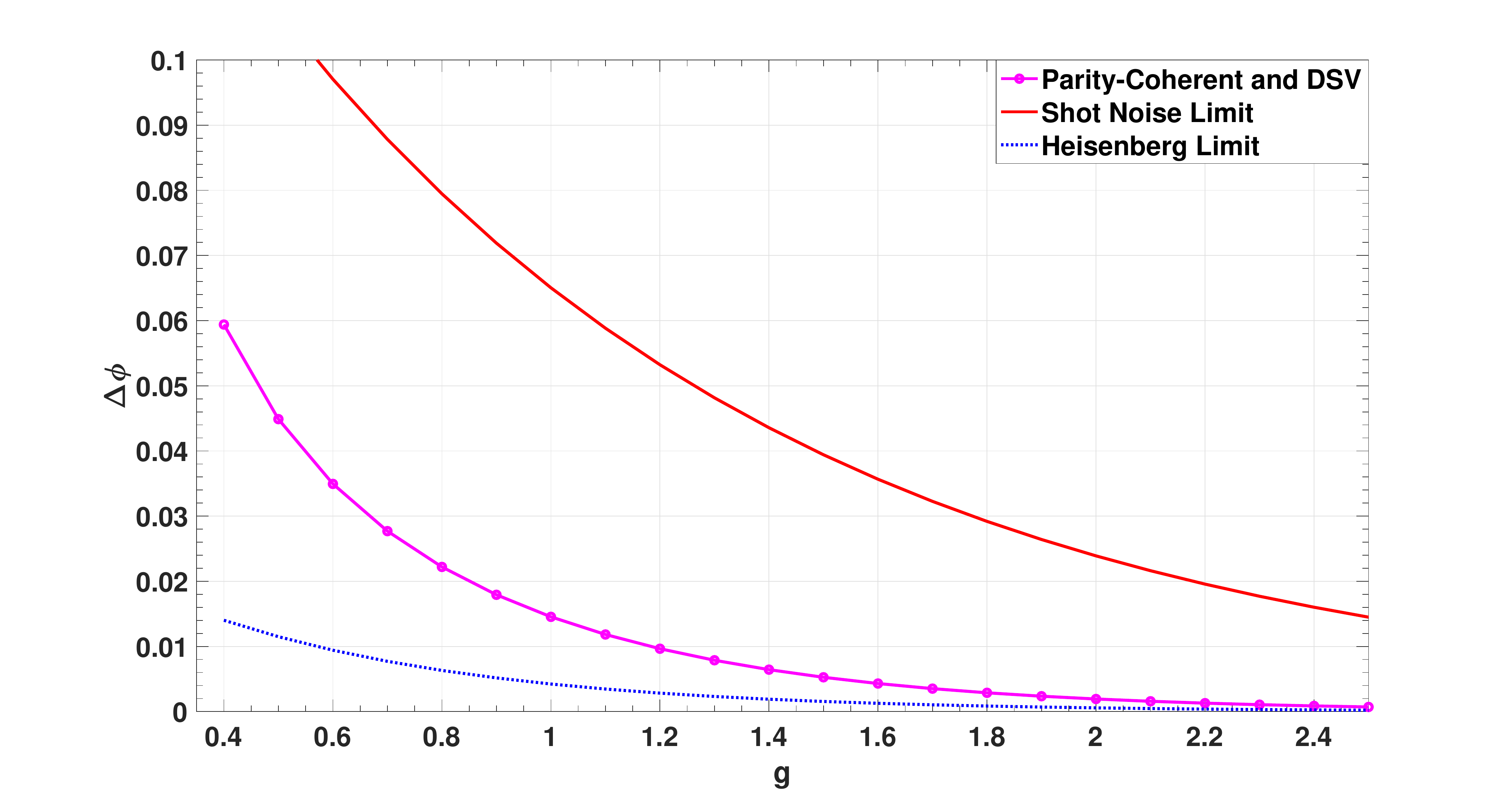}
    \caption{Phase sensitivity as a function of the squeezing strength g of the OPA. The sensitivity with coherent and DSV (pink) is obtained by numerically optimizing $\phi$. The HL (blue) and SNL (red) plotted for the sensitivity comparison. Plotted with $\bar{n}_1 = 16$ (mean photon number of the coherent state in mode A), $\bar{n}_2 = 4$ (mean photon number of the local oscillator used to produce the DSV), $r = 2$ (squeezing strength of the DSV state). This figure is reprinted from \textit{OSA Continuum. 2018 Oct 15;1(2):438-50}, with the permission of Optical Society of America Publishing.}
    \label{ch3_dsv}
\end{figure}

In Figure \ref{ch3_parity}, the phase sensitivity of our SU(1,1) scheme with the parity as the detection strategy is plotted. The figure shows the variation of the phase sensitivity against the squeezing parameter $r$ of the DSV. Remarkably, the sensitivity of phase estimation of our SU(1,1) scheme approaches the HL for higher values of $r>2$. Similarly, in Figure \ref{ch3_dsv}, the sensitivity of our scheme is plotted against $g$ and compared with the benchmarks SNL and HL as defined by equations \ref{SNL}, and \ref{HL}, respectively. The figure shows that the near-HL sensitivity is possible by tuning up the gain value of OPA. The phase sensitivity $\Delta \phi$ approaches the HL for values $g>2$.

\begin{figure}[ht!]
    \centering
    \includegraphics[width=0.95\textwidth]{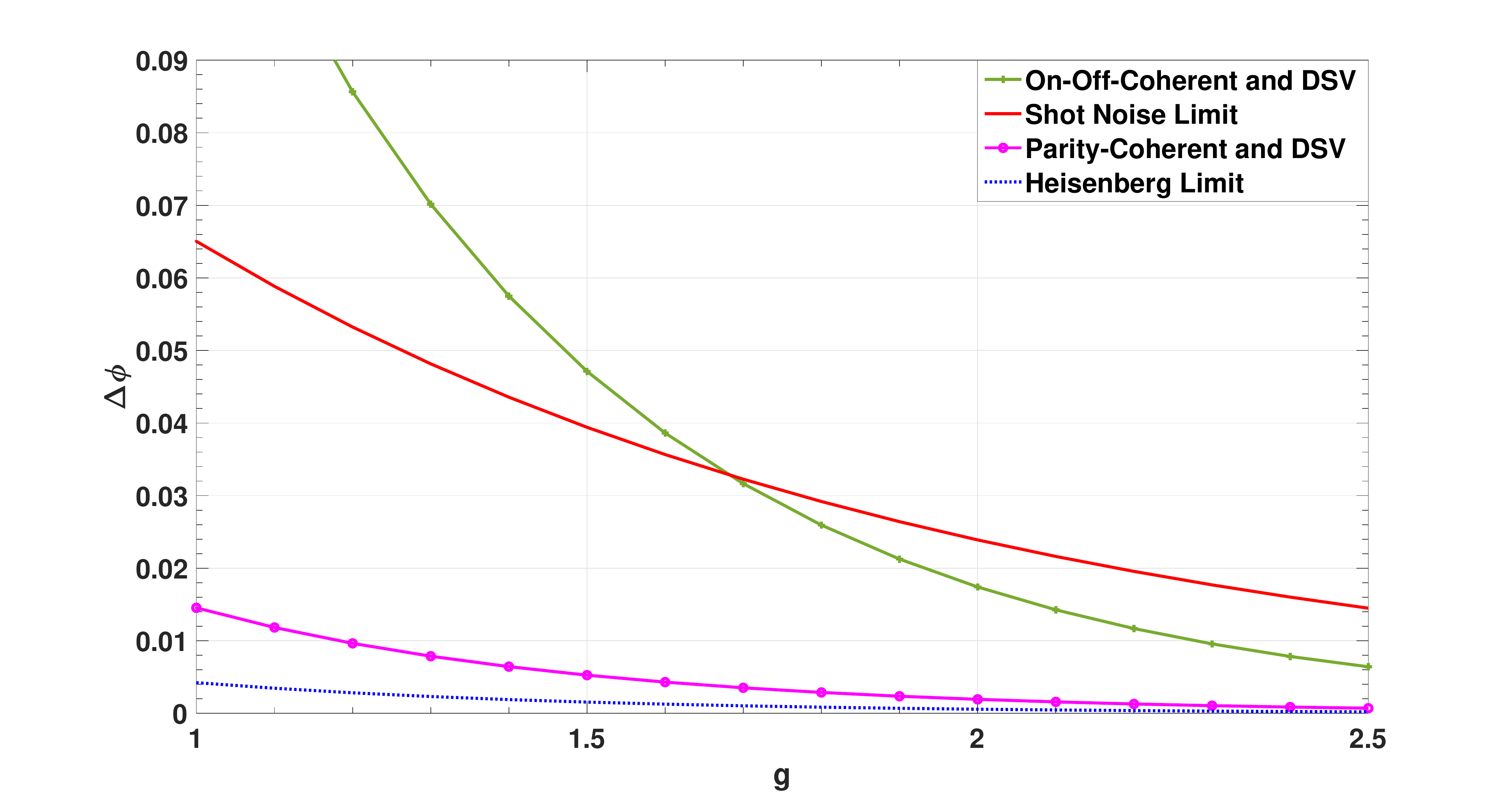}
    \caption{Phase sensitivity with coherent and DSV state with on-off detector. The sensitivity (green) is obtained by numerically optimizing $\phi$ from the classical Cram\`er-Rao Bound (CRB). The Shot-noise-limit (SNL) and Heisenberg Limit (HL) are shown in red and blue color respectively. Plotted with $\bar{n}_1 = 16$ (mean photon number of the coherent state in mode A), $\bar{n}_2 = 4$ (mean photon number of the local oscillator used to produce the DSV), $r = 2$ (squeezing strength of the DSV state). This figure is reprinted from \textit{OSA Continuum. 2018 Oct 15;1(2):438-50,} with the permission of Optical Society of America Publishing.}
    \label{ch3_onoff}
\end{figure}

Now, let us discuss our proposed SU(1,1) metrology scheme with on-off detection. The on-off detection is also called a bucket detector or click detector. It can only discriminate the presence of a photon from an absence. However, this detection can not differentiate one photon from more than one photon. The on-off detection can be applied to our existing mathematical setup by utilizing the output covariance matrix ($\Gamma_2$). We use the following equations to calculate the probability of detecting non-zero photons ($P_{on}$), and CFI ($F$) and hence the phase sensitivity ($\Delta \phi$) \cite{takeoka2015full,kay1993fundamentals}.
\begin{equation}
    P_{\textrm{on}} = 1 - \frac{2}{\sqrt{\textrm{det}(\Gamma + \textrm{I})}}
\end{equation}
\begin{equation}
F = \sum \frac{1}{P_{\textrm{on}}}\bigg(\frac{d{P_{\textrm{on}}}}{d \phi}\bigg)^2  \hspace{5mm} \textrm{ and } \hspace{5mm} \Delta \phi \geq 1/{\sqrt{F}}
\end{equation}

Despite having tremendous advantages in phase estimation, parity detection is experimentally challenging. Moreover, it is very sensitive to the effects like thermal noise, photon loss, scattering, etc. In addition, the parity measurement, in general, requires a photon-number-resolving detector (PNRD). These detectors are very costly and difficult to implement in an experimental setup. However, we have developed a simplified technique to resolve the photon numbers at a low-light-level only, which will be used in the next chapter. We implement a simple detection scheme that gives sub-shot-noise phase sensitivity using on-off detectors. In the literature, we have seen that a homodyne detection is superior in performance. But, it is a resource-intensive and complicated detection strategy. This further reiterates the importance of having a more realistic scheme such as ours. Figure \ref{ch3_onoff} shows the sensitivity of our scheme using an on-off detector. We can see that sub-shot-noise sensitivity can be achieved for $g<2$. Thus, if only sub-shot-noise sensitivity is desired for a particular application, our simple setup with an on-off detector will suffice, without having to use complicated detection strategies like PNRD, homodyne, and parity.

\section{Novel Squeezed Light Detection with Camera}
\label{ch32}
As discussed earlier in Chapter \ref{ch2}, squeezed light is a quantum state of light that shows suppressed or squeezed quadrature fluctuations. Due to the Heisenberg uncertainty principle, it is not possible to squeeze both the quadrature fluctuations simultaneously. The squeezing of one quadrature is done at the expense of the other quadrature. Numerous methods for the generation of squeezed light have been developed, based on a variety of nonlinear materials \cite{gerry2005introductory, walls1983squeezed}. The common ones utilize parametric down-conversion in nonlinear crystals \cite{loudon1987squeezed, lvovsky2015squeezed}, although atom-based sources based on a polarization self-rotation effect \cite{matsko2002vacuum, mikhailov2008low, ries2003experimental} and four-wave mixing \cite{slusher1985observation, embrey2015observation} are also being pursued.

\begin{figure}[ht!]
    \centering
    \includegraphics[width=0.7\textwidth]{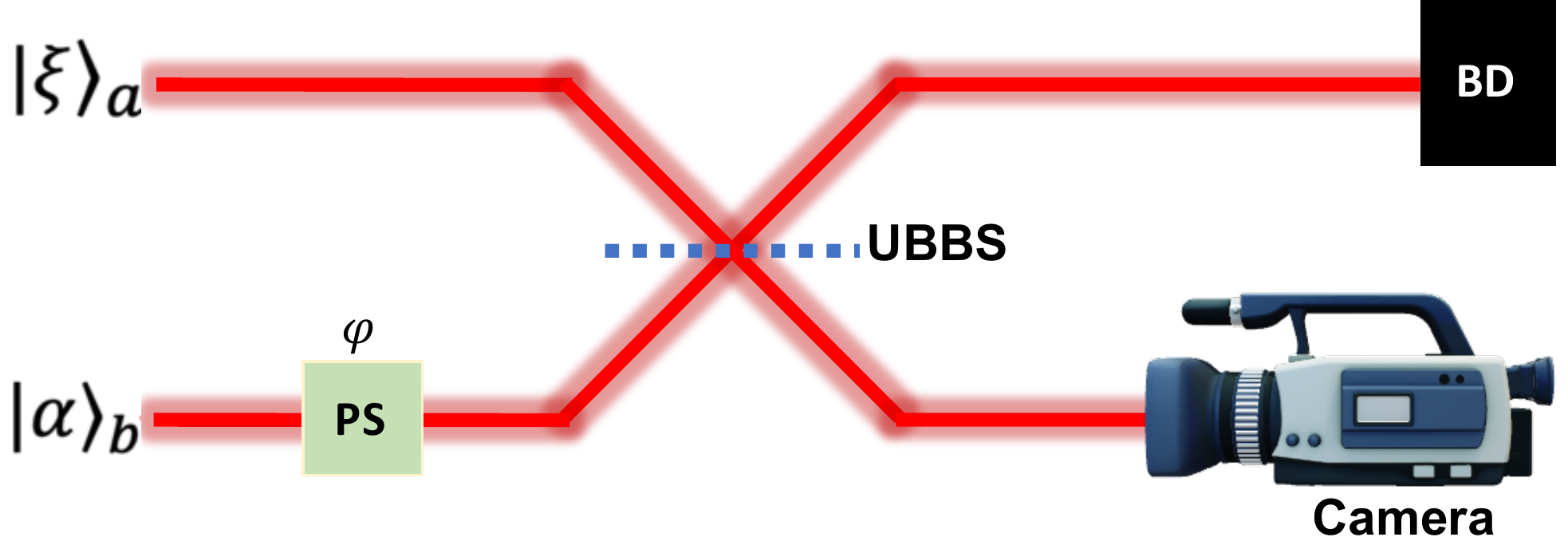}
    \caption{Schematic diagram of the proposed squeezed light detection scheme. A squeezed vacuum state $\ket{\xi}$ is displaced by a strong pump $\ket{\alpha}$ with the help of a highly unbalanced beamsplitter (UBBS). The phase of the resultant displaced-squeezed vacuum (DSV) state is controlled by a tunable phase-shifter (PS). One of the output beams from UBBS is dumped using a beam dumper (BD). A large number of spatial profiles of the other beam are recorded with a high-efficiency camera.}
    \label{ch3_sqzLightDetSch}
\end{figure}

Quadrature squeezing is an interesting quantum property of light fields which has been extensively utilized in quantum metrology (such as the one discussed earlier in this chapter) \cite{gerry2005introductory, bondurant1984squeezed, lee2002quantum}, continuous-variable quantum communication protocols \cite{ralph1998teleportation, ralph1999continuous}, large scale parameter estimation like LIGO \cite{barish1999ligo, chua2014quantum}, quantum imaging below the shot-noise limit \cite{brida2010experimental, omar:2019, genovese2016real}. However, the degree of utility of squeezed light depends on our ability to reliably measure them. Conventionally, squeezed light and the squeezing strength is measured using a balanced homodyne method \cite{breitenbach1997homodyne}. But the conventional methods to detect the squeezed light are cumbersome and resource-intensive. Indeed, this is identified as one of the hurdles to effectively utilize this intriguing quantum property of light in various applications.

\begin{figure}[ht!]
    \centering
    \includegraphics[width=1.0\textwidth]{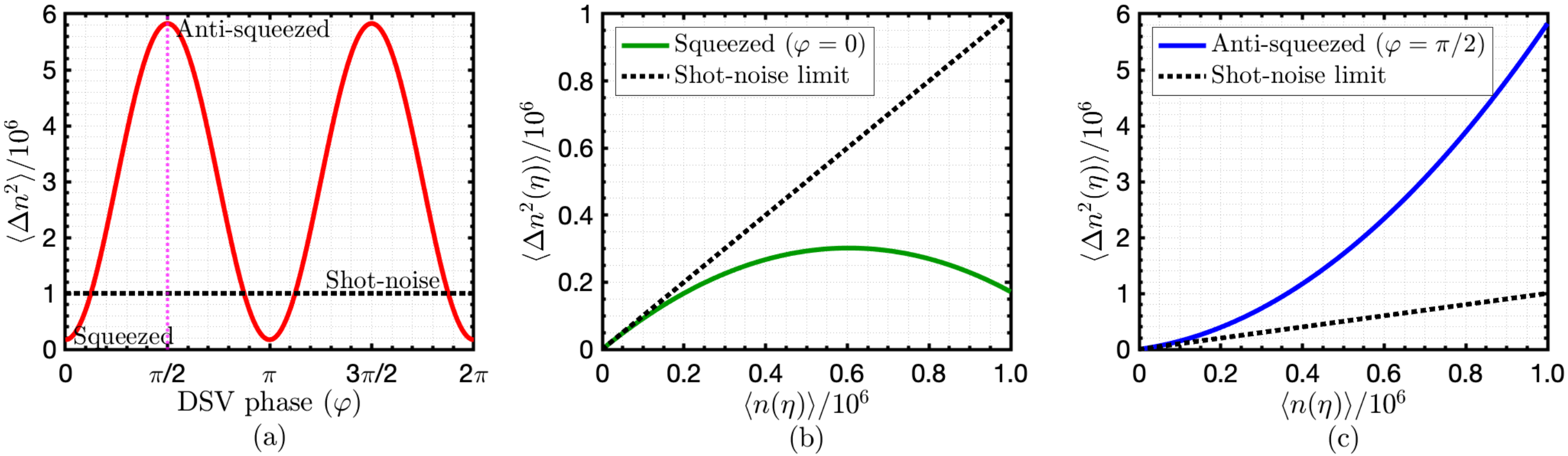}
    \caption{Theoretical plots of photon-number variance. (a) The photon number variance of displaced-squeezed vacuum (DSV) state is plotted agaist the phase $\varphi$. The phase angle $\varphi=0$ represents the squeezed light and $\varphi=\pi/2$ represents the anti-squeezed light. The shot-noise limit is shown with the dotted-black line. Parametric plot of variance versus mean-photon number (as functions of $\eta$) for squeezed light and anti-squeezed light are depicted in (b) and (c) respectively. The shot-noise limit is also shown in all the subplots for comparison.}
    \label{ch3_sqzLightDetTheory}
\end{figure}

In this section, I will discuss our novel technique to measure the squeezing strength of a quadrature squeezed light. First of all, we displace a squeezed-vacuum state ($\ket{\xi}_\textrm{a}$) by mixing a strong pump ($\ket{\alpha}_\textrm{b}$) in an unbalanced beamsplitter having small reflectivity as shown in Figure \ref{ch3_sqzLightDetSch}. We utilize a tunable phase shifter (PS) to induce a variable phase $\varphi$ in the output DSV state ($\ket{\alpha e^{i\varphi},\xi}$). The proposed scheme can measure the squeezing strength of the squeezed light ($\ket{\xi}$) by utilizing the correlations in the camera images. This novel technique is very simple to implement in an experiment unlike the conventional homodyne detection method. Let's begin by studying the photon fluctuations of the displaced-squeezed vacuum state theoretically. The following equation is derived for the photon-number fluctuation (variance) in a DSV state,
\begin{equation}
\label{vareqn1}
\left\langle\Delta \hat{n}^{2}\right\rangle=\bar{n}_{\alpha}+2 \bar{n}_{\alpha} \bar{n}_{s}+2 \bar{n}_{s}+2 \bar{n}_{s}^{2}
-2 \cos \left(2 \varphi\right) \bar{n}_{\alpha} \sqrt{\bar{n}_{s}\left(1+\bar{n}_{s}\right)},
\end{equation}
where $\bar{n}_s=\sinh^2(r)$ represents the mean-photon number in the squeezed vacuum state $\ket{\xi}$. Similarly, $\bar{n}_\alpha=|\alpha|^2$ denotes the mean-photon number in the coherent state (pump) that is used to displace the squeezed vacuum state. The symbol $\varphi$ represents the variable phase value induced in the DSV state. The equation \ref{vareqn1} is plotted in Figure \ref{ch3_sqzLightDetTheory}(a), which shows the variation of photon-number fluctuation with the phase of a DSV state. Since, we are only interested to measure the squeezing strength, for the further investigation, we choose the phase angles $\varphi=0$ and $\varphi=\pi/2$ that represent the squeezed and anti-squeezed light, respectively. In order to allow for the mean-photon number control, we place a tunable neutral density (ND) filter of transmissivity $\eta$ before the camera. The effect of ND filter in mean-photon $\langle \hat{n} \rangle$ number and photon fluctuation $\langle\Delta \hat{n}^{2}\rangle$ are given by following two equations, 
\begin{equation}
   \begin{aligned}
    \langle\hat{n}\rangle &=\eta\left(\bar{n}_{\alpha}+\bar{n}_{s}\right) \textrm{ and }\\
\left\langle\Delta \hat{n}^{2}\right\rangle &=\frac{1}{2} \eta\left\{2 \bar{n}_{\alpha}\left(1+\bar{n}_{s}\right)+\bar{n}_{s}\left(2+\bar{n}_{s}\right)\right.
&-4 \eta \bar{n}_{\alpha} \sqrt{\bar{n}_{s}\left(\bar{n}_{s}+1\right)} \cos (2 \varphi)\\
&\left.+\bar{n}_{s}\left(1+2 \bar{n}_{\alpha}+2 \bar{n}_{s}\right)(2 \eta-1)\right\},
    \end{aligned} 
\end{equation}
where the symbols denote the quantities just discussed. The parametric plots of variance versus the mean-photon number as a function of transmissivity ($\eta$) for $\varphi=0$ and $\varphi=\pi/2$ are depicted in Figure \ref{ch3_sqzLightDetTheory}(b) and 
\ref{ch3_sqzLightDetTheory}(c), respectively.

\begin{figure}[ht!]
    \centering
    \includegraphics[width=1.0\textwidth]{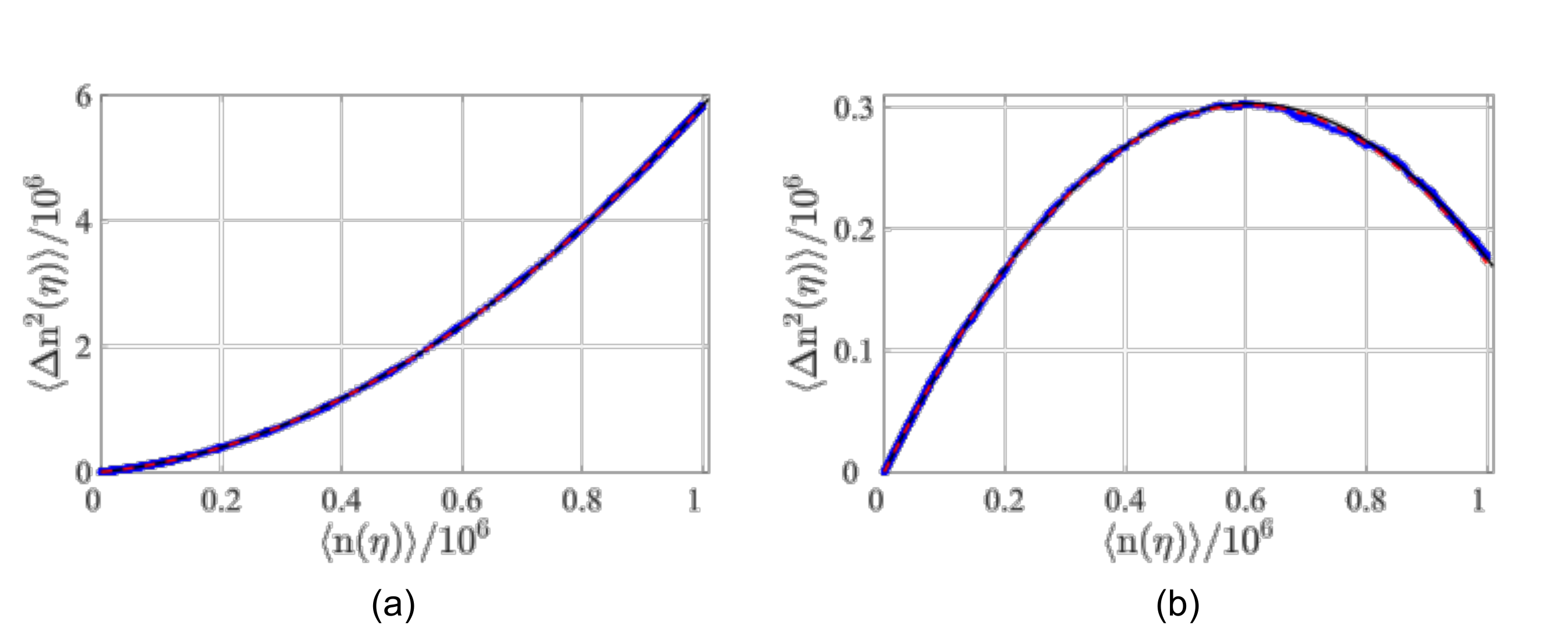}
    \caption{Theoretical prediction and simulation results for the squeezed light detection. (a) The simulation results of the photon number fluctuation (variance) in anti-squeezed light ($\varphi=\pi/2$) as a function of the intensity. The plotted results are obtained after the $10^4$ iterations. (b) The simulation results for squeezed light ($\varphi=0$). In both cases, the results are shown after integrating over pixels. The corresponding theoretical lines are also plotted for comparison. The state parameters are; $\bar{n}_\alpha=10^6$ and $\bar{n}_s=1$. The solid black line is a fit to second-order polynomial, and the dashed red line is the theoretical relation. The figure demonstrates an excellent agreement between our theory and simulations. This figure is reprinted from \textit{Physical Review Letters. 2020 Sep 10;125(11):113602} \cite{matekole2020quantum}, with the permission of APS Publishing.}
    \label{ch3_sqNsq}
\end{figure}

I will now discuss our simulation methods to measure the squeezing strength. In order to demonstrate our method, we simulate the experiment of measuring squeezed state with a camera. In short, a photon number is randomly picked, according to the DSV state photon statistics. The photons are distributed to $32\times32$ camera pixels. After repeating the simulation N times, the intensity and variance are computed for each pixel. The variance can be plotted as a function of intensity where each point in the plot is represented by a different pixel. To increase the precision of the results (without adding more data), one can integrate or group pixels. It can be done in many ways, and here we choose to integrate over many pixels such that the first point is the first pixel, the second point sums over the first two pixels, the third on three, and so on. The last point sums over all of the pixels. The simulated values of photon-number variance are plotted against the mean-photon number in Figure \ref{ch3_sqNsq}. Figure \ref{ch3_sqNsq}(a) and \ref{ch3_sqNsq}(b) show the cases of anti-squeezed and squeezed light, respectively. We can see that our numerical simulation results agree with the analytical results very well. This demonstrates that the amount of squeezing can be estimated from the first two moments (mean and variance) of the photon statistics obtained from the camera. We show that our method does very well, and is quantum-limited.

\section{Summary}
For several centuries, scientists have been trying to increase the precision of measurements. In the field of optical metrology, there has been an enormous interest in improving the sensitivity of parameter estimation beyond the classical limit called the shot-noise limit. Several seminal papers have been published since the 1970s that are aimed at pushing the limits of phase estimation using optical interferometry. However, the proposed schemes were not experimentally feasible with the available technology at the time. Since 2000, the focus has significantly shifted towards building applications that require experimentally feasible input states and detection strategies. In an attempt to make SU(1,1) metrology experiment-friendly and sensitive, I discussed our theoretical study on a SU(1,1) metrology scheme which uses two bright sources (coherent state, and displaced-squeezed vacuum state) and a simple detection strategy on-off.   

In this chapter, I first discussed the historical context of optical metrology and methods employed in MZI to push the phase sensitivity beyond SNL and HL. I laid out a detailed account of symplectic formalism for the Gaussian states. Furthermore, I showed how the quadrature mean and covariance can be used to simulate an optical system. Applying the symplectic formalism, I calculated the output Wigner function using the evolved mean and covariance. I utilized a property of the Wigner function to perform parity measurement, estimating the phase sensitivity of SU(1,1) interferometer with coherent states and displaced-squeezed vacuum states as the inputs. In addition to parity measurement, I presented the results of our SU(1,1) scheme with the on-off detection strategy. The on-off detection is very simple and less susceptible to effects like photon loss and thermal noise. Remarkably, we show a sub-shot-noise sensitivity of phase estimation by just using a click detector. In addition, I discussed our novel squeezed light detection technique which is simpler and faster compared to the conventional balanced homodyne techniques. Our technique only requires an unbalanced beamsplitter and a high-efficiency charge-coupled device (CCD) camera with high frame rates. This novel method exploits correlations in camera images to measure the squeezing strength of a squeezed light field. The theoretical and numerical simulation results presented in this chapter show an excellent agreement. The proposed technique is not only simple but also remarkably accurate.

%% file: chapter4.tex
\label{ch4}
The classification and characterization of light sources are very important capabilities for multiple photonic technologies. As discussed in Chapter \ref{ch2}, various states of light are characterized by unique Wigner functions, photon statistics, statistical fluctuations, and correlation properties in multipartite systems. Even though it seems relatively straightforward theoretically, the experimental characterization of states of light is a challenging problem that requires a large number of measurements. More precisely, the conventional techniques to classify the light sources requires hundreds of thousands of measurements. The identification of light sources becomes even harder when the mean photon numbers of the light fields are very low. The performance of many quantum photonic applications depends on our ability to perform single-photon-level measurements and efficient characterization of light sources. With the emergence of important quantum photonic technologies, the problem of identifying light sources has been even more relevant lately. In this chapter, I discuss our efforts to take a smart approach to this challenging problem, where we utilize the self-learning and self-evolving features of artificial neural networks to identify the coherent and thermal light sources at the single-photon level.  We demonstrate a dramatic reduction in the number of measurements required to efficiently discriminate the two light sources. In addition, we demonstrate similar performance with naive Bayes classifiers. Our work has important implications for multiple photonic technologies such as LIDAR and microscopy operating in the low-photon regime. 
This work has attracted a lot of attention from the quantum photonics and quantum optics community which is evident from the fact that a lot of similar and follow-up works have been published recently. We believe that this new research direction will eventually yield several smart quantum photonic technologies operating at the single-photon level.  

\section{Background and Motivation}
The underlying statistical fluctuations of the electromagnetic field have been widely utilized to identify diverse sources of light \cite{glauber1963quantum,mandel1995optical}. In this regard, the Mandel parameter constitutes an important metric to characterize the excitation mode of the electromagnetic field and consequently to classify light sources \cite{mandel1979sub}. Similarly, the degree of optical coherence has also been extensively used to identify light sources \cite{mandel1979sub, mandel1965coherence, liu2009n}. Despite the fundamental importance of these quantities, they require large amounts of data which impose practical limitations \cite{dovrat2012measurements,dovrat2013direct,zambra2005experimental}. This problem has been partially alleviated by incorporating statistical methods, such as bootstrapping, to predict unlikely events that are hard to measure experimentally \cite{dovrat2013direct, zambra2005experimental}. Unfortunately, the constraints of these methods severely affect photonic technologies for metrology, imaging, remote sensing, and microscopy \cite{dowling2015quantum, sher2018low, wang2016optimal, dowling2008quantum, omarm:2019}.

The potential of machine learning has motivated novel families of technologies that exploit self-learning and self-evolving features of artificial neural networks to solve a large variety of problems in different branches of science \cite{lecun:2015, biamonte:2017}. Conversely, quantum mechanical systems have provided new mechanisms to achieve quantum speedup in machine learning \cite{biamonte:2017,dunjko2016quantum}. In the context of quantum optics, there has been an enormous interest in utilizing machine learning to optimize quantum resources in optical systems \cite{hentschel2010machine,lumino2018experimental, melnikov2018active}. As a tool to characterize quantum systems, machine learning has been successfully employed to reduce the number of measurements required to perform quantum state discrimination, quantum separability and quantum state tomography \cite{sanjaya:2018, gao:2018, torlai2018neural}.

Here, we demonstrate the potential of machine learning to perform discrimination of light sources at extremely low-light levels. This is achieved by training artificial neural networks with the statistical fluctuations that characterize coherent and thermal states of light. The self-learning features of neural networks enable the dramatic reduction in the number of measurements and the number of photons required to perform the identification of light sources. Our experimental results demonstrate the possibility of using less than ten measurements to identify light sources with mean photon numbers below one. In addition, we demonstrate similar experimental results using the naive Bayes classifier, which are outperformed by our artificial neural network approach. Finally, we present a discussion on how artificial neural networks can dramatically reduce, by several orders of magnitude, the number of measurements required to discriminate signal photons from ambient photons. This possibility has strong implications for the realistic implementation of LIDAR, remote sensing, and microscopy.

\section{Experimental Setup}
\begin{figure}[ht!]
    \centering
    \includegraphics[width=0.80\textwidth]{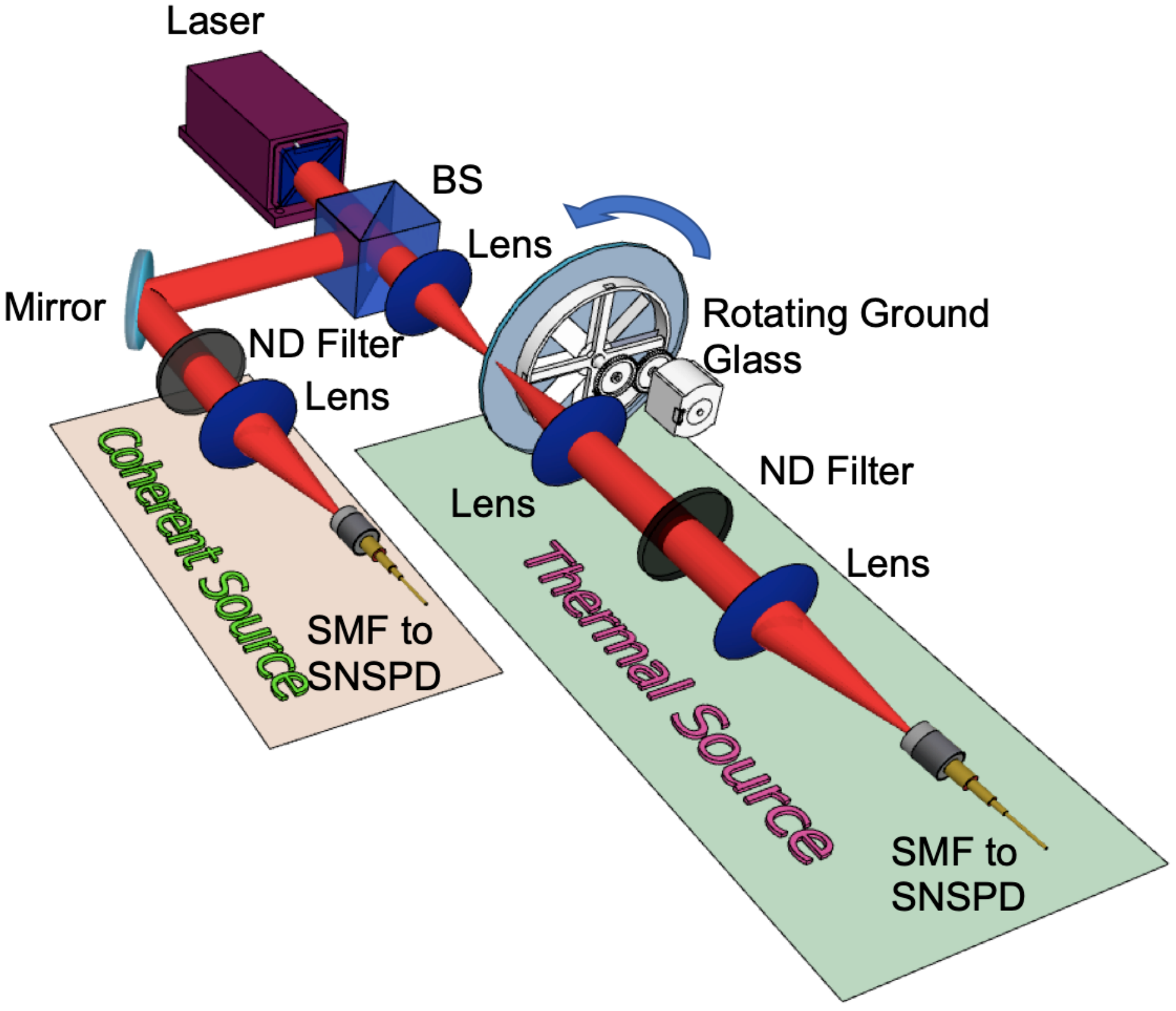}
    \caption{A schematic representation of the experimental setup. A laser beam is divided by a beam splitter (BS); the two replicas of the beam are used to generate light with Poissonian (coherent) and super-Poissonian (thermal) statistics. The thermal beam of light is generated by a rotating ground glass. Neutral density (ND) filters are utilized to attenuate light to the single-photon level. Coherent and thermal light beams are measured by superconducting nanowire single-photon detectors (SNSPDs). This figure is reproduced from reference \textit{Applied Physics Reviews 7.2 (2020): 021404} \cite{cyou:2020}, with the permission of AIP Publishing.}
    \label{ch4_sch}
\end{figure}

As shown in Figure \ref{ch4_sch}, we utilize a continuous-wave (CW) laser beam that is divided by a 50:50 beam splitter. The transmitted beam is focused onto a rotating ground glass which is used to generate pseudo-thermal light with super-Poissonian statistics. The beam emerging from the ground glass is collimated using a lens and attenuated by neutral-density (ND) filters to mean photon numbers below one. The attenuated beam is then coupled into a single-mode fiber (SMF). The fiber directs photons to a superconducting nanowire single-photon detector (SNSPD). The beam reflected by the beam splitter is used as a source of coherent light. This beam, characterized by Poissonian statistics, is also attenuated, coupled into an SMF, and detected by an SNSPD. The brightness of the coherent beam is matched to that of the pseudo-thermal beam of light.

In order to perform photon counting from our SNSPDs, we use the surjective photon counting method described in ref. \cite{rafsanjani2017quantum}. The TTL pulses produced by our SNSPDs were detected and recorded by an oscilloscope. The data was divided into time bins of 1 \textmu s, which corresponds to the coherence time of our CW laser. Voltage peaks above \textasciitilde0.5 V were considered as one photon event. The number of photons (voltage peaks) in each time bin was counted to retrieve photon statistics. These events were then used for training and testing our one-dimensional convolutional neural network (1D CNN) and naive Bayes classifier.

The probability of finding $n$ photons in coherent light is given by $P_{\text{coh}}(n)= e^{-\bar{n}} (\bar{n}^n/n!)$, where $\bar{n}$ denotes the mean photon number of the beam. Furthermore, the photon statistics of thermal light is given by $P_{\text{th}}(n)=\bar{n}^n / (\bar{n}+1)^{n+1}$.
It is worth noting that the photon statistics of thermal light is characterized by random intensity fluctuations with a variance greater than the mean number of photons in the mode. As described by their photon number probability distributions, coherent light and thermal light are different. For coherent light, the maximum of the photon-number probability sits around $\bar{n}$. For thermal light, the maximum is always at vacuum. However, when the mean photon number is low, the photon number distribution for both kinds of light becomes similar. Consequently, it becomes extremely difficult to identify one source from the other. The conventional approach to discriminate light sources makes use of histograms generated through the collection of millions of measurements \cite{dovrat2012measurements,zambra2005experimental,burenkov2017full,montaut2018compressive}. Unfortunately, this method is not only time-consuming, but also imposes practical limitations.

We establish the baseline performance for our 1D-CNN by using the naive Bayes classifier. This is a simple classifier based on Bayes' theorem \cite{gelman2013bayesian}. Throughout this letter, we assume that each measurement is independent. Moreover, we represent the measurement of the photon number sequence as a vector $\mathbf{x}=(x_1,...,x_k)$. Then, the probability of this sequence generated from coherent or thermal light is given by $p(C_j|x_1,...,x_k)$, where $C_j$ could denote either coherent or thermal light. Using Bayes' theorem, the conditional probability can be decomposed as $p(C_j|\mathbf{x}) = p(C_j) p(\mathbf{x}| C_j)/p(\mathbf{x})$. By using the chain rule for conditional probability, we have $p(C_k|x_1,...,x_k) = p(C_j) \prod_{i=1}^k p(x_i | C_j)$. Since our light source is either coherent or thermal, we assume $p(C_j)=0.5$. Thus, it is easy to construct a naive Bayes classifier, where one picks the hypothesis with the highest conditional probability $p(C_j|\mathbf{x})$. We used theoretically generated photon-number probability distributions as the prior probability $p(x_i | C_j)$, and used the experimental data as the test data.

\section{Computational Methods}
\begin{figure}[ht!]
    \centering
    \includegraphics[width=0.8\textwidth]{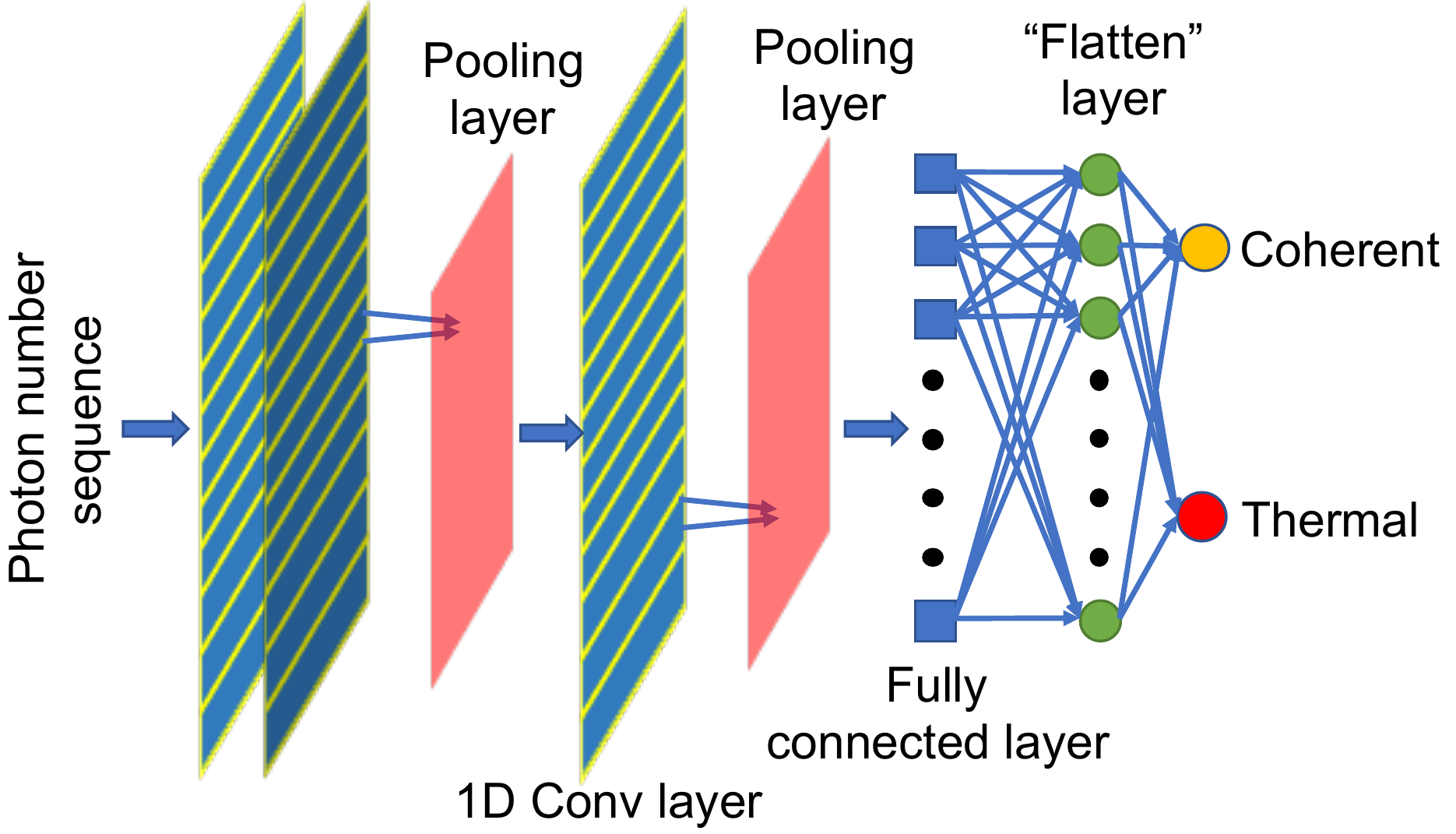}
    \caption{Schematic diagram of the computational algorithm: A schematic representation of the one-dimensional convolutional neural network (1D CNN) used in light source identification. This consists of a input layer, two convolution layers, two max-pooling layers, a fully connected flattening layer, and a output softmax layer. ReLU (rectified linear unit) is used as an activation function in each layer of this neural networks.This figure is reproduced from reference \textit{Applied Physics Reviews 7.2 (2020): 021404}, with the permission of AIP Publishing.}
    \label{ch4_ml}
\end{figure}

The schematic diagram of the machine learning algorithm utilized in this smart quantum technology is shown in Figure \ref{ch4_ml}. This architecture of the neural networks is called 1D convolutional neural networks (1D CNN), which is a deep learning algorithm. The beauty of this algorithm is that it can extract features or trends from a large volume of data without having to specify any theoretical model. More specifically, we are utilizing the 1D CNN to extract or recognize the features contained in a short segment of photon number measurements, which are not accessible through the conventional methods. First of all, our computational algorithm contains an input layer that accepts the photon statistics obtained experimentally. In order to ease the training process, we divide the training data into smaller batches of data points. In addition, we allocate a small fraction of the entire dataset for the testing purpose. The architecture consists of two convolutional layers each of which is immediately followed by a max-pooling layer. In the convolutional layers, different filters are applied which help to highlight different features from the given numerical sequences. The max-pooling layers are utilized to downsample the outputs from convolutional layers, leading to a computational simplification by removing redundant and unnecessary information. Finally, a fully-connected and a flattening layer precede the output layer consisting of two softmax functions, which output the probability distribution of the two labels \enquote{Coherent} and \enquote{Thermal}. It is important to note that we utilized ReLU (rectified linear unit) as the activation function in each layer of the neural networks. In all cases, we train the neural networks until 50 epochs.

\section{Photon Statistics of Light Sources}
\begin{figure}[hb!]
    \centering
    \includegraphics[width=1.0\textwidth]{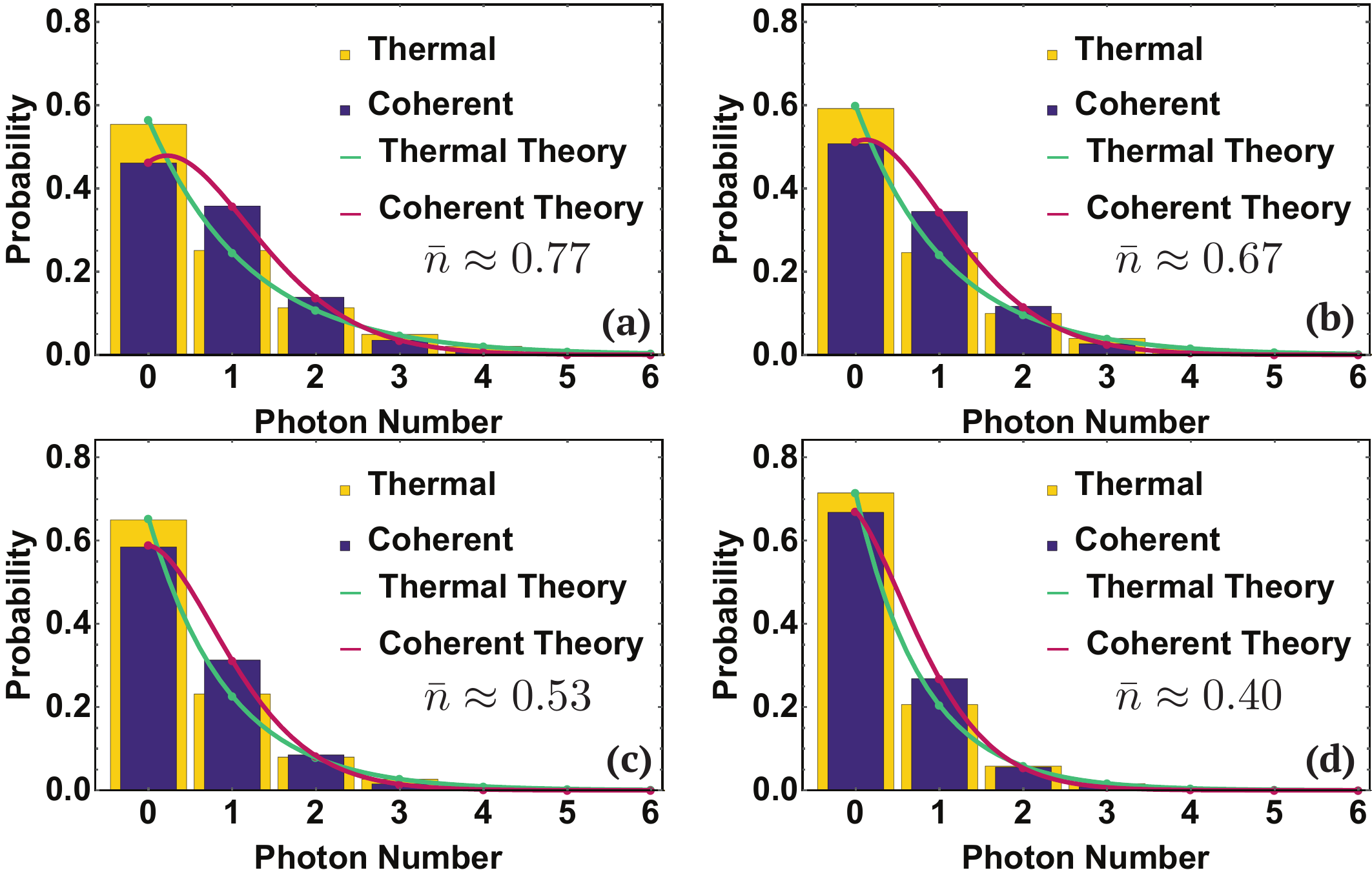}
    \caption{A set of histograms displaying theoretical and experimental photon number probability distributions for coherent and thermal light beams with different mean photon numbers. Our experimental results are in excellent agreement with theory. The photon number distributions illustrate the difficulty in discriminating light sources at low-light levels even when large sets of data are available}
    \label{ch4_phStat}
\end{figure}

As discussed in Chapter \ref{ch2}, different light sources are characterized by their photon statistics, Wigner functions, $g^2(0)$, quadrature fluctuations, photon number fluctuations, etc. In this chapter, I am limiting our discussion on discriminating the thermal light from the coherent light source. Naturally, they have different photon statistics and photon number fluctuations. One would think that this should make the task of identification relatively straightforward. However, the probability distribution of thermal light closely resembles the Poisson statistics for a weak light field having a mean photon number below 1. The histograms reconstructed using the experimental data are shown in Figure \ref{ch4_phStat}. In this figure, We compare the histograms for the theoretical and experimental photon number distributions for different mean photon numbers $\bar{n}=$ 0.40, 0.53, 0.67, and 0.77. The bar plots are generated with one million measurements for each source; the curves in each of the panels represent the expected theoretical photon number distributions for the corresponding mean photon numbers. Fig. \ref{ch4_phStat} shows excellent agreement between theory and experiment which demonstrates the accuracy of our surjective photon counting method. Furthermore, from Fig. \ref{ch4_phStat} (a)-(d), we can also observe the effect of the mean photon number on the photon number probability distributions.

\begin{figure}[hb!]
    \centering
    \includegraphics[width=0.60\textwidth]{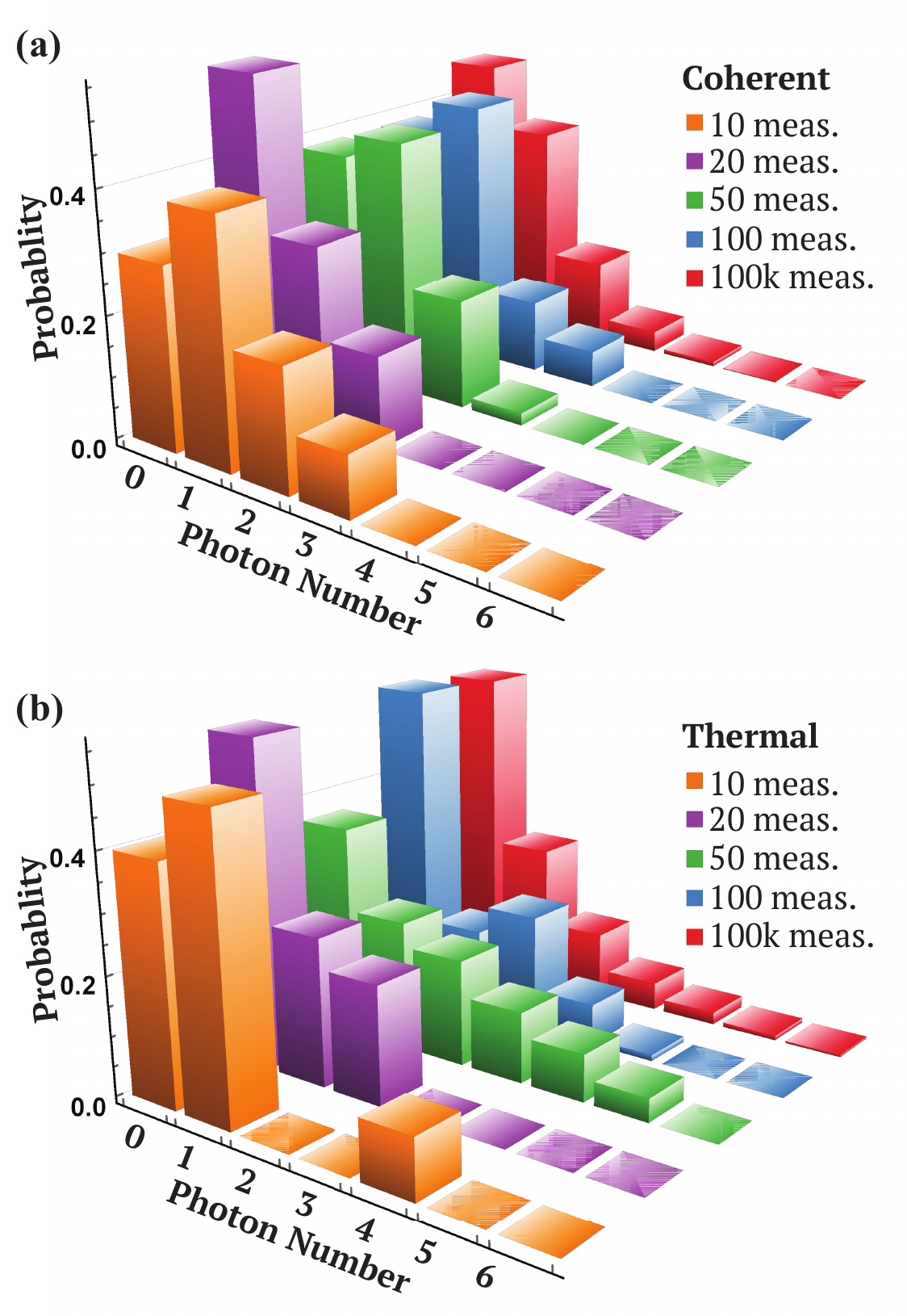}
    \caption{Probability distributions of coherent and thermal light,for varying dataset sizes (10, 20, 50, 100, 10k). Data used hereis randomly selected from of the measurement presented in Figure \ref{ch4_phStat}(a). This figure is reproduced from reference \textit{Applied Physics Reviews 7.2 (2020): 021404}, with the permission of AIP Publishing.}
    \label{ch4_samp}
\end{figure}

As shown in Figure \ref{ch4_phStat} (a), it is evident that millions of measurements enable one to discriminate two light sources. On the other hand, Figure \ref{ch4_phStat} (d) shows a situation in which the source mean-photon number is very low, which makes the discrimination of two light sources very cumbersome even with millions of measurements. In realistic applications, we always have limited resources. In order to further illustrate the difficulty of identifying light sources with the limited sets of data at low mean photon numbers, we plot the histograms obtained with a small subset of the data. We pick the random subsets of 10, 20, 50, 100, and 100000 data points from the large data set used in Figure \ref{ch4_phStat}(a). As shown in Figure \ref{ch4_samp}, the photon number distributions obtained with a limited number of measurements do not resemble those in the histograms shown in Fig. \ref{ch4_phStat} (a), for both coherent and thermal light beams.

\section{Source Discrimination with Naive Bayes}
\begin{figure}[hb!]
    \centering
    \includegraphics[width=0.90\textwidth]{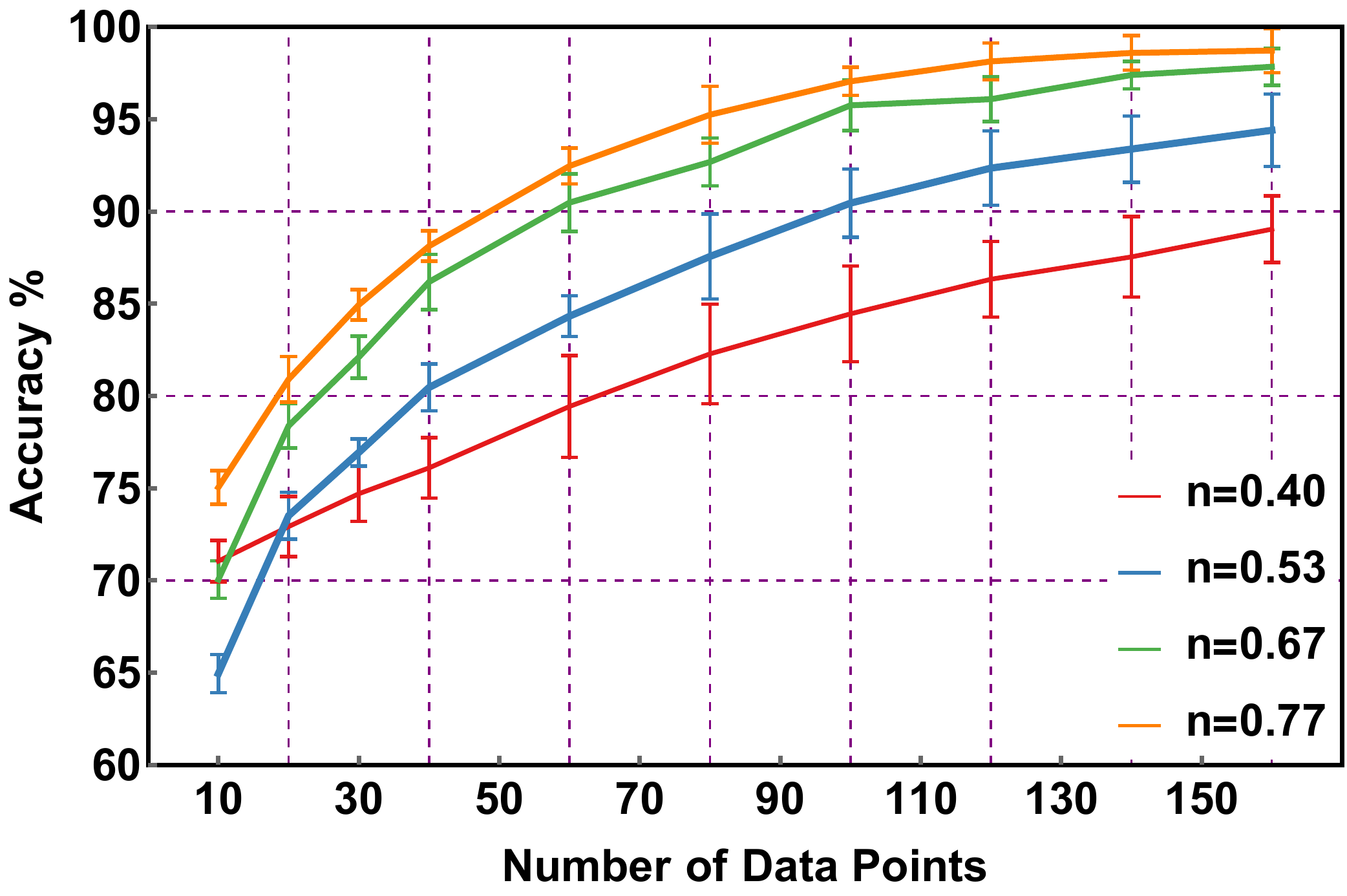}
    \caption{Overall accuracy of light discrimination vs the number of data points used in a naive Bayes classifier. The curves represent the accuracy of light discrimination for $\bar{n}:0.40$ (red line), $\bar{n}:0.53$ (blue line), $\bar{n}:0.67$ (green line), and $\bar{n}:0.77$ (orange line). The error bars are generated by dividing the test dataset into ten subsets. This figure is reproduced from reference \textit{Applied Physics Reviews 7.2 (2020): 021404}, with the permission of AIP Publishing.}
    \label{ch4_bayes}
\end{figure}

In Fig. \ref{ch4_bayes}, we show the overall accuracy for light source discrimination using the naive Bayes classifier. As expected, the accuracy increases with the number of data points used in a test sample. For example, when $\bar{n}=0.40$, the accuracy of discrimination increases from approximately 72\% to 88\% as we increase the number of data points from 10 to 160. It is worth noting that even with a small increase in the number of measurements, the naive Bayes classifier starts to capture the characteristic feature of different light sources, given by distinct sequences of photon number events. This is obvious since larger sets of data contain more information pertaining to the probability distribution. Furthermore, the mean photon number of the light field significantly changes the discrimination accuracy profile. As the mean photon number increases, the overall accuracy converges faster towards 100\% as expected. This is due to the fact that the photon number probability distributions become more distinct at higher mean photon numbers.

\section{Source Discrimination with 1-D CNN}
\begin{figure}[ht!]
    \centering
    \includegraphics[width=0.90\textwidth]{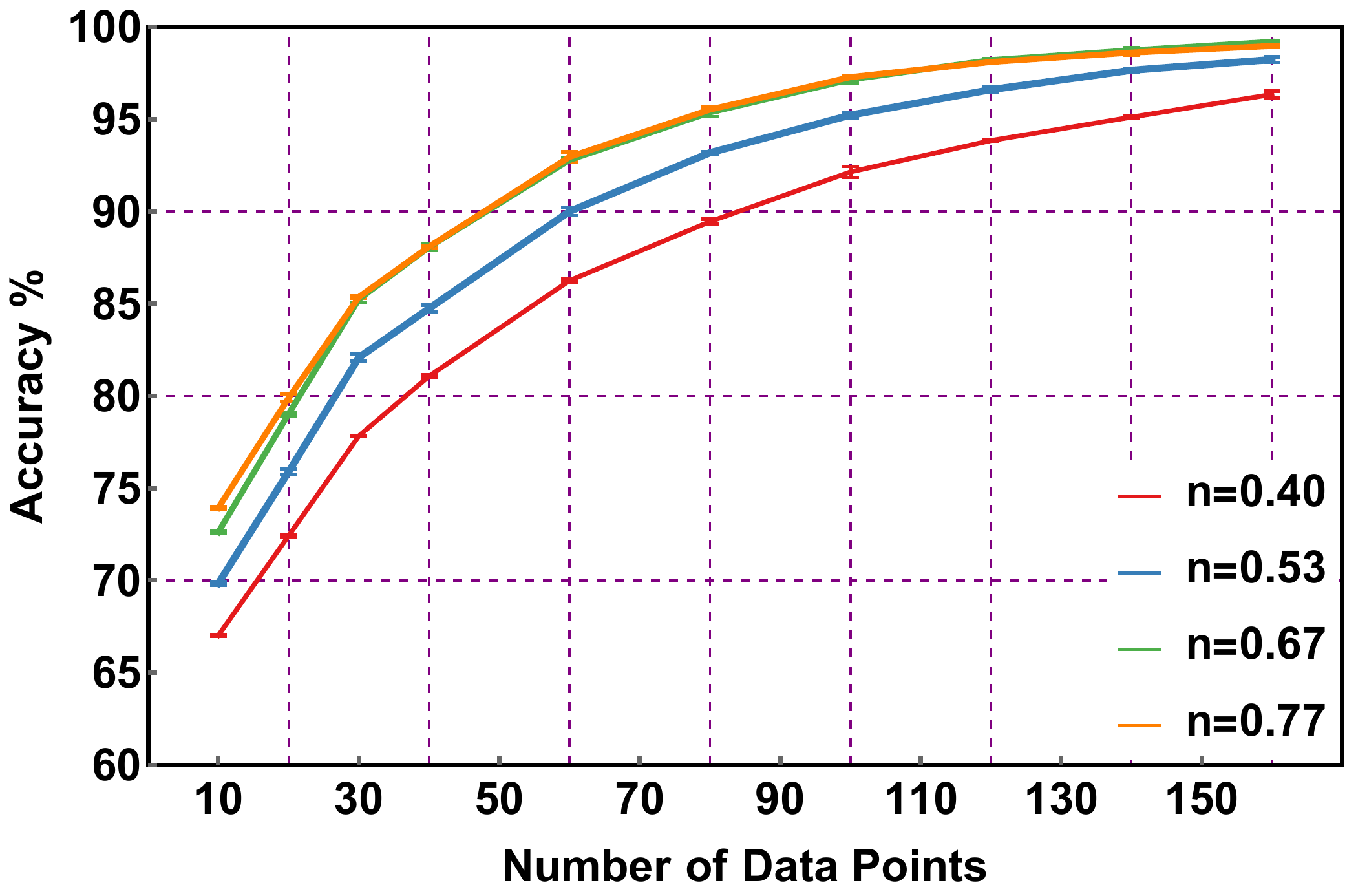}
    \caption{Overall accuracy of light discrimination vs the number of data points used in a 1D-CNN. The curves represent the accuracy of light discrimination for $\bar{n}:0.40$ (red line), $\bar{n}:0.53$ (blue line), $\bar{n}:0.67$ (green line), and $\bar{n}:0.77$ (orange line). The error bars represent the standard deviation of the training epochs for 1D-CNN. This figure is reproduced from reference \textit{Applied Physics Reviews 7.2 (2020): 021404}, with the permission of AIP Publishing.}
    \label{ch4_nn}
\end{figure}

The overall accuracy of light source discrimination with respect to the number of data points is shown in Figure \ref{ch4_nn}. Using only 10 data points, the 1D-CNN leads to an average accuracy between 65\%-75\% for $\bar{n}=0.40$, whereas when using 160 points of data, the accuracy is greater than 95\%. The comparison of Fig. \ref{ch4_bayes} and Fig. \ref{ch4_nn} reveals that the 1D-CNN outperforms naive Bayes classifier in general. Similar to naive Bayes classifier, 1D-CNN classifier accuracy increases with the number of data points and mean photon numbers. However, there are some clear distinctions between the 1D-CNN and naive Bayes classifier. The rate of convergence for the 1D-CNN classifier is significantly higher than that of the naive Bayes classifier. For low mean photon numbers such as $\bar{n}=0.40$, the improvement in accuracy scales linearly for naive Bayes classifier, as opposed to almost logistic growth that shows our 1D-CNN. Surprisingly, the accuracy for $\bar{n}=0.67$ and $\bar{n}=0.77$ overlaps; this shows that for a low mean photon number regime, the peak performance for 1D-CNN saturates much faster than naive Bayes classifier. Despite the vital differences between the performances of these two techniques at low mean photon numbers, they demonstrate similar overall accuracy at $\bar{n}=0.77$. These results suggest that for light discrimination at relatively high mean photon numbers, one could resort to the naive Bayes classifier, which requires less computational resources. However, when the light field has substantially low mean photon numbers, 1D-CNNs outperform naive Bayes classifier.

\section{Summary}
For more than twenty years, there has been an enormous interest in reducing the number of photons and measurements required to perform imaging, remote sensing, and metrology at extremely low-light levels. In this regard, photonic technologies operating at low-photon levels utilize weak photon signals that make them vulnerable to the detection of unwanted environmental photons emitted from natural sources of light such as sunlight. In fact, this limitation is identified as the biggest challenge for LIDAR technology operating at the ultra-low light level. Furthermore, a low mean photon number is inevitable while working with a photosensitive sample, which imposes experimental limitations on how long one can expose the sample without damaging it. Unfortunately, the conventional approaches to characterizing photon-fluctuations rely on the acquisition of a large number of measurements that impose constraints on the identification of light sources.

In this chapter, I presented a quantum light source discrimination technique based on machine learning. We demonstrated smart discrimination of light sources at mean photon numbers below one. Our protocol shows an improvement, in terms of the number of measurements, of several orders of magnitude with respect to conventional schemes for light identification. Our results indicate that 1D-CNN outperforms naive Bayes classifier at low-light levels. We believe that our work has important implications for multiple photonic technologies, such as LIDAR, low-light imaging, quantum state characterization, and microscopy of biological materials.

%% file: chapter5.tex
\label{ch5}
Spatial modes of light constitute valuable resources for a variety of quantum technologies ranging from quantum communication and quantum imaging to remote sensing. Nevertheless, their vulnerabilities to phase distortions, induced by random media, impose significant limitations on the realistic implementation of numerous quantum-photonic technologies. Unfortunately, this problem is exacerbated at the single-photon level. Over the last two decades, this challenging problem has been tackled through conventional schemes that utilize optical nonlinearities, quantum correlations, and adaptive optics. Here, we exploit the self-learning features of artificial neural networks to correct the complex spatial profile of distorted Laguerre-Gaussian modes at the single-photon level. Furthermore, we demonstrate the possibility of boosting the performance of an optical communication protocol through the spatial mode correction of single photons using machine learning. 
The results presented in this chapter have important implications for real-time turbulence correction of structured photons and single-photon images. In this chapter, I first lay out the background information and motivation for this research project. Then, I describe the details of our experimental setup of the free-space communication protocol. We assume a realistic strength of atmospheric turbulence and demonstrate how a smart feedback-based algorithm can be utilized to cope with the spatial mode distortions. The experimental setup is followed by the computational methods and relevant details of the optimization algorithm utilized in this project. Then the results of spatial mode correction are presented for the classical source of light. Furthermore, I present the spatial mode correction of heralded single photons generated through spontaneous parametric down-conversion. Then, I present a series of techniques like OAM state tomography, correlation matrix construction to quantitatively assess the performance of our communication protocol with and without the spatial mode correction. In addition, the channel capacities of the free-space communication channel for three different scenarios are calculated and compared.

\section{Background and Motivation}
\label{ch5Intro}
Spatially structured beams of light have been extensively used over the last two decades for multiple applications ranging from 3D surface imaging to quantum cryptography \cite{dunlop:2016,bell:1999,geng:2010}. In this regard, Laguerre-Gaussian (LG) modes represent an important family of spatial modes possessing orbital angular momentum (OAM) \cite{andrew:2012}. The OAM of photons is due to a helical phase front given by an azimuthal phase dependence of the form $e^{i\ell\phi}$, where $\ell$ represents the OAM number and $\phi$ represents the azimuthal angle. These beams have enabled the encoding of many bits of information in a single photon, a possibility that has enabled new communication and encryption protocols \cite{omar:2019,willner:2015,yao:2011,wang:2012,milione:2015,erhard2018twisted,giordani2019experimental}. In the past, these optical modes have been exploited to demonstrate high-speed communication in fiber, free-space, and underwater \cite{krenn2015twisted, baghdady:2016}. Furthermore, structured light beams have enabled increased levels of security against eavesdroppers, a crucial feature for secure communication applications \cite{bouren:2002, mirho:2015, grobla:2006, willner:2015, malik:2012}. Last but not least, structured spatial profiles of single photons have been proven to be extremely useful for remote sensing technologies and correlated imaging \cite{omar:2016, yang2017,cvijetic:2015, pattar:2013, omarm:2019, milione:2017, magana2014amplification,lavery:2013,chen:2014,jack:2009, malik:2014, scarcelli2006phase}.

Unfortunately, the spatial profile of photons can be easily distorted in realistic environments \cite{brandon:2012}. Indeed, random phase distortions and scattering effects can destroy information encoded in structured beams of light \cite{roden:2014, paterson:2005, tyler:2009}. Consequently, these spatial distortions severely degrade the performance of protocols for communication, cryptography, and remote sensing \cite{yao:2011, malik:2012}. These problems are exacerbated at the single-photon level, imposing important limitations on the realistic implementation of quantum photonic technologies. Hitherto, these limitations have been alleviated through conventional schemes that use adaptive optics, quantum correlations, and nonlinear optics \cite{hugo:2018,thomas:2018, nick:2019, roden:2014}. However, an efficient and fast protocol to overcome undesirable turbulence effects, at the single-photon level, has not yet been experimentally demonstrated.

Recently, artificial intelligence (AI) has gained popularity in optics due to its unique potential for handling complex classification and optimization tasks \cite{lin:2018, dunjko:2018, lecun:2015, biamonte:2017, sush:2020,cai2015entanglement,taira2020,romero2020variational}. Indeed, machine learning has been used to engineer quantum states of light \cite{Krenn:2016, cui:2019}, and to identify their properties in different degrees of freedom \cite{gao:2018, cyou:2020}. The use of machine learning is rapidly growing in multiple areas like quantum state tomography, quantum metrology, and optical communication \cite{polino2020photonic,torlai2018neural,carleo2019machine,lohani2020generative}. Moreover, convolutional neural networks have been demonstrated to be efficient in learning and characterizing the topographical features of images \cite{gu:2018}. An important number of recent articles have demonstrated the potential of artificial neural networks for efficient pattern recognition and identification of the spatial modes \cite{krenn2014communication, krenn2016twisted}. In addition to mode classification, artificial intelligence has enabled spatial mode de-multiplexing, which is important for harnessing multiple bits of information per photon \cite{doster2017machine,park2018multiplexing}. Furthermore, the self-evolving and self-learning features of artificial neural networks have been exploited to prepare, classify, and characterize quantum optical systems. Remarkably, for these particular tasks, machine learning techniques have outperformed conventional approaches \cite{turpin2018, sanj:2018, sanjaya:2018,liu2019deep}.

Here, we experimentally demonstrate a smart communication protocol that exploits the self-learning features of convolutional neural networks to correct the spatial profile of single photons. The robustness and efficiency of our scheme is tested in a communication protocol that utilizes LG modes. Our results dramatically outperform previous protocols that rely on conventional adaptive optics \cite{roden:2014, paterson:2005, tyler:2009, hugo:2018}. Furthermore, we demonstrate near-unity fidelity in time periods that are comparable to the fluctuation of atmospheric turbulence. Our results have significant implications for various  technologies that exploit spatial modes of single photons \cite{yao:2011,omar:2019,willner:2015,jin2020spectrally}. In addition, our work shows a potential to enable the possibility of overcoming phase distortions induced by thick atmospheric turbulence.

\section{Experimental Setup}
\begin{figure}[hb!]
    \centering
    \includegraphics[width=1.0\textwidth]{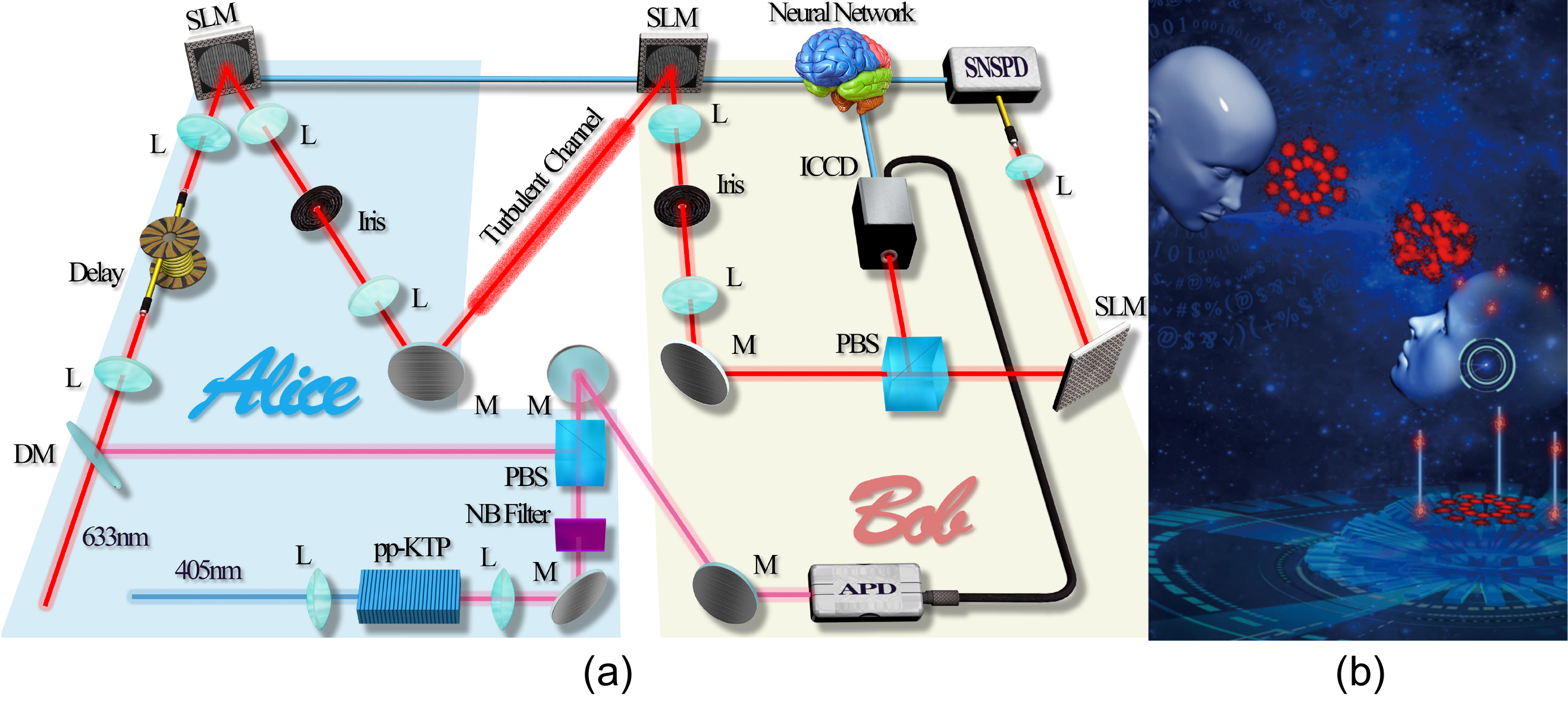}
    \caption{(a) Schematic diagram of the experimental setup for the turbulence correction of Laguerre-Gaussian (LG) modes. The experiment is carried out for a Helium-Neon (He-Ne) laser and spontaneous parametric down-conversion (SPDC) source which are switched using a dichroic mirror (DM). The first part of the setup, labeled as Alice, is dedicated to the generation of the LG modes using a spatial light modulator (SLM). The second part, labeled as Bob,  is set up to perform the experimental simulation of turbulence, turbulence correction, and orbital angular momentum (OAM) tomography with the help of a superconducting nanowire single-photon detector (SNSPD). We use a charged coupled device (CCD) and intensified charged coupled device (ICCD) cameras to acquire images while using laser and SPDC sources respectively. (b) A concept diagram of the communication scheme showing information exchange between Alice and Bob who is equipped with artificial intelligence (AI) to correct the distorted spatial modes. The figure (a) is reprinted from the preprint version of \textit{Advanced Quantum Technologies. 2021 Jan; 2000103} \cite{bhusal2020spatial}, with the permission of Wiley Publishing. Similarly, the concept diagram in (b) is from the front cover of the March 2021 issue \cite{bhusal2021front}.}
    \label{ch5_schExp}
\end{figure}
Atmospheric turbulence is one of the most influential sources of phase distortion in free-space propagation. We demonstrate a turbulence correction protocol for LG modes. The schematic diagram of our experimental setup is depicted in Figure \ref{ch5_schExp}(a). The experimental setup consists of two parties -- Alice and Bob. Here, Alice prepares spatial modes that are transmitted to Bob through a turbulent communication channel. The atmospheric turbulence in the communication channel induces aberrations in the optical beams that degrade the quality of the information encoded in their phase. This undesirable effect compromises Bob's ability to correctly decode and make measurements on the spatial modes. Bob overcomes this problem by training an artificial neural network with multiple turbulence-distorted beams that allow him to correct the spatial profile of photons. Figure \ref{ch5_schExp}(b) depicts the concept diagram of our protocol in which Bob uses AI to boost the communication efficiency by correcting the modes sent by Alice through a turbulent free-space communication channel.

In our experiment, we use a spatial light modulator (SLM) and computer-generated holograms to produce LG modes \cite{forbes2016creation}. This technique allows us to generate any arbitrary spatial mode in the first-diffraction order of the SLM. The generated modes are filtered and collimated using a 4f-optical system and then projected onto a second SLM. We use this second spatial modulator to display random phase screens that simulate turbulence \cite{roden:2014}. The beam reflected by the second SLM is then split into two beams using a polarizing beam splitter (PBS). The spatial profiles of the beams reflected by the PBS are recorded by a CCD camera. Bob collects 50 distorted modes for one specific superposition of spatial modes transmitted through a turbulent channel. The communication channel is characterized by a standard refractive index $C_n^2$. Then, 45 of these images are used as a training set and the remaining 5 are used as a test set. Each of the experimental images has a resolution of 400$\times$400 pixels, then each image is down-sampled to form a matrix of 128$\times$128 pixels before the CNN. Once the neural network is optimized, Bob utilizes the CNN to predict the turbulence strength and the initial correction phase masks. The initial phase masks are then optimized by minimizing the MSE using the GDO algorithm. Furthermore, Bob utilizes the same correction masks for the single-photon and high-light level implementation of our protocol. This is possible given the fact that the turbulence of the communication channel is independent of the number of transmitted photons. Naturally, turbulence characterization using single photons requires longer integration times. On the other hand, the beam transmitted by the PBS is characterized through quantum state tomography with the help of a superconducting nanowire single-photon detector (SNSPD).

Over the past two decades, the possibility of performing image correction at the single-photon level has represented one of the main goals of the quantum imaging community \cite{omar:2019}. Due to the relevance of single-photon imaging for multiple applications \cite{omar:2019,mirhosseini2016wigner,chen:2014,jack:2009, malik:2014}, we also perform a proof-of-principle experiment using heralded single photons produced by a process of spontaneous parametric down-conversion (SPDC). We utilize a dichroic mirror (DM) to ease the transition from one source to another as shown in Fig. \ref{ch5_schExp}(a). We use an SPDC source generated by pumping a potassium titanyl phosphate (ppKTP), a $\chi^{(2)}$ nonlinear crystal with type-II phase matching, with a high-intensity laser beam of the wavelength 405 nm. We utilize a narrowband (NB) filter to remove the pump photons from the downconverted photon pairs generated at the wavelength 810 nm which are quantum correlated. A PBS is used to separate the correlated photon pairs in our experiment as they are found in relative orthogonal polarization.

Imaging at the single-photon level is experimentally a challenging task due to a very low signal-to-noise ratio. In addition to building a shielded housing for the intensified charged coupled device (ICCD) camera, it is important to utilize the triggering mechanism of the camera. The electrical signal generated by avalanche photodiode detector (APD) in response to the photons incident from the transmission arm of the PBS is utilized as the external trigger to the ICCD camera. A high transistor-transistor logic (TTL) pulse in the trigger signal heralds the arrival of the photon in the camera. Full width half maximum (FWHM) for the signals generated by APD was recorded to be 17 ns which determines the time that the camera shutter remains open for each trigger signal.  However, we account for the internal delay of the ICCD camera by adding additional optical fiber to perform gated imaging. This technique of gating is crucial in acquiring images at the single-photon level as it can filter out a significant number of background photons arriving at the random time window \cite{fickler2013real}. This improves the signal to noise ratio (SNR) of the acquired images which is highly desirable especially when you are trying to perform imaging at very low light levels.

\section{Computational Methods}
\begin{figure}[ht!]
    \centering
    \includegraphics[width=0.75\textwidth]{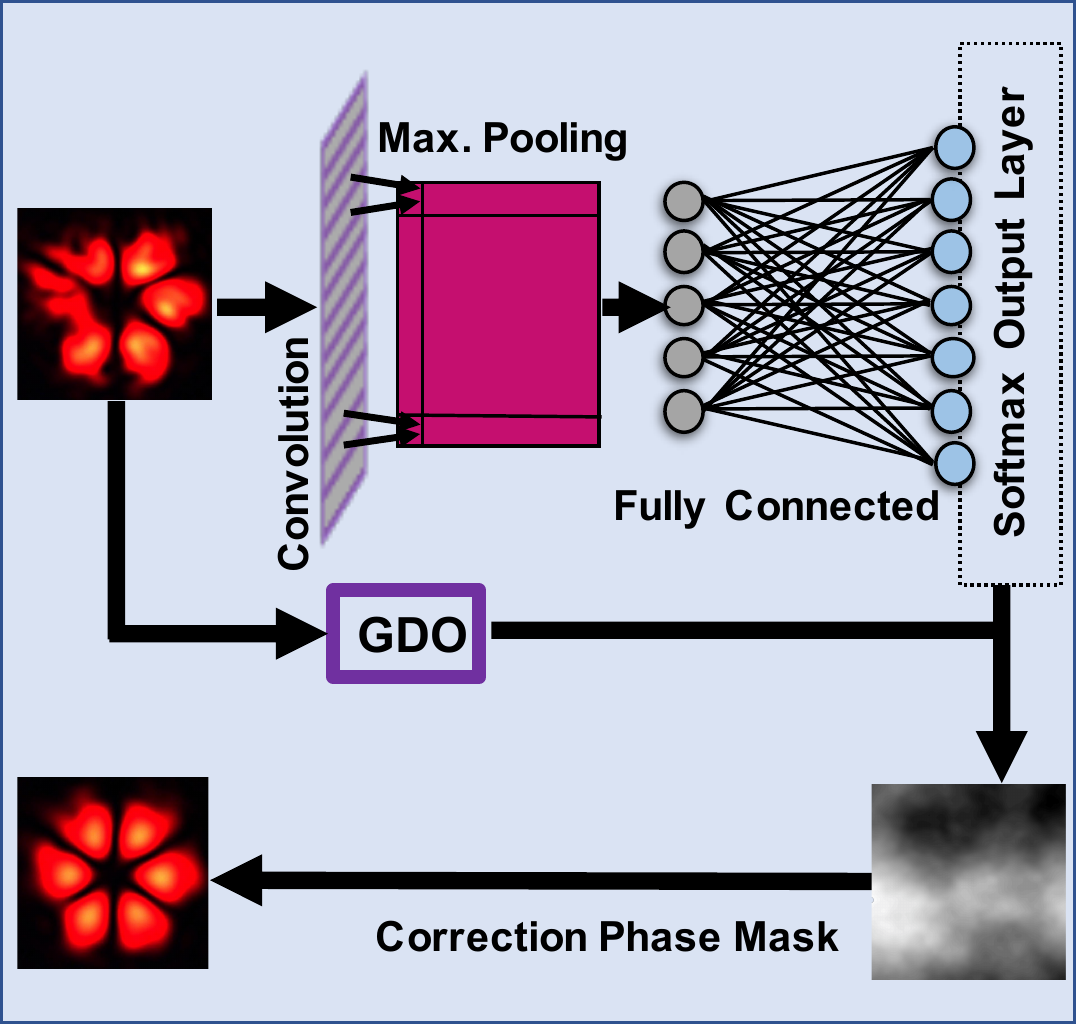}
    \caption{Computational model for generating turbulence correction phase mask. This comprises a 5-layer convolutional neural network (CNN), and a feedback loop with gradient descent optimizer (GDO). This figure is reprinted from \textit{Advanced Quantum Technologies. 2021 Jan; 2000103}, with the permission of Wiley Publishing.}
    \label{ch5_schML}
\end{figure}
After the experimental setup, let us now discuss the computational algorithm we utilize in this smart quantum technology. Figure \ref{ch5_schML} illustrates our machine learning algorithm for the correction of structured photons. This is based on a convolutional neural network followed by a gradient descent optimizer \cite{sanjaya:2018, nielsen:2015}. The optimizer consists of a 5-layer CNN and a gradient descent optimization (GDO) algorithm \cite{ruder2016overview}. The CNN takes turbulent images of LG modes and convolves them with a $5\times5$ filter. The step is immediately followed by a $2\times2$ max-pooling layer before feeding them into 100 fully connected neurons. Finally, the network contains a softmax output layer. We utilize hundreds of realizations of distorted images for multiple turbulence strengths to train the neural networks. The function of the trained CNN is to predict the strength of turbulence in terms of standard refractive index ($C_n^2$) values. The predicted values of $C_n^2$ are fed into the additional GDO which forms a feedback loop in our experiment. The function of the GDO loop is to optimize the correction phase masks over many realizations of random matrices that simulate turbulence. The phase masks are then encoded in the second SLM to obtain the corrected spatial modes at the image plane of the SLM.

We prepare symmetric superpositions of LG modes to demonstrate smart optical communication. This family of modes are solutions to the Helmholtz equation in cylindrical coordinates \cite{omar:2019}. In our experiment, we distort the communication modes by using atmospheric turbulence simulated in an SLM \cite{brandon:2016}. We use the Kolmogorov model of turbulence to simulate the turbulent communication channel \cite{roden:2014,sanjaya:2018,Bos:2015}. Turbulence induces a random modulation of the index of refraction that results from inhomogeneities of temperature and pressure of media. This, in turn, leads to distortions of the phase front of the spatial profile of optical modes. The degree of distortion is quantified through the Fried's parameter $r_0$, which is defined in terms of the standard refractive index $C_n^2$,\begin{equation}
\label{eq:1}
    \Phi(p,q)=\mathbb{R}\left\{\mathcal{F}^{-1}\left(\mathbb{M}_{NN}\sqrt{\phi_{NN}(k)}\right )\right\},
\end{equation}
with $\phi_{NN}(k)$=$0.023 r_0^{-5/3} \left(k^2+k_0^2\right)^{-11/6} e^{-k^2/k_m^2}$ and the Fried's parameter in terms of the standard refractive index as $r_0$=$\left(0.423 k^2 C_n^2d\right)^{-3/5}$. The mathematical symbol $\mathbb{R}$ represents the real part of the complex field, 
whereas $\mathcal{F}^{-1}$ indicates the inverse Fourier transform operation. Furthermore, $k$, $d$, and $\mathbb{M}_{NN}$ denote the wave number ($2\pi/\lambda$), the propagation distance, and the encoded random matrix, respectively. Even though the strength of phase distortion can be varied using $d$ and $C_n^2$, we choose to vary its strength using $C_n^2$. Furthermore, we perform the phase mask optimization iteratively using the GDO algorithm
\begin{equation}
\begin{aligned}
\label{eq:2}
\Phi^{j}(p, q)= \angle\left[\mathcal{F}^{-1}\left\{\frac{1}{H} \times \mathcal{F}\left[\mathcal{F}^{-1}\left(\mathcal{F}\left(G\left(p, q, w_{0}\right) \exp\left(i \Theta^{(\ell,-\ell)}\right)\right) H\right)\exp \left(-i \Phi_{\mathrm{est}}^{j}(p, q)\right)\right]\right\}\right].
\end{aligned}
\end{equation}

\noindent The mean squared error (MSE) between the predicted intensity and the corresponding simulated target intensity is used as the cost function. In this case, the symbol $\angle$ represents the complex phase defined by $\arctan\left(\mathbb{I}/\mathbb{R}\right)$, with $\mathbb{I}$ describing the imaginary part of the complex field. Moreover, $\mathcal{F}$ indicates a Fourier transform operation, and $\Phi^{j}(p,q)$ the phase mask at the $j^{\text{th}}$ iteration. The Gaussian beam $G(p,q,w_0)$ is characterized by a waist $w_0$, and the transfer function describing the SLM transformation together with the propagation function of the beam is represented by $H$. The phase mask used to generate the original LG superposition mode is described by $\Theta^{(\ell,-\ell)}$ in Equation (\ref{eq:2}).

\section{Spatial Mode Correction}
\begin{figure}[ht!]
    \centering
    \includegraphics[width=1.0\textwidth]{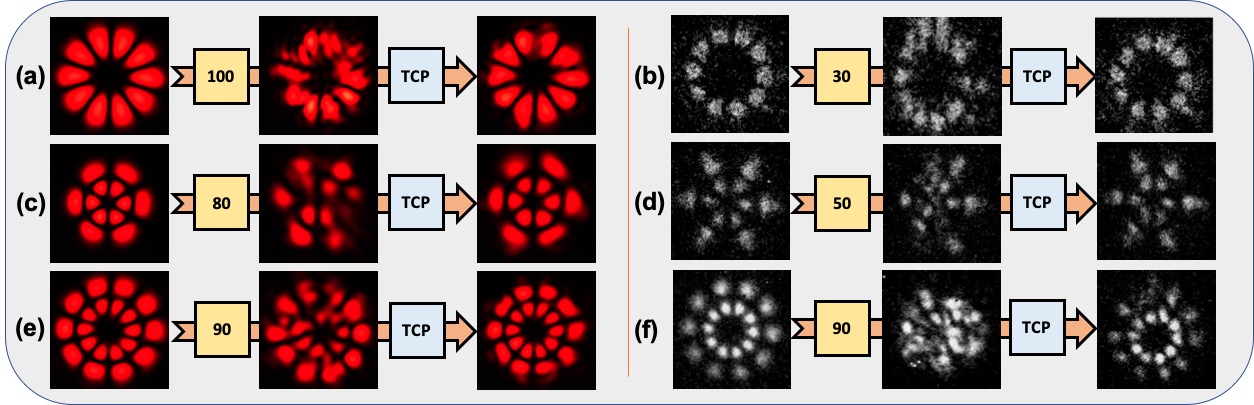}
    \caption{Spatial profiles of LG modes at high- and single-photon levels for different turbulence conditions. The first column in each of the panels shows the states prepared by Alice without distortions. The second columns display the distorted beams measured by Bob. The strength of turbulence is characterized by $C_n^2$ ($\times 10^{-13} \text{mm}^{-2/3})$, these numbers are reported in the yellow rectangle. The spatial profiles after our turbulence correction protocol (TCP) are shown in the third column. (a), (b), and (c) show high-light-level demonstrations of our protocol for multiple LG superpositions,  $\ket{\psi}=\frac{1}{\sqrt{2}}\left( \ket{\text{LG}_{+5,0}}+\ket{\text{LG}_{-5,0}}\right)$, $\ket{\psi}=\frac{1}{\sqrt{2}}\left( \ket{\text{LG}_{+3,1}}+\ket{\text{LG}_{-3,1}}\right)$, and $\ket{\psi}=\frac{1}{\sqrt{2}}\left( \ket{\text{LG}_{+5,1}}+\ket{\text{LG}_{-5,1}}\right)$ respectively. The corresponding single-photon demonstrations of (a), (b) and (c) are shown in (d), (e), and (f), respectively. This figure is reprinted from \textit{Advanced Quantum Technologies. 2021 Jan; 2000103}, with the permission of Wiley Publishing.}
    \label{ch5_corrImgs}
\end{figure}

\begin{figure}[ht!]
    \centering
    \includegraphics[width=0.90\textwidth]{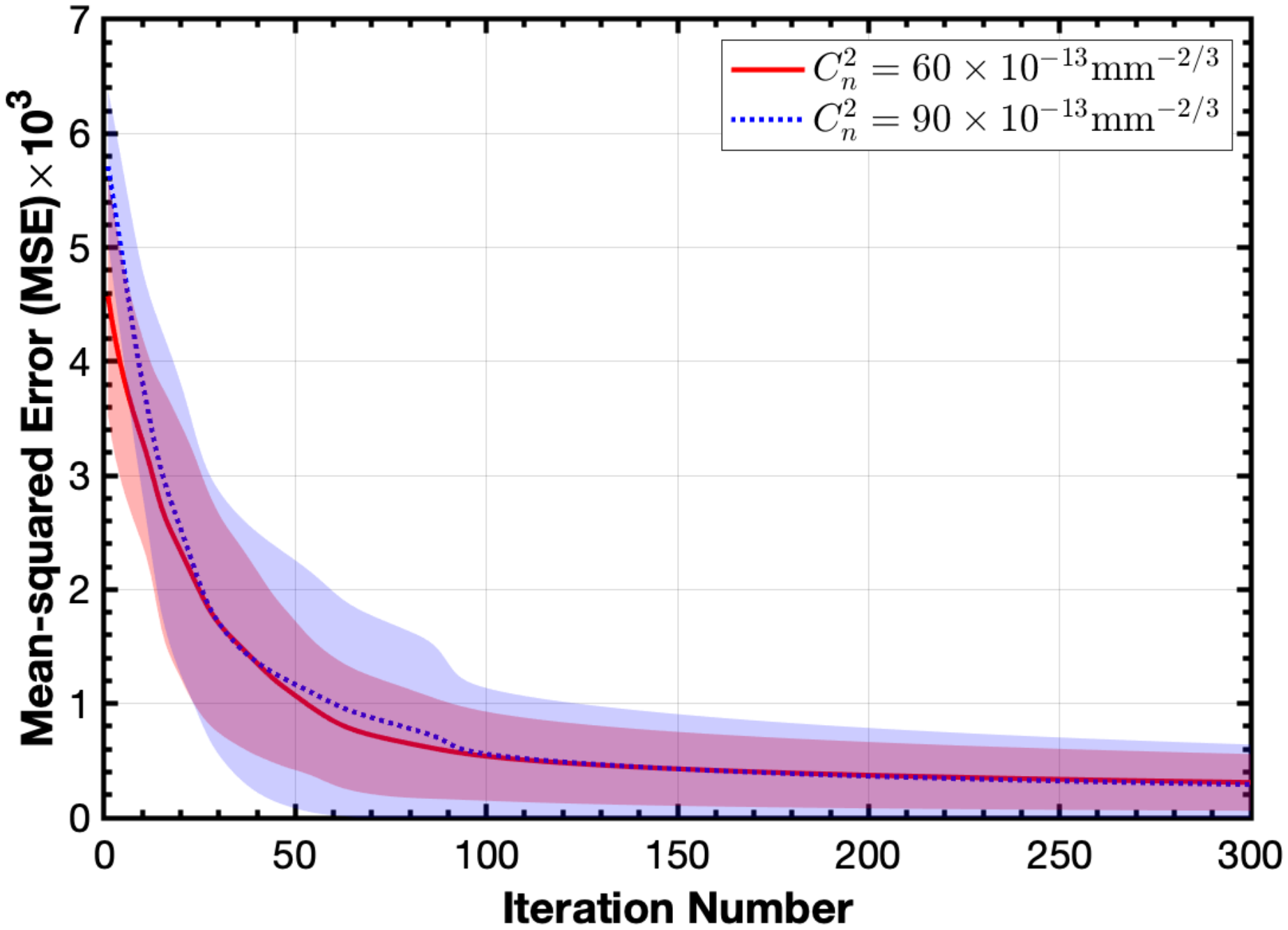}
    \caption{Mean-squared error (MSE) versus the iteration number in GDO. The red and blue lines indicate the MSE values for turbulence strengths $C_n^2$ = 60$\times 10^{-13} \text{mm}^{-2/3}$ and 90$\times 10^{-13} \text{mm}^{-2/3}$ respectively. The number of iterations needed for the algorithm to achieve convergence depends on the strength of turbulence. In general, stronger turbulence requires longer times to converge. This figure is reprinted from \textit{Advanced Quantum Technologies. 2021 Jan; 2000103}, with the permission of Wiley Publishing.}
    \label{ch5_mse}
\end{figure}

In Figure \ref{ch5_corrImgs}(a)-(c), we present experimental results obtained with a He-Ne laser. The first column in each of the panels shows the spatial profile of the undistorted modes prepared by Alice. The spatial profiles of the modes are distorted due to atmospheric turbulence in the communication channel. The aberrated modes are shown in the second column of Figure \ref{ch5_corrImgs}. In the experiment, Bob collects hundreds of realizations of the aberrated beams to train the artificial neural network in Figure \ref{ch5_schML}. The strength of turbulence predicted by our CNN was utilized to perform the phase mask optimization by means of a feedback GDO loop. Thus, the CNN in combination with the GDO loop generates the correction phase masks which are then encoded in the second SLM to alleviate turbulence effects. We indicate this process with the blue box labeled as \enquote{TCP} in Figure \ref{ch5_corrImgs}. The CNN was trained in a high-performance computing cluster. The pre-trained CNNs are used to estimate the turbulence strength and initial phase distribution in few milliseconds. The pre-trained CNNs and GDO are run on a computer with an Intel(R) Core(TM) i7-8750H CPU $@$ 2.20GHz and 16 GB of RAM to generate optimized turbulence correction phase masks. In order to show the performance of our artificial neural network, in Figure \ref{ch5_mse} we plot the mean-squared error (MSE) as a function of the iteration number. This plot allows for a qualitative comparison of our protocol with other adaptive optics techniques \cite{roden:2014, xie2015phase}. Naturally, the number of iterations required for convergence depends on the strength of turbulence. Our protocol shows a similar performance to other adaptive optics protocols, see \cite{roden:2014, xie2015phase}. Nevertheless, the standard refractive index ($C_n^2$) values are orders of magnitude higher. The MSE starts to converge near 100 iterations for the turbulence strengths used in the experiment, see Figure \ref{ch5_mse}. This process enables Bob to obtain optimized phase masks which are used to correct turbulence-induced distortions. The corrected intensity profiles measured by Bob are depicted in the last column of each panel in Figure \ref{ch5_corrImgs}. In Figure \ref{ch5_corrImgs}(a), we show the spatial profile of a structured beam corrected by our protocol for the superposition of LG modes $\ket{\psi}=\frac{1}{\sqrt{2}}\left( \ket{\text{LG}_{+\ell,0}}+ \ket{\text{LG}_{-\ell,0}}\right)$ with $\ell=5$. In Figure \ref{ch5_corrImgs}(b) and \ref{ch5_corrImgs}(c), we show experimental results for complex LG modes, with radial structure, described by $\ket{\psi}=\frac{1}{\sqrt{2}}\left( \ket{\text{LG}_{+\ell,1}}+ \ket{\text{LG}_{-\ell,1}}\right)$ for $\ell=3$ and $\ell=5$ respectively.

We also demonstrate the robustness of our technique to correct the spatial profile of heralded single photons. In Figure \ref{ch5_corrImgs}(d), \ref{ch5_corrImgs}(e), and \ref{ch5_corrImgs}(f), we display turbulence correction of single photons prepared in LG superpositions with different azimuthal and radial quantum numbers, expressed as $\ket{\psi}=\frac{1}{\sqrt{2}}\left( \ket{\text{LG}_{+5,0}}+\ket{\text{LG}_{-5,0}}\right)$, $\ket{\psi}=\frac{1}{\sqrt{2}}\left( \ket{\text{LG}_{+3,1}}+\ket{\text{LG}_{-3,1}}\right)$, and $\ket{\psi}=\frac{1}{\sqrt{2}}\left( \ket{\text{LG}_{+5,1}}+\ket{\text{LG}_{-5,1}}\right)$ respectively. These images were acquired using an ICCD camera. Each of the background-subtracted images are formed by accumulating photons over a time period of 20 minutes. These images demonstrate an excellent mitigation of the turbulence at the single-photon level. It is important to note that the implementation of the protocol in real-time is the ultimate goal. However, we would like to emphasize the fact that speed and collection time in our experiment are limited by our computational resources and the performance of our equipment. Thus, the overall reported speed in our manuscript is not a fundamental constraint nor a problem of our protocol. Indeed, it is possible to speed-up our scheme by replacing our commercial ICCD camera with a fast single-photon camera with nanosecond resolution such as the one described in ref. \cite{montaut2018compressive}.

\section{OAM State Tomography}
\label{ch5Tomo}
\begin{figure}[ht!]
    \centering
    \includegraphics[width=0.65\textwidth]{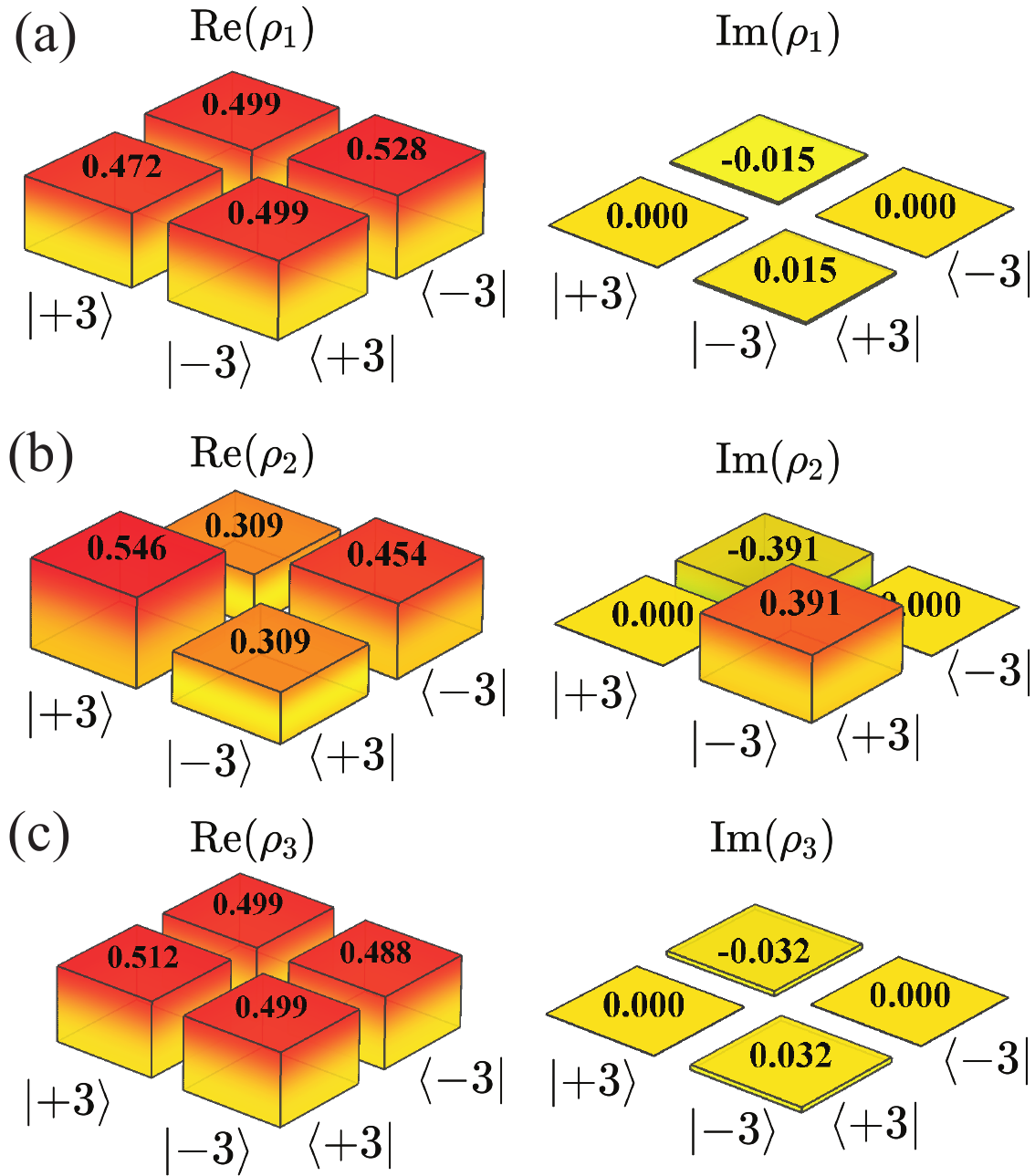}
    \caption{Real and imaginary parts of the density matrices for the qubits encoded in the OAM basis. In this case, we prepared $\ket{\psi_3}=\frac{1}{\sqrt{2}}\left(\ket{\text{LG}_{+3,0}}+\ket{\text{LG}_{-3,0}}\right)$. (a) shows the real and imaginary parts of the density matrix for the undistorted state. In (b) we show the density matrix for the aberrated qubit. In this case, the strength of the simulated turbulence is characterized by $C_n^2$ = 80$\times 10^{-13} \text{mm}^{-2/3}$. In (c) we show the density matrix for the corrected qubit after applying our turbulence correction protocol. We measured fidelity of 99.9\% for the prepared state, 81.7\% for the distorted state, and 99.8\% for the state corrected through our machine learning protocol. This figure is reprinted from \textit{Advanced Quantum Technologies. 2021 Jan; 2000103}, with the permission of Wiley Publishing.}
    \label{ch5_tomo}
\end{figure}
In order to certify the spatial correction of single photons and the recovery of spatial coherence, we perform quantum state tomography of the spatial modes \cite{mirhosseini2016wigner}. Quantum state tomography is a process of reconstructing a density matrix that uniquely identifies the quantum state in the Bloch/Poincaré sphere. This process involves a series of projective measurements on the six principal states along the three axes. For this purpose we use superpositions of the following form, $\ket{\psi_{\ell}}=\alpha \ket{\text{LG}_{+\ell,0}}+\beta \ket{\text{LG}_{-\ell,0}}$, where $\alpha$, and $\beta$ represent complex amplitudes \cite{kwiat:2005,daniel:2001,adrien:2015,brandon:2016}. For simplicity, in our experiment we use the following spatial qubit $\ket{\psi_3}=\frac{1}{\sqrt{2}}\left(\ket{\text{LG}_{+3,0}}+\ket{\text{LG}_{-3,0}}\right)$. In Figure \ref{ch5_tomo}(a), \ref{ch5_tomo}(b), and \ref{ch5_tomo}(c), we show the real and imaginary parts of the reconstructed density matrices in the absence of turbulence, with turbulence, and after applying turbulence correction, respectively. As shown in Figure \ref{ch5_tomo}(a), in this case, all the elements of the real part of the density matrix should be equal to $\frac{1}{2}$, and the matrix elements of the imaginary part should be 0. The presence of any deviation from that is attributed to experimental imperfections. Furthermore, Figure \ref{ch5_tomo}(b) shows the detrimental effects produced by turbulence. The strength of turbulence in this case is $C_n^2$ = 80$\times 10^{-13} \text{mm}^{-2/3}$. Remarkably, after applying our machine learning protocol, we recover the original state almost perfectly. The density matrices in Figure \ref{ch5_tomo} certify the robustness of our technique. We quantify the fidelity using $F=\left(\text{Tr}\sqrt{\sqrt{\rho_{1}}\rho_{3} \sqrt{\rho_1}}\right)^2$, where $\rho_1$ and $\rho_3$ represent the density matrices of the original and turbulence corrected spatial qubits. In fact, the fidelity between two quantum states measures the overlap between them. The measured fidelity for the prepared state is 99.9\%, whereas that of the distorted state is 81.7\%. Remarkably, the fidelity for the state corrected through our machine learning protocol is 99.8\%. These numbers reaffirm the robustness of our machine-learning-based approach to tackle the problem of random phase distortion in spatial modes.

\section{Correlation Matrix Reconstruction}
\begin{figure}[ht!]
    \centering
    \includegraphics[width=1.0\textwidth]{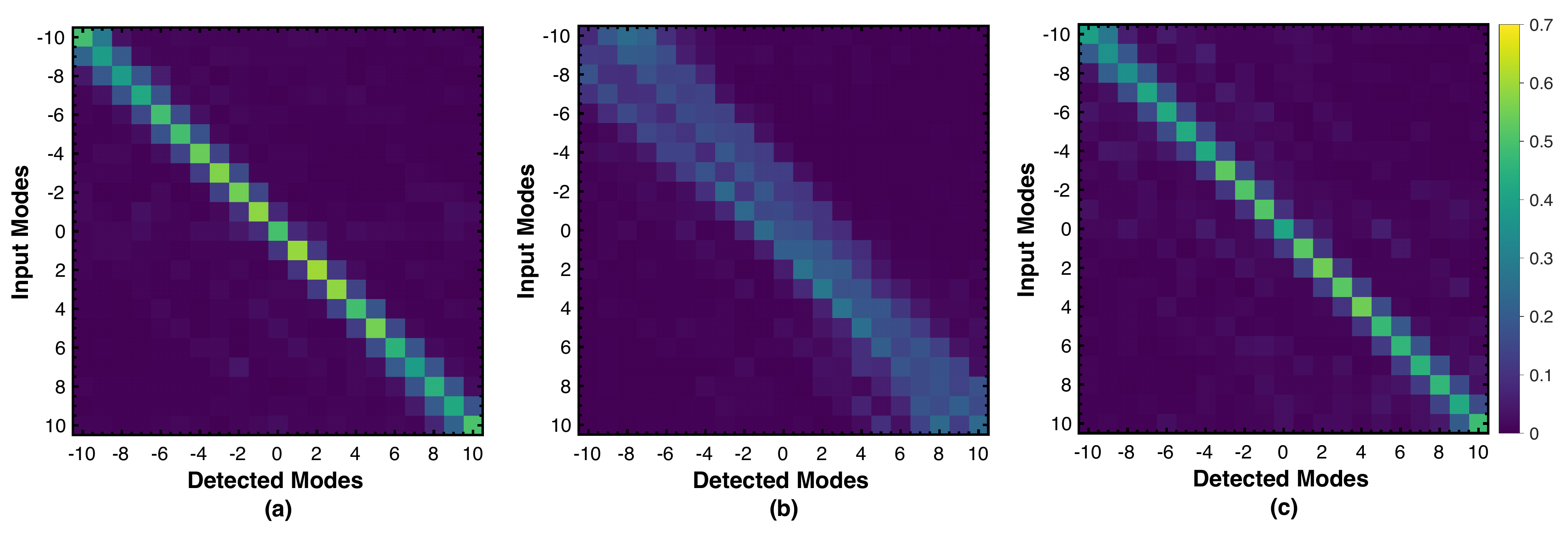}
    \caption{The cross-correlation matrices represent conditional probabilities between sent and detected modes in the OAM basis. (a) shows the cross-correlation matrix obtained for our communication protocol in the absence of turbulence. In (b), we plot the cross-correlation matrix for in the presence of turbulence characterized by $C_n^2$ = 90$\times 10^{-13} \text{mm}^{-2/3}$. In this case, it is almost impossible to correctly identify the spatial modes. (c) shows the cross-correlation matrix after applying our turbulence correction protocol. This figure is reprinted from \textit{Advanced Quantum Technologies. 2021 Jan; 2000103}, with the permission of Wiley Publishing.}
    \label{ch5_cormat}
\end{figure}

The OAM state tomography discussed in Section \ref{ch5Tomo} is limited in the sense that it only reconstructs the density matrices in 2-dimensional Hilbert space. In realistic turbulent environments, there is always a possibility of inducing cross-talk between the modes outside just the nearest neighbors. Therefore, it is important to evaluate the performance of our turbulence correction protocol on the expanded Hilbert space. For this purpose, a series of projective measurements are performed on the original, distorted, and corrected OAM states. Figure \ref{ch5_cormat}(a) shows the cross-correlation matrix for different transmitted modes in the absence of turbulence. In order to generate this matrix, Bob performs a series of projective measurements on the modes sent by Alice. The cross-correlation matrix represents the conditional probabilities between the modes sent and detected in the communication protocol. A small spread around the diagonal elements even in the absence of turbulence is caused due to diffraction, the finite size of the optical fibers, and experimental misalignment. The cross-correlation matrix obtained in the presence of atmospheric turbulence is shown in Figure \ref{ch5_cormat}(b). In this case, the spatial distortion induces modal cross-talk that degrades the performance of the communication protocol. These undesirable effects increase with the strength of turbulence in the communication channel. Indeed, this represents an important limitation of free-space communication with spatial modes of light \cite{omar:2019}. In Figure \ref{ch5_cormat}(c), we show our experimental results for the cross-correlation matrix after applying our turbulence correction protocol. In this case, the cross-correlation matrix is nearly diagonal, showing a dramatic improvement in the performance of our communication protocol. 

\section{Mutual Information of the Channel}
\begin{figure}[hb!]
    \centering
    \includegraphics[width=0.95\textwidth]{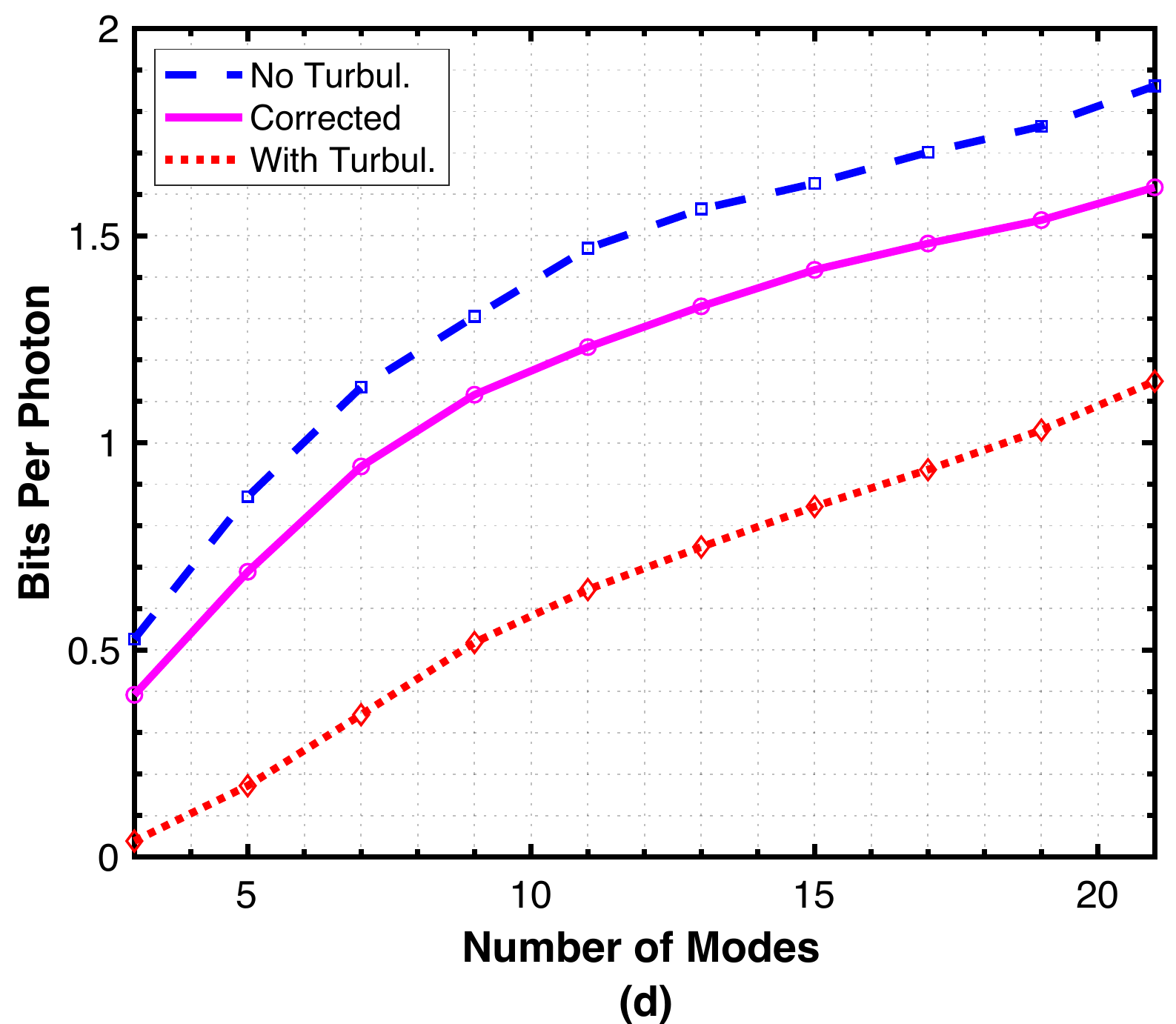}
    \caption{Our turbulence correction protocol significantly improves the performance of the communication system. We display channel capacity in terms of bits per photon in this figure. This figure is reproduced from \textit{Advanced Quantum Technologies. 2021 Jan; 2000103}, with the permission of Wiley Publishing.}
    \label{ch5_MI}
\end{figure}
As discussed in Section \ref{ch5Intro}, one of the important perks of using the spatial modes of light is the ability to encode multiple bits of information per photon. Theoretically, there is no upper bound to the capacity of such a free-space channel. However, the experimental limitations like the growing size of the beam vortex, susceptibility to phase distortions prevent us from utilizing very high $\ell$ values. In order to quantify the performance of our correction protocol through the channel capacity of our optical communication system, we calculate the normalized mutual information. This allows us to quantify the channel capacity in terms of bits per photon \cite{roden:2014} as shown in Figure \ref{ch5_MI}. We use the conditional probabilities of the cross-correlation matrices to calculate the mutual information for a high-dimensional Hilbert space according to the following equation 
\begin{equation}
    \text{MI}=\frac{1}{N} \sum_{d,s} P(d \mid s) \log _{2}\left(\frac{P(d \mid s) N}{\sum_{s} P(d \mid s)}\right),
\end{equation}
where the dimension is described by the parameter $N$, and the subscripts $d$ and $s$ represent the detected and sent modes respectively. Here, $P(d \mid s)$ denotes the conditional probability of detecting the state in spatial mode $d$, given mode $s$ is sent by Alice. The channel capacity plot demonstrates the potential of our technique to correct spatial modes of light.

\section{Summary}
Spatial photonic modes have been in the spotlight for the past few decades due to their enormous potential as quantum information resources. However, these modes are fragile and vulnerable to random phase fluctuations induced by turbulence.  Unfortunately, these problems are exacerbated at the single-photon level. Indeed, the fragility of spatial modes of photons imposes important limitations on the realistic implementation of optical technologies in free-space. In this work, we have experimentally demonstrated the first smart communication protocol that exploits the self-learning features of convolutional neural networks to correct the spatial profile of single photons. This work represents a significant improvement over conventional schemes for turbulence correction. The high fidelities achieved in the reconstruction of the spatial profile of single photons make our technique a robust tool for free-space quantum technologies. We believe that our work has important implications for the realistic implementation of photonic quantum technologies.

%% file: chapter6.tex
\label{ch6}
I began this dissertation by presenting a discussion about the history of optics and quantum technologies. In Chapter \ref{ch1}, I described our motivations to incorporate artificial intelligence in photonic technologies operating at the single-light-level. In Chapter \ref{ch2}, I extensively reviewed the fundamental concepts behind the quantum technologies discussed in the subsequent chapters. The chapters \ref{ch3}, \ref{ch4}, and \ref{ch5} are reproduced from our recent publications. In the final chapter, Chapter \ref{ch6}, I wrap up this dissertation by summarizing our main results and their implications for the future of quantum technologies.

Photonic quantum metrology is one of the most successful and promising quantum technologies. The interference of electromagnetic fields is the most commonly used approach to estimate the phase shift, which is inaccessible through the direct intensity measurement. Measuring physical parameters as precisely as possible is the cornerstone of metrology. In Chater \ref{ch3}, I laid out the theoretical details of a nonlinear metrology scheme that surpasses the classical limit called shot-noise. In Section \ref{ch31}, I presented a sub-shot-noise limited quantum metrology scheme utilizing two different measurement protocols. We demonstrated that by injecting a displaced-squeezed vacuum state in the vacuum port of the SU(1, 1) interferometer, we can substantially improve the sensitivity of phase estimation. In this nonlinear metrology scheme, we utilized parity measurements and on/off measurements as the detection strategy to achieve sub-shot-noise limited sensitivity. Furthermore, in the case of parity measurement, the phase sensitivity was shown to approach the Heisenberg limit. However, it is important to note that the parity measurement is very sensitive to photon loss. Therefore, a simple on/off detection, which is more resilient to photon loss, was also investigated as a more experimentally feasible alternative. Interestingly, with the displaced-squeezed vacuum and coherent state as inputs to SU(1, 1) interferometer, phase sensitivity was shown to surpass the classical limit \enquote{shot-noise limit} for reasonably achievable squeezing strengths. In addition, in Section \ref{ch32}, I discussed our novel efforts to detect squeezed light by utilizing spatial correlations on hundreds of camera images separated in time. The conventional methods to measure the squeezing strength of the squeezed vacuum state such as Wigner function reconstruction and balanced homodyne detection are considered cumbersome. Our technique provides a smart and convenient alternative to these techniques. Remarkably, our technique makes squeezing detection effortless without compromising the accuracy of the measurement.

The efficiency of quantum photonic technologies is heavily dependent on the innovations in the generation, control, and measurement of the quantum states of light. In the last few decades, we have made significant progress on all three fronts. Nevertheless, there are still a significant number of challenges to truly realizing efficient, and reliable quantum photonic applications. In Chapter \ref{ch4}, I presented our novel efforts to exploit artificial intelligence to improve discrimination of coherent light and thermal light. Our ability to discriminate thermal light from coherent light is crucial for applications like quantum microscopy, and LIDAR which are operated ideally at the single-photon level. The presence of thermal light, such as sunlight, in the probe signal, is considered one of the important hurdles to the realistic implementation of quantum LIDAR. The conventional technique to discriminate a thermal light from a coherent light source relies on the reconstruction of photon statistics from a large number of measurements. In our widely appreciated work, we demonstrated that the self-evolving and self-learning features of artificial neural networks can dramatically reduce the number of measurements necessary to identify the light sources. Remarkably, we have shown that even a few tens of measurements are sufficient to achieve the near-perfect discrimination of thermal light source and coherent light source.

Spatial modes of light are valuable resources for a variety of quantum technologies like quantum communication, quantum sensing, quantum imaging, and remote sensing. Spatial modes, orbital angular momentum modes, in particular, are viewed as an important dimension to encode quantum information because they allow us to encode multiple bits of information per photon. The access to infinite-dimensional Hilbert space enables the increased level of security against eavesdropper in quantum cryptography. The higher dimensional quantum states prepared in OAM superposition are utilized in high-speed optical communication protocols. However, the spatial modes of light are extremely susceptible to distortions. Therefore, controlling the state of spatial modes of light is a challenging task. Conventionally, these challenges are tackled by utilizing optical nonlinearities, quantum correlations, and adaptive optics. In Chapter \ref{ch5}, I presented our efforts to smartify the correction of spatial modes of single photons, where we experimentally demonstrate that the self-learning and self-evolving features of artificial neural networks can efficiently correct the complex spatial profiles of distorted Laguerre-Gaussian modes at the single-photon level. Our results have important implications for a variety of quantum technologies based on spatial modes of light. The results become even more important in the single-photon regime because the effects of distortions are extremely detrimental at the low-light-level.

In a nutshell, photonic technologies have been playing a crucial role in our lives through multiple avenues like science, technology, innovation, healthcare. Quantum photonics has a huge potential to revolutionize optical technologies we rely on daily. Effective implementation of quantum communication, quantum cryptography, quantum sensing, and quantum imaging can constitute a paradigm shift in the ways we communicate with each other and measure certain things. However, as of now, our ability to generate, control, and measure the quantum states of light impose significant limitations. In this dissertation, I presented our series of efforts to integrate ultra-fast cameras and artificial intelligence in quantum detection and quantum control strategies. We demonstrated that these newly introduced smart quantum photonic technologies substantially outperform conventional methods.